\documentclass[sigconf]{acmart}

\makeatletter
\def\@ACM@checkaffil{%
    \if@ACM@instpresent\else
    \ClassWarningNoLine{\@classname}{No institution present for an affiliation}%
    \fi
    \if@ACM@citypresent\else
    \ClassWarningNoLine{\@classname}{No city present for an affiliation}%
    \fi
    \if@ACM@countrypresent\else
        \ClassWarningNoLine{\@classname}{No country present for an affiliation}%
    \fi
}
\makeatother

\renewcommand\footnotetextcopyrightpermission[1]{}

\usepackage{algorithm}
\usepackage[noend]{algpseudocode}

\usepackage{subfig}
\usepackage{graphicx}

\usepackage{tabularx}

\usepackage{bbm}
\usepackage{hyperref}

\usepackage{dirtytalk}
\usepackage{lipsum}

\usepackage{enumitem}

\usepackage{tcolorbox}
\usepackage{tikz}

\usepackage{xcolor,xspace}

\newcommand{\edit}[1]{#1}

\algnewcommand\algorithmicinput{\textbf{Input:}}
\algnewcommand\algorithmicoutput{\textbf{Output:}}
\algnewcommand\Input{\item[\algorithmicinput]}%
\algnewcommand\Output{\item[\algorithmicoutput]}%

\usepackage[normalem]{ulem}

\newcommand{\tabitem}{~~\llap{\textbullet}~~}

\AtBeginDocument{%
  \providecommand\BibTeX{{%
    \normalfont B\kern-0.5em{\scshape i\kern-0.25em b}\kern-0.8em\TeX}}}

\newcommand{\norm}[1]{\left\lVert#1\right\rVert}

\newcommand{\R}{\mathbb{R}}
\newcommand{\Z}{\mathbb{Z}}

\newcommand{\prob}{\mathbb{P}}
\newcommand{\E}{\mathbb{E}}
\newcommand{\bolds}[1]{\boldsymbol{#1}}

\newtheorem{fact}{Fact}[section]

\begin{document}

\newpage
\twocolumn

\title{Federated Boosted Decision Trees with Differential Privacy}
\titlenote{Full version of a paper to appear at ACM CCS'22}

\author{Samuel Maddock}
\authornote{Author correspondence to \href{mailto:s.maddock@warwick.ac.uk}{\nolinkurl{s.maddock@warwick.ac.uk}}}
\affiliation{%
  \institution{University of Warwick}
}

\author{Graham Cormode}
\affiliation{%
  \institution{Meta AI}
}

\author{Tianhao Wang}
\authornote{Work was done while part-time at Meta AI}
\affiliation{%
 \institution{University of Virginia}
}

\author{Carsten Maple}
\affiliation{%
 \institution{University of Warwick}
}

\author{Somesh Jha}
\authornotemark[3]
\affiliation{%
 \institution{University of Wisconsin-Madison}
}

\begin{abstract}
There is great demand for scalable, secure, and efficient privacy-preserving machine learning models that can be trained over distributed data. While deep learning models typically achieve the best results in a centralized non-secure setting, different models can excel when privacy and communication constraints are imposed. Instead, tree-based approaches such as XGBoost have attracted much attention for their high performance and ease of use; in particular, they often achieve state-of-the-art results on tabular data. Consequently, several recent works have focused on translating Gradient Boosted Decision Tree (GBDT) models like XGBoost into federated settings, via cryptographic mechanisms such as Homomorphic Encryption (HE) and Secure Multi-Party Computation (MPC).  
However, these do not always provide formal privacy guarantees, or consider the full range of hyperparameters and implementation settings. In this work, we implement the GBDT model under Differential Privacy (DP). We propose a general framework that captures and extends existing approaches for differentially private decision trees. Our framework of methods is tailored to the federated setting, and we show that with a careful choice of techniques it is possible to achieve very high utility while maintaining strong levels of privacy.
\end{abstract}


\maketitle
\pagestyle{plain}

\section{Introduction}\label{sec:introduction}
It is well known that machine learning models can leak private information about \edit{individuals in the training set} %
\cite{shokri2017membership, carlini2021membership}. 
Differential privacy (DP) \cite{dwork2014algorithmic} is a popular definition that has been developed to mitigate such privacy risks and has become the dominant notion of privacy in recent years. 
Much of the current research on private machine learning is focused on training deep learning models with differential privacy \cite{abadi2016deep, kairouz2021practical, kurakin2022imagenet, stock2022defending}. 
DP is often combined with federated learning, where data resides on client devices, and only small information about model updates is collected from clients, in order to further minimize the privacy risk \cite{kairouz2021advances}. 

While deep learning models are powerful for a range of real-world tasks in a centralized setting, they are sometimes beaten by \say{simpler} models on tabular datasets. 
One such competitor is Gradient Boosted Decision Trees (GBDTs) %
\cite{shwartz2022tabular, elsayed2021we, gorishniy2021revisiting} \edit{\cite{grinsztajn2022tree}}. 
GBDT methods build an ensemble of weak decision trees that incrementally correct for past mistakes in training to \edit{improve} %
predictions. 
Many GBDT frameworks such as XGBoost \cite{xgboost2016}, LightGBM \cite{ke2017lightgbm}, and CatBoost \cite{dorogush2018catboost} have seen widespread industry adoption \cite{BoE2019, Bracke2019, KPMG2020, salesforce2022}. 
GBDT methods are an attractive alternative to deep learning due to their speed, scalability,  ease of use, and impressive performance on \edit{tabular datasets}. %

Recent works have studied GBDT implementations such as XGBoost under secure training in the federated setting \cite{cheng2021secureboost, feng2019securegbm, deforth2021xorboost, meng2020privacy}. 
These methods typically rely on cryptographic techniques such as Homomorphic Encryption (HE) or Secure Multi-Party Computation (MPC). 
While this allows secure joint training of a GBDT model without any participant directly releasing their data, the end model may not necessarily be private and will not guarantee formal differential privacy (DP) \cite{fang2021large}.
For instance, in the case of decision trees, split decisions in a tree can directly reveal sensitive information regarding the training set. 
Moreover, such reliance on heavyweight cryptographic techniques such as HE or MPC often makes methods computationally intensive or require a large number of communication rounds, making them impractical to scale beyond more than a few participants \cite{cheng2021secureboost, law2020secure}.

In parallel, many works have studied decision tree models under the central model of DP \cite{fletcher2015differentially, rana2015differentially, zhao2018inprivate}. 
Most studies focus on training random forest (RF) models and there has been little research to explore trade-offs between gradient boosting and DP; those that do often use central DP mechanisms that cannot easily be extended to federated settings \cite{grislain2021dp}.
It therefore remains an open problem to implement GBDTs in the federated DP setting, and show how to obtain utility comparable to their centralized non-private counterparts. 

Our focus is on DP-GBDT methods that operate within the federated setting via lightweight MPC methods such as secure aggregation \cite{bonawitz2017secureAgg, bell2020secure}. 
This setting has recently risen to prominence, as it promises an attractive trade-off between the computational efficiency of central DP techniques and the security of cryptographic methods. 
Recent federated works that consider GBDTs have proposed methods under the local model of DP, but due to the use of local noise, incur a significant loss in utility \cite{tian2020federboost, le2021fedxgboost, wang2021feverless}.

In this paper, we bring together existing methods under a unified framework where we propose techniques to satisfy DP that are well suited to the federated setting. 
We find that by dissecting the GBDT algorithm into its constituent parts and carefully considering \edit{the} %
options for each component of the algorithm we can identify specific combinations that achieve the best balance of privacy and utility. 
We also emphasise variants that can train such private GBDT models in only a small number of communication rounds, which is of particular importance to the federated setting.

\edit{Our high-level finding is that it is possible to achieve high performance with GBDT models, even comparable to that of non-private methods.  
In order to do so, one must allocate privacy budget to the quantities that are most important for the learning process. For example, we show that spending such budget on computing split decisions of trees is not as important as spending it on the leaf weights. %
Using our findings under the efficient privacy accounting of R\'enyi Differential Privacy (RDP) leads to performance that is far closer to the non-private setting than seen in previous works.}

Our main contributions are as follows:
\begin{itemize}
    \item A clear and concise framework for differentially private gradient boosting with decision trees. We deconstruct the GBDT algorithm into five main components, showing how to federate each component while satisfying R\'enyi Differential Privacy (RDP). We present a unifying approach, capturing recently proposed DP tree-based models as special cases.
    \item A new set of techniques for improving the utility of private federated GBDT models. For example, we propose a private method for discretising continuous features that makes as much use of the private training information as possible, incurring little additional privacy cost. Additionally, we %
    explore batching weight updates\edit{,} showing it is possible to maintain competitive model performance while reducing the number of communication rounds needed.
    \item An extensive set of experiments on a range of benchmark datasets exploring the trade-offs between various options in our framework. 
    By evaluating \edit{the choices in each of the components of our framework},%
    we find a clear dominant approach is formed by adapting and simplifying the GBDT algorithm while combining it with our improved split candidate method. 
    We show it is possible to achieve higher utility than state-of-the-art (SOTA) DP-RF and DP-GBDT methods on a range of datasets with reasonable levels of privacy. %
    \item We provide open-source code at
    \url{https://github.com/Samuel-Maddock/federated-boosted-dp-trees}
\end{itemize}
\noindent
\textit{Roadmap.} 
In Section \ref{sec:prelim} we outline technical preliminaries required to understand differentially private GBDTs before covering related works in Section \ref{sec:related-work}. In Section \ref{sec:core} and \ref{sec:additional} we describe our framework for DP-GBDTs, fitting existing methods within this and proposing combinations to study. In Section \ref{sec:experiments} we provide extensive experimental evaluations, comparing our methods to existing baselines within our framework before concluding with Section \ref{sec:conclusion}. %

\section{Preliminaries}\label{sec:prelim}

\subsection{Gradient Boosted Decision Trees (GBDT)}\label{sec:prelim:xgb}

Tree-based ensemble methods form a collection of $T$ decision trees that predict $\hat{y}_i$ for each input $\bolds{x}_i$: 
\begin{equation*} \textstyle
    \hat{y}_i := f(\bolds{x}_i) = \sum_{t=1}^T f_t(\bolds{x}_i)
\end{equation*}
For a specific tree $f_t$ let $L_t$ denote the number of leaf nodes. Each leaf node of a tree contains a weight, which will be the output of the tree for observations that are classified into that leaf. We denote $\bolds{w}^{(t)} \in \R^{L_t}$ as the vector of leaf node weights for a tree $f_t$.

GBDT methods train trees sequentially making use of past predictions to correct for mistakes. This is in contrast to random forest (RF) methods that train $T$ trees in parallel, averaging the weights of trees for the final prediction.

For a set of examples $D = \{(\bolds{x}_i, y_i)\}_{i=1}^n$ with corresponding predictions $\{\hat{y}_i\}_{i=1}^n$ the GBDT objective function is defined as
\begin{align}
    \label{eq:ob1}
    \textstyle
    \mathcal{L}(f) = \sum_{i=1}^n \ell(y_i, \hat{y}_i) + \sum_{t=1}^T \Omega(f_t)
\end{align}
\noindent
where $\ell$ is a twice-differentiable loss function, typically the cross-entropy loss (binary classification) or squared-error loss (regression). 
The term $\Omega(f_t) = \gamma L_t +  \frac{\lambda}{2}||\bolds{w}^{(t)}||_2^2$ is a form of regularisation such that $\gamma \geq 0$ penalises the size of the tree and $\lambda \geq 0$ penalises the magnitude of weights. 
This regularisation term is present in \edit{the popular} XGBoost \edit{algorithm} %
but is often omitted in other GBDT variants; we adopt it for our experimental study. 

Equation \eqref{eq:ob1} evades direct optimization. 
Rather, GBDT models are trained sequentially based on previous models. 
At step $t$ we can define the model prediction $\hat{y}_i^{(t)} = \sum_{k=1}^t f_k(\bolds{x}_i) = \hat{y}_{i}^{(t-1)} + f_t(\bolds{x}_i)$. The objective for optimising $f_t$ becomes
\begin{align}\label{eq:iterobj}
\textstyle
    \mathcal{L}^{(t)}(f_t) = \sum_{i=1}^n \ell(y_i, \hat{y}_i^{(t-1)} + f_t(\bolds{x}_i)) + \Omega(f_t)
\end{align}
For step $t$ we are concerned with finding a tree $f_t$ that minimises \eqref{eq:iterobj}. 
Since $f_t$ is not differentiable we can use a Taylor approximation. Taking the first-order approximation leads to the standard Gradient Boosting Machine (GBM) method. Taking a second-order approximation leads to Newton boosting as used by XGBoost \edit{\cite{xgboost2016}}.

When taking a first-order approximation we obtain
\begin{align*}
\textstyle
    \mathcal{L}^{(t)}(f_t) \approx \sum_{i=1}^n \left(\ell(y_i, \hat{y}^{(t-1)}) + g_i^{(t)} f_t(\bolds{x}_i)\right) 
    + \Omega%
\end{align*}
\noindent
where $g_i^{(t)} = \frac{\partial}{\partial \hat{y}^{(t-1)}_{i}} \ell(y_i, \hat{y}_i^{(t-1)})$ is the gradient of the loss function at the start of step $t$. By considering the index sets of examples mapped to leaf node $l$ i.e., $I_l = \{i| \bolds{x}_i \text{ belongs to leaf $l$ of } f_t \}$ one can show by expanding the above and differentiating with respect to $w_l^{(t)}$ that the optimal leaf weight is
\begin{align}\label{eq:gbm_weights}
\textstyle
    w_l^{(t)} = - \frac{\sum_{i \in I_l} g_i^{(t)}}{|I_l| + \lambda}
\end{align}
We denote this as a gradient weight update. Taking a second-order approximation of \eqref{eq:iterobj} instead gives
\begin{align*}
\textstyle
    \mathcal{L}^{(t)}(f_t) \approx \sum_{i=1}^n \left(\ell(y_i, \hat{y}^{(t-1)}) + g_i^{(t)} f_t(\bolds{x}_i) + h_i^{(t)}\frac{f^2_t(\bolds{x}_i)}{2}\right) + \Omega(f_t)
\end{align*}
\noindent
where $h_i^{(t)} = \frac{\partial^2}{\partial (\hat{y}^{(t-1)}_{i})^2} \ell(y_i, \hat{y}_i^{(t-1)})$ is the Hessian of the loss at the start of step $t$. As before one can show that the optimal weights this time are
\begin{align}\label{eq:xgb_weights}
\textstyle
    w_l^{(t)} = - \frac{\sum_{i \in I_l} g_i^{(t)}}{\sum_{i \in I_l} h_i^{(t)} + \lambda}
\end{align}
which we denote as a Newton weight update.
Substituting optimal weights from either the first or second-order approximation into Equation \eqref{eq:iterobj} leads to quantities that can be used to measure a split score.
In other words, when considering a split option that partitions examples into disjoint index sets $I  = I_1 \cup I_2$, the split score is a measure of how useful a split is for classification. The split score for Newton updates can be computed as
\begin{align}\label{eq:split_score}
\textstyle
    SS(I_1, I_2) = \frac{1}{2} \left[\frac{(\sum_{i \in I_1} g_i)^2}{\sum_{i \in I_1} h_i + \lambda} + \frac{(\sum_{i \in I_2} g_i)^2}{\sum_{i \in I_2} h_i + \lambda} - \frac{(\sum_{i \in I}g_i)^2}{\sum_{i \in I} h_i + \lambda}\right] - \gamma
\end{align}
In practice to form such split options, GBDT methods often discretize continuous features (e.g., via quantiles) into $Q$ split candidates. In order to handle categorical features, GBDT methods like XGBoost typically transform them e.g., via a one-hot encoding. In either case, this leads to splits of the form $I_{\leq} = \{i : x_{ij} \leq s_q^j\}$ for a split candidate $s_q^j$. Equation \eqref{eq:split_score} can then be used to greedily choose the feature split-candidate pair with the largest score when growing the tree structure during training.

\subsection{Differential Privacy}
Differential Privacy (DP) is a formal definition of privacy that guarantees the output of a data analysis does not depend significantly on a single individual's data item. Such a definition can be based on the notion of privacy loss.

\begin{definition}[Privacy Loss Random Variable]
    Given a randomised mechanism $\mathcal{M}: \mathcal{X} \rightarrow \mathcal{Y}$ we define the privacy loss random variable $L_{\mathcal{M}, x, x^\prime}$ over \say{neighbouring} datasets $x,x^\prime \in \mathcal{X}$ as
    \begin{align*}
    \textstyle
        L_{\mathcal{M}, x, x^\prime} = \log \left( \frac{p_{\mathcal{M}(x)}(X)}{p_{\mathcal{M}(x^{\prime})}(X)} \right)
    \end{align*}
    where $X \sim \mathcal{M}(x)$ and $p_{\mathcal{M}(\cdot)}$ is the density of the mechanism applied to the respective dataset.
\end{definition}
We take neighbouring datasets $x, x^\prime \in \mathcal{X}$ to mean that $x$ and $x^\prime$ differ on a single individual. The privacy loss allows us to succinctly describe differential privacy.
\begin{definition}[Differential Privacy \edit{in terms of privacy loss \cite{balle2018improving}}]
We say that a randomised mechanism $\mathcal{M}: \mathcal{X} \rightarrow \mathcal{Y}$ satisfies $(\epsilon, \delta)$-DP if for any adjacent datasets $x, x^\prime \in \mathcal{X}$ 
\begin{align*}
\textstyle
    \prob(L_{\mathcal{M}, x, x^\prime} \geq \epsilon) \leq \delta
\end{align*}
The privacy parameter $\epsilon$ is referred to as the privacy budget. 
When $\delta = 0$, we say that $\mathcal{M}$ satisfies $\epsilon$-DP. 
In this work we only consider privacy guarantees where $\delta > 0$ i.e., the case of approximate-DP. 
\end{definition}

While $(\epsilon, \delta)$-DP is a useful definition of privacy it does not allow us to tightly quantify the privacy loss from the composition of multiple mechanisms \cite{kairouz2015composition}. 
This is particularly important in machine learning where we wish to use mechanisms many times over the same dataset to train models. Instead, the notion of R\'enyi Differential Privacy (RDP) provides a succinct way to track the privacy loss from a composition of multiple mechanisms by representing privacy guarantees through moments of the privacy loss.

\begin{definition}[R\'enyi Differential Privacy \cite{mironov2017renyi}]
A mechanism $\mathcal{M}: \mathcal{X} \rightarrow \mathcal{Y}$ is said to satisfy $(\alpha, \tau)$-RDP if the following holds for any two adjacent datasets $x, x^\prime \in \mathcal{X}$ 
\begin{align*}
\textstyle
    \E \left[L_{\mathcal{M}, x, x^\prime}^{(\alpha-1)} \right] \leq \exp((\alpha-1)\tau)
\end{align*}
\end{definition}

One of the simplest and most widely-used mechanisms to guarantee $(\alpha, \tau)$-RDP is the Gaussian mechanism.

\begin{fact}[Gaussian Mechanism \cite{dwork2014algorithmic, mironov2017renyi}]
    The Gaussian mechanism $\mathcal{M}: \mathcal{X} \rightarrow \R^m$ of the form
    \begin{align*}
    \textstyle
        \mathcal{M}(x) = q(x) + N(0, \Delta_2(q)^2\sigma^2 I_m)
    \end{align*}
    satisfies $(\alpha, \tau)$-RDP with $\tau = \frac{\alpha}{2\sigma^2}$ and 
    \begin{align*}
    \textstyle
        \Delta_2(q) = \max_{x, x^\prime} \norm{q(x) - q(x^\prime)}_2
    \end{align*}
\end{fact}
The quantity $\Delta_2(q)$ is the $L_2$-sensitivity of the query $q$. The above shows that in order to make a real-valued query $q$ differentially private we just need to add suitably calibrated Gaussian noise.

An attractive property of this formulation of DP is that it is easy to reason about the privacy of an analysis where mechanisms are used multiple times on the same dataset.
\begin{fact}[Parallel Composition]
    Given a dataset $X$, a disjoint partition $X = X_1 \cup X_2 \dots \cup X_k$ and a mechanism $\mathcal{M}$ that satisfies $(\alpha, \tau)$-RDP. Then the mechanism $\mathcal{M}^\prime(X) := (\mathcal{M}(X_1), \dots, \mathcal{M}(X_k))$ satisfies $(\alpha, \tau)$-RDP.
\end{fact}
\begin{fact}[Sequential Composition]
    If $\mathcal{M}_1$ and $\mathcal{M}_2$ are $(\alpha, \tau_1)$-RDP and $(\alpha, \tau_2)$-RDP respectively then the mechanism that releases $(\mathcal{M}_1(\cdot), \mathcal{M}_2(\cdot))$ is $(\alpha, \tau_1 + \tau_2)$-RDP.
\end{fact}
\begin{fact}[Post-Processing]
    If $\mathcal{M}$ is an $(\alpha, \tau)$-RDP mechanism and $f$ is any function that does not depend on any private data then $f(\mathcal{M}(\cdot))$ is also $(\alpha, \tau)$-RDP. 
\end{fact}
Sequential composition tells us that using a mechanism multiple times on the same data leads to an increase in privacy loss. 
In the case of composing $k$ Gaussian mechanisms, we must increase the noise added through $\sigma$ by the order of $\sqrt{k}$ under RDP.

In practice, we care about obtaining the more meaningful notion of $(\epsilon, \delta)$-DP. When working with RDP we can rely on conversion lemmas such as those presented in \cite{canonne2020discrete} to convert between $(\alpha, \tau)$-RDP and $(\epsilon, \delta)$-DP. In our implementations, we use the analytical moment accountant developed by Wang et al. to provide tight numerical accounting of the privacy loss under RDP \cite{wang2019subsampled}. 

It is common to fix the privacy parameters $(\epsilon, \delta)$ before the analysis and then minimises $\sigma$ over a range of $\alpha$ values to obtain the smallest such noise needed to guarantee the chosen level of privacy. We use the autodp package\footnote{\url{https://github.com/yuxiangw/autodp}} to verify our accounting provides the correct $(\epsilon, \delta)$-DP guarantee.
An additional benefit of working with RDP is then that our framework easily extends to other mechanisms that satisfy RDP such as the Skellam mechanism which may be more suited to distributed settings~\cite{agarwal2021skellam}.

\begin{table*}[t]
  \small
  \caption{\edit{Summary of our Private Federated GBDT Framework}}
  \label{tab:framework}
  \begin{tabular}{lll}
    \toprule
     Component & Methods & Privacy Cost (in terms of $\kappa_s, \kappa_w, \kappa_c$)  \\
    \midrule
    \hyperref[framework:split_methods]{\textbf{(C1)}} Split Method & \tabitem Histogram-based (Hist) (\S\ref{framework:split_methods:hist}) & $\kappa_s = Tmd $ \\ & \tabitem Partially Random (PR) (\S\ref{framework:split_methods:random}) & $\kappa_s = Tmd$, does not require construction of a histogram \\ 
    & \tabitem Totally Random (TR) (\S\ref{framework:split_methods:random}) & $\kappa_s = 0$
     \\ \midrule
    \hyperref[framework:weight_update]{\textbf{(C2)}} Weight Update & \tabitem Averaging (\S\ref{framework:weight_update:averaging}) & If using a Hist or PR $\kappa_w=0$ otherwise $\kappa_w = T$ \\
    & \tabitem Gradient (\S\ref{framework:weight_update:gradient})&  \\
    & \tabitem Newton (\S\ref{framework:weight_update:newton}) & 
    \\ \midrule
    \hyperref[framework:split_candidates]{\textbf{(C3)}} Split Candidate & \tabitem Uniform, Log (\S\ref{framework:split_candidates:baseline})& Data-independent, $\kappa_c = 0$ \\
    & \tabitem Quantiles (non-private) (\S\ref{framework:split_candidates:baseline})& N/A \\
    & \tabitem Iterative Hessian (IH) (\S\ref{framework:split_candidates:IH}) & If using Hist, $\kappa_c = 0$. If using TR, with $s$ rounds of IH, $\kappa_c = sm$
    \\ \midrule
   \hyperref[framework:feature_interactions]{\textbf{(A1)}} Feature Interactions & \tabitem Cyclical $k$-way (\S\ref{framework:feature_interactions})& If using Hist or PR, $\kappa_s = Tkd$, if $k=1$ then $\kappa_s = T$.  \\ & \tabitem Random $k$-way (\S\ref{framework:feature_interactions}) & If using TR with IH then $\kappa_c = sk$
    \\ \midrule
    \hyperref[framework:batched_updates]{\textbf{(A2)}} Batched Updates & \tabitem $B=1$ (Boosting) (\S\ref{framework:batched_updates}) & Post-processing, no effect on privacy \\
    & \tabitem $B=T$ (RF-type predictions) (\S\ref{framework:batched_updates})&  \\
    & \tabitem $B=p \cdot T$ for some $p \in (0,1)$ (\S\ref{framework:batched_updates}) &
    \\ 
    \bottomrule
\end{tabular}
\end{table*}

\subsection{The Federated Model of Computation}\label{sec:fed}
Federated Learning (FL) has become a popular paradigm for large-scale distributed training of machine learning models \cite{kairouz2021advances}. 
In this work, we consider the horizontal setting, where a set of participants each hold a local dataset over the same space of $m$ features. We assume that there are $n$ data items in total and we consider the problem of training a differentially private GBDT model over the distributed dataset.
A powerful tool is secure aggregation, which allows the computation of a sum without revealing any intermediate values~\cite{bonawitz2017secureAgg, bell2020secure}. 
Specifically, 
when each participant $P_k$ has a number $x_k \in \Z$,  
secure aggregation computes the result $\sum_k x_k$ securely without any participant directly sharing their $x_k$. 

Our focus is on a framework that combines secure aggregation with DP to securely and privately train GBDT models. 
For the rest of this paper, we present algorithms as if the data were held centrally, with the understanding that all the operations we use can be performed in the federated model (with rounding to fixed precision)\footnote{The rounding introduces a small amount of imprecision in representing values, but this is overwhelmed by the noise added for privacy.}. 
This means that we avoid techniques designed for central evaluation such as the exponential mechanism 
\cite{li20201privacypreserving,zhao2018inprivate, grislain2021dp}.

\smallskip\noindent
\edit{\textbf{Threat Model:} In this work, in common with many other works in the federated setting, we assume an honest-but-curious model, where the clients do not trust others with their raw data.
We study the aggregating server’s knowledge based on the information gathered from clients. 
While there is potential for clients to attempt to disrupt the protocol, 
we leave the detailed study of more malicious threat models and model poisoning to future work.} 
In order to combine secure aggregation with DP, we act as if
there were a trusted central server that securely aggregates quantities and adds the required DP noise before sending the updated (private) model back to participants (as assumed in~\cite{mcmahan2017learning}). 
In practice, we can eliminate the need for a central server by well-established implementations of secure computation that rely on techniques from secure multi-party computation, either among a small number of honest-but-curious servers, or via clients working with small groups of neighbors and a single untrusted server~\cite{bell2020secure}. 
\edit{Sufficient noise for DP guarantees can be 
added by honest-but-curious servers, or introduced by each client adding a small amount of discrete noise, such that the total noise across clients adds up to the desired volume}~\cite{shiprivacy,roy2020crypt,bohler2020secure,bohler2021secure,roth-2019-honeycrisp,champion2019securely,roth-2021-mycelium,roth-2020-orchard}.

\section{Related Work} \label{sec:related-work}
Differentially private decision trees have been well studied in the central setting with a strong focus on random forest (RF) models \cite{fletcher2015differentially, fletcher2017differentially, zhao2018inprivate}. 
However, the boosted approach (i.e., private GBDT models) has been less well-explored. 
Recently, federated XGBoost models have been presented, with most works focused on secure training via cryptographic primitives such as Homomorphic Encryption (HE) and Secure Multi-Party Computation (MPC) and with no DP guarantees \cite{cheng2021secureboost, feng2019securegbm, deforth2021xorboost}. 

Some related works (e.g., \cite{tian2020federboost}) study XGBoost in a federated setting with local DP (LDP) guarantees. 
The closest work to ours in this regard is the FEVERLESS method~\cite{wang2021feverless}, which translates the XGBoost algorithm into the vertical federated setting using secure aggregation and the Gaussian mechanism. 
In particular, FEVERLESS securely aggregates gradient information into a private histogram which is used to compute split scores and leaf weights (Equations~\eqref{eq:xgb_weights} and \eqref{eq:split_score}).
A certain subset of the participants are chosen as \say{noise leaders} to add Gaussian noise to their gradients information before aggregating to achieve an overall DP guarantee after securely aggregating across all participants.
As we will see, %
the main disadvantage of directly translating the XGBoost algorithm in this way is the high privacy cost of repeatedly computing split scores.  This results in having to add more noise into split score/leaf weight calculations and a lower utility model.

To reduce this privacy cost, one can consider making split decisions independently of the data. These so-called totally random (TR) trees have been studied in both the non-private and private settings with random forests \cite{geurts2006extremely, fletcher2015tr}. 
In the private setting, proposed methods often use central DP mechanisms that are hard to federate \edit{\cite{fletcher2017differentially, asadi2022private}}. For example, Fletcher and Islam~\cite{fletcher2017differentially} propose a DP-RF method that utilises the exponential mechanism to output the majority label in leaf nodes under the notion of smooth sensitivity, which is unsuited to the federated setting. 

In this work, we also consider TR trees as an option under our framework but for a federated and private GBDT model. 
To the best of our knowledge, the only other work that considers private boosting with random trees is that of Nori et al. \cite{nori2021accuracy}. 
They consider a central DP setting with a focus on training \edit{private} explainable models via Explainable Boosting Machine (EBMs). 
We compare the technical differences in Section \ref{framework:feature_interactions} and empirically in Section \ref{exp:comp}. 

\section{Private GBDT Framework}\label{sec:core}
In this section, we perform a comprehensive investigation of the main components needed to train GBDT models in the federated setting. We propose a framework of methods for training DP-GBDT models by identifying three main components that require DP noise and two additional components that interact with these. \edit{The full framework is summarized in Table \ref{tab:framework}}.

We explain the various options in each component and how they affect privacy guarantees and conclude by instantiating related work into the framework before  empirically evaluating methods in Section~\ref{sec:experiments}.
A particular strategy we highlight is replacing data-dependent choices with random or uniform choices. 
Although counter-intuitive, it often holds that the privacy ``cost'' of fitting the choices to the data is not made up for by the utility gain, and picking among a set of random options is sufficient for good results. 
This is evaluated in our experimental study.

For simplicity, we assume that each participant holds a single data item $(\bolds{x}_i, y_i)$ with $n$ participants (data items) in total. 
We additionally assume that we have (publicly) known bounds on each feature.
All of these assumptions can be easily removed, potentially with some additional privacy cost. Table \ref{tab:notation} in the \edit{a}ppendix displays commonly used notation for convenience. %

\begin{algorithm}[t]
\caption{General GBDT}\label{alg:gbdt}
\begin{algorithmic}[1]
\Input Number of trees $T$, maximum depth $d$, number of split candidates $Q$, privacy parameters $\epsilon, \delta$
    \State %
    \colorbox{pink!60}{\parbox{0.43\textwidth}{For each feature $j=1,\dots,m$ generate $Q$ split candidates $S_j := \{s_{1}^j, \dots, s_{Q}^j\}$ \hyperref[framework:split_candidates]{\textbf{(C3)}}}}
    \State Initialise the forest $\mathcal{T} \gets \emptyset$
    \For{$t=1, \dots, T$}
        \State  %
        \colorbox{white!0}{\parbox{0.4\textwidth}{For each $(x_i, y_i) \in D$ compute the required gradient information $(g_i, h_i)$ based on $\hat{y}_i^{(t-1)}$ \hyperref[framework:weight_update]{\textbf{(C2)}}}}
        \State \colorbox{blue!25}{\parbox{0.405\textwidth}{Choose a subset of features $F^{(t)} \subseteq \{1, \dots, m\}$ with $|F^{(t)}| = k$ for the current tree $f_t$  \hyperref[framework:feature_interactions]{\textbf{(A1)}}}}
        \While{depth of the current node (in $f_t$) is $\leq d$}
            \State  \colorbox{pink!60}{\parbox{0.38\textwidth}{Choose a feature split candidate pair $(j, s_{q}^j)$ from $F^{(t)}$ \hyperref[framework:split_methods]{\textbf{(C1)}}}}
            \State \colorbox{white!0}{\parbox{0.38\textwidth}{Split the current node with observations $I$ into two child nodes with index sets $I_{\leq} = \{i: x_{ij} \leq s_{q}^j\}$ and $I_{>} = I \setminus I_{\leq}$}}
            \State \colorbox{white!0}{\parbox{0.38\textwidth}{Repeat (6)-(9) recursing separately on the child nodes %
            }}
        \EndWhile
        \State \colorbox{pink!60}{\parbox{0.405\textwidth}{For each leaf $l$ calculate a weight $w^{(t)}_l$ from the examples in the leaf according to the chosen update method \hyperref[framework:weight_update]{\textbf{(C2)}}}}
        \State \colorbox{blue!25}{\parbox{0.405\textwidth}{ Update predictions $\hat{y}^{(t)}_i$ or batch updates \hyperref[framework:batched_updates]{\textbf{(A2)}}}}
        \State Add the $t$th tree $f_t$ to the ensemble, $\mathcal{T} = \mathcal{T} \cup \{f_t\}$
    \EndFor
    \State  \Return the trained forest $\mathcal{T}$
\end{algorithmic}
\end{algorithm}

\subsection{A General Recipe}

In order to train the GBDT algorithm outlined in Section~\ref{sec:prelim:xgb} we only need to specify a few core choices:
 How to pick split candidates (for discretizing continuous features), calculate the split scores at each internal node, and compute the leaf weights for prediction. 
One can note from Equations~\eqref{eq:xgb_weights} and \eqref{eq:split_score} that the leaf weights and split scores only depend on the sum of gradients and Hessians at an internal or leaf node of a decision tree. 
It is therefore natural to utilise secure aggregation as a tool to federate the GBDT algorithm. 
In Algorithm~\ref{alg:gbdt} we present the general GBDT algorithm assuming these quantities can be %
gathered.
Looking closely at Algorithm \ref{alg:gbdt}, the only time we need to directly query participants' data is when we compute the three quantities just mentioned.

Based on this we divide the general algorithm into 3 core components that require some form of DP noise: 
Split Methods \hyperref[framework:split_methods]{\textbf{(C1)}}, 
Weight Updates \hyperref[framework:weight_update]{\textbf{(C2)}}, and
 Split Candidates \hyperref[framework:split_candidates]{\textbf{(C3)}}. 
 These are the core components required for training a GBDT model.
We also consider two additional aspects to specify when training a GBDT model: Feature Interactions \hyperref[framework:feature_interactions]{\textbf{(A1)}} and
 Batched Updates \hyperref[framework:batched_updates]{\textbf{(A2)}}. 
These are aspects that interact with the core components but do not require any additional noise. 
To reason about the privacy guarantees of our GBDT framework, we introduce some variables to count the number of queries needed when training a GBDT model with $T$ trees. 
Let $\kappa_{c}$ denote the number of queries needed to calculate split candidates; $\kappa_{s}$ for the queries needed to calculate inner node splits; and $\kappa_{w}$ for the queries to calculate leaf weights.
Counting the number of queries needed for each component is enough to give a privacy guarantee for Algorithm \ref{alg:gbdt}.
\begin{theorem}
    Suppose that each mechanism for the framework components satisfies $(\alpha, \tau_c), (\alpha, \tau_s), (\alpha, \tau_w)$-RDP respectively. 
    Then the GBDT algorithm satisfies $(\alpha, \tau)$-RDP with $\tau = \kappa_c \tau_c + \kappa_s \tau_s + \kappa_w \tau_w$.
\end{theorem}
The above simply follows from the sequential composition properties of RDP. In our experimental study we utilise the Gaussian mechanism for each core component, hence $\tau = (\kappa_c + \kappa_s + \kappa_w) \frac{\alpha}{2\sigma^2}$ and so $\sigma = O_{\epsilon}(\sqrt{\kappa_c + \kappa_s + \kappa_w})$. 
This shows that if we can minimise the number of queries that each main component requires, then we reduce the amount of noise we add to the learning process while still maintaining privacy. \edit{Various methods affect the privacy cost in different ways. 
The privacy implications for different choices in terms of $\kappa_c, \kappa_s, \kappa_w$ are shown in Table \ref{tab:framework}.}

\subsection{Federating GBDTs}
\label{sec:fedgdbt}
At the start of building the $t$-th tree, each participant calculates gradient information $(g^{(t)}_i, h^{(t)}_i)$ for their examples. 
Throughout the training of a single tree, to calculate the desired components we can rely on querying data in the form $q(I) = \left(\sum_{i\in I} g_{i}^{(t)}, \sum_{i \in I} h_i^{(t)}\right)$ over some set of observations $I$, e.g., all observations in a specific tree node. 
To do this securely, we can apply secure aggregation to aggregate gradient information at the various stages that require it in Algorithm \ref{alg:gbdt} (\textbf{C1- C3}).

In order to apply the Gaussian mechanism, we must bound the sensitivity of such a query function. In this case, we need bounds on the gradient quantities $g_i^{(t)}, h_i^{(t)}$. Our focus in this paper is on binary-classification problems. %
In binary-classification our loss function is of the form $\ell(y_i, \hat{y}_i) = y_i\log(\hat{y}_i) + (1-y_i)\log(1-\hat{y}_i)$ (i.e., binary cross-entropy) and has gradients $g_i \in [-1,1]$ and Hessians $h_i \in [0, \frac{1}{4}]$. Hence the sensitivity of aggregating gradient information is $\Delta_2(q) = \sqrt{1 + \frac{1}{16}} = \frac{\sqrt{17}}{4}$.%
If the chosen loss function has unbounded gradient information (e.g., regression problems) we can employ gradient clipping (similarly to DP-SGD) to obtain a bounded sensitivity \cite{abadi2016deep}.

\edit{The computational and communication costs of these steps are low. 
Decision tree-based methods are often preferred for their ease of construction, and this translates to the federated setting: 
each client computes its local updates (e.g., gradients and Hessians) and shares these through secure aggregation. 
The communication costs are linear in the size of the updates computed, which are fairly low dimensional: 
we quantify this in the subsequent sections.} 

\subsection{Component 1: Split Methods}\label{framework:split_methods}

\subsubsection{Greedy Approach: Histogram-Based}\label{framework:split_methods:hist}
As described in Section \ref{sec:prelim:xgb}, the standard GBDT algorithm will calculate $Q$ split-scores for every feature $j$. This forms feature-split %
pairs $(j, s^j_q)$ and at each internal node the pair with the highest score is chosen to grow the tree. 
This split score depends on aggregating gradient and Hessian values. 
The most suitable way to do this in a federated setting is to form a histogram over the split candidates for every feature. 
This requires (securely) aggregating the gradient and Hessian values into bins partitioned by the split candidate values. 
Hence the $q$-th gradient histogram bin for feature $j$ contains $G_q^j =  \sum_{i \in \{i : s_{q-1}^j < x_{ij} \leq s_{q}^j\}} g_i^{(t)}$, and similarly for Hessians.

We can apply our generic aggregation query $q(I)$ %
with $I = \{i: s_{q-1}^j < x_{ij} \leq s_q^j\}$ to aggregate bins of both the gradient and Hessian histograms. 
Each participant's data item will fall into exactly one histogram bin, so via parallel composition we just need to count the number of times a histogram is computed during training.
At each internal node of a tree, we must compute split-scores and thus gradient histograms. When considering all $m$ features per split, this requires $\kappa_s = Tmd$ queries for a model with $T$ trees of maximum depth $d$. %
This incurs a high privacy cost for large ensembles\footnote{Default XGBoost parameters take $d=6$ and $T=100$ which implies a high privacy cost on any dataset with a moderate number of features $m$}.
\edit{Each client can quickly compute and send their histograms of size $Q$ for each feature considered for a split.}

\subsubsection{Randomised Approach: Partially and Totally Random}\label{framework:split_methods:random}
In \cite{geurts2006extremely}, Geurts et al. initiate the study of \say{Extremely Randomised Trees} (ERTs) in the non-private setting. 
In ERTs the idea is to add randomness into the split choices when growing the tree. 
The motivation was to show that accuracy comparable to that of greedy tree-building models could be obtained for large enough ensembles. 
ERTs are potentially much faster to train as there is no need to compute split scores for each internal node. 
This leads to two pragmatic choices for splitting nodes:
\begin{itemize}[leftmargin=*]
    \item \textbf{Partially Random (PR)}: For each feature $j$ pick a split candidate $s^j_q \in S_j$ uniformly at random, \edit{where $S_j$ is the set of split candidates for $j$}. 
    The split score of $(j, s^j_q)$ is computed for each feature and the pair with the highest score is chosen. This still requires $\kappa_s = Tmd$ queries but does not require building histograms.
    \item \textbf{Totally Random (TR)}: Pick a feature $j \in [m]$ and a split candidate $s^j_q$, both uniformly at random. 
    This does not require any queries for internal nodes ($\kappa_s = 0$) as it is data independent.
\end{itemize}

Since TR trees do not access data to build tree structure they are attractive from a privacy perspective. 
All trees in the ensemble can be pre-computed by choosing random splits, \edit{which can be communicated to clients at a cost linear in the size of the tree}. 
Hence building a TR ensemble requires far fewer queries than histogram-based methods. 
However, a TR ensemble often requires a much larger number of trees to achieve similar model performance as histogram-based counterparts.
We explore such trade-offs between TR and histogram-based methods in Section~\ref{exp:split_methods}.

\subsection{Component 2: Weight Updates}\label{framework:weight_update}
Once a tree has been built, the records in the dataset will be partitioned among the leaf nodes of the tree. 
In the following we consider the $l$-th leaf of tree $t$ with weight $w_l^{(t)}$ which contains records $I_l^{(t)} = \{i \in [n]: i \text{ belongs to leaf } l \edit{\text{ of tree } t}\}$.
\edit{Each client needs to compute and send the weights of leaf nodes, at cost proportional to the number of leaves, $2^d$ if there are $d$ binary splits}. 
Both RF and GBDT methods update these leaf nodes with a weight that contributes to prediction.  
As we noted in Section \ref{sec:prelim:xgb}, taking a first-order or second-order approximation to Equation~\eqref{eq:iterobj} leads to two different weight updates. 
Note that by setting $h_i = 1$ in both \eqref{eq:xgb_weights} and \eqref{eq:split_score} we recover the gradient weight update of Equation~\eqref{eq:gbm_weights} and also obtain a split score for gradient updates. 
Hence when $h_i=1$ both approaches are equivalent and so Newton updates can be seen as generalising the standard gradient approach. While RFs do not calculate gradient information, we can still view them as a special case within our framework. RF trees typically compute the class probabilities in leaf nodes which are averaged across all trees in the ensemble. This leads to three main weight updates: zeroth-order (Averaging), first-order (Gradient), and second-order (Newton).

\subsubsection{Averaging Updates}\label{framework:weight_update:averaging}
For random forests the leaf nodes store the class distribution. 
For regression problems this is the average value of $y$ in the leaf node. 
With binary classification, the weight update is simply the proportion of positive examples in the leaf node i.e., $w_l^{(t)} = 
\frac{1}{|I_l^{(t)}|} \sum_{i\in I_l^{(t)}} \mathbbm{1}\{y_i = 1\}$. 

Although RF models do not compute gradients we can still utilise our generic aggregation query by having participants send $g_i = \mathbbm{1}\{y_i = 1\}$ and $h_i = 1$. In this case $\sum_{i \in I}g_i$ counts the number of class 1 examples and $\sum_{i \in I}h_i$ counts the number of examples in a node. This changes the sensitivity of our query to $\Delta(q) = \sqrt{2}$.

In RF models the trees are independent from one another with final predictions formed from the average of weights across all trees. We denote this as an averaging update from now on.

\subsubsection{Gradient Updates}\label{framework:weight_update:gradient}
Each participant calculates $g_i = \frac{\partial}{\partial \hat{y}_i} \ell(y_i, \hat{y}_i)$ and $h_i = 1$ and uses this in the weight update defined in Equation~\eqref{eq:gbm_weights}, i.e., the weights are the average negative gradient values in the leaf node. 
This can be viewed as a gradient descent step over the batch of observations in leaf node $j$. The sensitivity of the query also changes to $\Delta(q) = \sqrt{2}$.

\subsubsection{Newton Updates}\label{framework:weight_update:newton}
Participants calculate both first-order and second-order gradients of the form $g_i = \frac{\partial \ell}{\partial \hat{y}_i}, h_i = \frac{\partial^2 \ell}{\partial (\hat{y}_i)^2}$ and use the weight update in Equation~\eqref{eq:xgb_weights}.

In classification problems the total weight across trees for an observation $i$ is aggregated and the sigmoid function $\sigma(\cdot)$ is applied. %
It is also standard in GBDT methods to perform post-processing on leaf weights. In practice we consider updates of the form %
$-\eta \cdot \max\{w_l^{(t)}, \beta \cdot \operatorname{sgn}(w_l^{(t)})\}$
where $\eta > 0$ is the learning rate and $\beta \geq 0$ is a clipping factor to control the magnitude of updates%
.

For histogram-based splitting, the final gradient histograms from the parent of a leaf node can be used to calculate weights, meaning $\kappa_w = 0$. 
For TR splitting, participants do not calculate histograms so they must directly aggregate the required gradient information in each leaf node. 
This is a single query per leaf node that happens once per tree, and so $\kappa_w = T$. 

\subsection{Component 3: Generating Split Candidates}\label{framework:split_candidates}

One major step needed to train GBDT models is to identify split candidates for each (continuous) feature. 
In traditional GBDT models such as XGBoost, split candidates are chosen by computing the quantiles of a feature. 
Computing quantiles is a succinct way to describe a feature's distribution but can be slow in practice for large datasets. The original XGBoost paper proposes a weighted quantile sketch to make this process faster, using the Hessian information as weights. %
While this is suitable in non-private settings, it is difficult to calculate such quantiles (or quantile sketches) accurately without incurring an appreciable privacy cost. 
Existing work on DP-GBDTs has computed split candidates either with LDP quantiles in the local setting~\cite{le2021fedxgboost}, DP quantiles in the central setting~\cite{grislain2021dp} or with MPC methods (without DP guarantees) in distributed settings~\cite{tian2020federboost}.

\subsubsection{Data-Independent Split Candidates}\label{framework:split_candidates:baseline}
The simplest and cheapest (from a privacy perspective) approach is to propose split candidates independently of the data, such as via uniform splits. 
For a feature $j$ with values in $[a, b]$, one can generate a split candidate for each $q \in [Q]$ uniformly over $[a,b]$ as $s_q^j = a + (q-1)(b - a)/(Q-1)$. 
As we assume bounds on features are public knowledge, we do not need to query participants' data, and hence $\edit{\kappa_c} = 0$.

A disadvantage of this approach arises when features are heavily skewed as uniform splits are unlikely to cover important areas of the feature's distribution. %
One possible approach would be to take a $\log$ transform of skewed features and then split uniformly over the transformed feature. 
In the non-private setting, one can manually check features or use statistical skewness tests to determine when to transform features. 
This poses a problem in the private setting as we may not know a priori which features are skewed and privately computing such a test may be expensive privacy-wise. 

\subsubsection{Iterative Hessian \edit{(IH)} Splitting}\label{framework:split_candidates:IH}
We propose an alternative method based on making use of information that is usually calculated during the training process. 
We will verify for datasets with heavily skewed features that we can achieve similar AUC to optimal non-private split candidate methods for little to no additional privacy cost.
Specifically, the Hessian information in Newton boosting captures the certainty of predictions and is often used in the non-private setting to guide quantile finding. 
We can take a similar approach in the private setting provided we estimate aggregated Hessian values at each split candidate bin i.e., a Hessian histogram. 
We propose the following intuition to propose new split candidates at each round:
\begin{itemize}[leftmargin=*]
    \item \textbf{Merge bins} with low \edit{(or zero)} Hessian since this indicates a split is too fine-grained to be useful.
    \item \textbf{Split bins} that have large Hessian value as this indicates a large number of observations lie in the bin. 
    To refine a bin we can split by taking the midpoint of adjacent bin edges.
\end{itemize}
\edit{In practice, we split a bin if its Hessian value is greater than the total Hessian uniformly divided over the $Q$ bins. If at the end of a round we end up with fewer than $Q$ bins, then we fill the remaining bins by uniformly splitting. The full algorithm for IH is given in Algorithm \ref{alg:hess} in the appendix.}
Carrying out IH splitting is a form of post-processing on the Hessian histogram and thus has no extra privacy cost beyond the cost to compute the histogram. \edit{However, the} choice of split method \edit{may incur additional} privacy cost:
\begin{itemize}[leftmargin=*]
    \item \emph{Hist:} In histogram-based methods, a Hessian histogram is computed at the start of every tree for all features. 
    We can use the previous tree's Hessian information to inform our split candidates for each new tree. 
    We incur no additional privacy cost and hence $\edit{\kappa_c} = 0$
    \item \emph{Totally random:} As TR trees are built independently of the data, Hessian histograms are never computed. 
    We propose to calculate a Hessian histogram for the first $s$ rounds of training and thus the number of queries we need for split candidates is $\edit{\kappa_c} = sm$. 
    For the first $s$ rounds we refine our split candidates using IH, after which we use the final set of candidates found in round $s$ for the remaining $T-s$ trees.
\end{itemize}

\section{Additional Considerations}\label{sec:additional}
\subsection{Feature Interactions}\label{framework:feature_interactions}
Explainable Boosting Machines (EBMs) are a popular method for training GBDTs to ensure explainability of the resulting model~\cite{lou2012intelligible}. 
The main idea is to construct an additive model of the form $f(\bolds{x}) = \sum_{j=1}^m \alpha_j f_j(x)$ where each $f_j(x)$ is a boosted decision forest with trees that are trained only on the $j$-th feature.
Nori et al.~\cite{nori2021accuracy} consider the problem of training DP-EBM models in the central setting. Their method relies on training many  very shallow trees with totally-random (TR) splits. In order to ensure explainability, each tree of the ensemble is restricted to a single feature at a time. This results in a \say{cyclical} boosting method where tree $t$ is trained only on feature $j= t\mod m$.
Although the focus of our work is not on explainability, Nori et al. note that the cyclical training method of EBMs actually results in more accurate models (with DP) when compared with models that can freely split on all features per tree. 
This presents another design decision---whether to train trees cyclically (so that each tree only splits on a single feature at a time), or to train trees that consider a subset of $k$ features to choose from when splitting a node. 
We define $k$-way feature splitting as considering $k$ features at a time per tree. This can be done in two different ways:

\textbf{1. Cyclical $k$-way:} Consecutive trees train on a subset of the next $k$ features and repeat in cycles every $m/k$ trees.

\textbf{2. Random $k$-way:} The $k$ features are chosen randomly at the start of each tree.

When $k=1$ with cyclical training we recover the method used in EBM. 
When $k=m$ we recover the standard GBDT splitting method. The choice of $k$ determines the maximum number of feature interactions that are possible within a tree. %
We note that the random $k$-way method is also commonly used in the non-private setting to reduce computation time and act as model regularisation~\cite{friedman2002stochastic}.
\edit{The computation and communication costs for each client scale proportionally to $k$.}
For histogram-based methods with $k > 1$ the number of queries required to form internal node splits for $k$-way splitting is $\kappa_s = Tkd$. When $k=1$ this reduces to $\kappa_s = T$ and $\kappa_w = 0$ since gradient histograms can be computed once at the root node and this same histogram can be used to calculate split-scores for every level in the tree. For totally-random trees the value of $k$ does not affect the number of queries and $\kappa_s = 0$ remains. \edit{In Appendix \ref{appendix:interactions} we present experiments detailing the effect of feature interactions. We observe that cyclical $k=1$ (i.e, EBM) performs the best and we focus on this in our end-to-end comparison in Section \ref{exp:comp}.}

\subsection{Batched Updates}\label{framework:batched_updates} 
One advantage of random forests (RF) in distributed settings is that trees can be trained in parallel. 
In the case of totally random (TR) trees the model orchestrator can precompute the structure of all trees and participants can compute  gradient statistics for leaf weights over the entire forest in a single round of communication.  
On the other hand, gradient boosting methods are inherently sequential---results of the previous ensemble determine the gradient calculations for the next tree. This is a bottleneck for weight updates. One way to parallelise this is to consider batching updates.

Suppose we use a batch size of $B$ and are training $T$ trees. A batched update is of the form 
\begin{displaymath}
\textstyle
    \hat{y}_{i}^{(t)} = \hat{y}_i^{(t-B)} + \eta\sigma\left(\frac{1}{B}\sum_{k=t-B}^B w^{(k)}_i\right) 
\end{displaymath}
where we abuse notation to let $w_i^{(k)}$ denote the weight of the leaf in tree $k$ that $\bolds{x}_i$ is partitioned into. 

Batched updates require  participants update their predictions every $B$ rounds based on the average leaf weight of the trees in the batch. 
At the start of the $(B+1)$-th round the gradient information is recomputed so that the next batch is boosting predictions from the previous batch. One can think of this as boosting a set of $B$-sized random forests. When $B=T$ we recover RF-type predictions but note that if $w_l^{(t)}$ uses gradient or Newton weights then the model updates are different from averaging updates (which instead use class probabilities as weights). Batching updates also has no extra privacy cost as it is a form of post-processing.
    
If we wish to train $T$ trees with a batch size $B$ then the communication rounds of the boosting process reduce from $O(T)$ to $O(T/B)$\footnote{Ignoring the constant number of rounds required for secure aggregation. \edit{See Appendix \ref{sec:computationcommunicationcost} for more details}}. 
In Section \ref{exp:bb} we consider batching gradient and Newton updates for different batch sizes and compare to DP-RF which requires $O(1)$ rounds of communication \cite{fletcher2015tr}. 

\subsection{Instantiating the GBDT Framework}
\subsubsection{Instantiating Components} \label{sec:instantitate}
In the previous sections, we have deconstructed the GBDT algorithm into various core components that require us to add noise to guarantee DP. We also noted two additional \edit{considerations} that \edit{interact with the core components}. When instantiating our framework in experiments we will combine the options for each component \edit{as listed in Table \ref{tab:framework}. 
A key contribution of this work is in the comprehensive study of both existing (Section \ref{sec:baselines}) and new (Section \ref{sec:new}) methods %
as follows:}
\noindent
    \hyperref[framework:split_methods]{\textbf{(C1)}} \textbf{Split Methods:} \edit{The histogram over split candidates is how centralized algorithms like XGBoost structure the problem \cite{xgboost2016}. It is also friendly to federation and has been used by prior works so forms a natural baseline \cite{wang2021feverless, cheng2021secureboost, feng2019securegbm}. Totally random trees have been widely used in non-private RF models but have not been well-studied in private, federated and GBDT settings \cite{geurts2006extremely}. DP-EBM is the only prior example we are aware of here \cite{nori2021accuracy}}
    
\noindent
    \hyperref[framework:weight_update]{\textbf{(C2)}} \textbf{Weight Updates:} \edit{We consider standard update methods used in GBDTs/RFs noting that Hessian updates have not been as well-studied under privacy or the federated setting.}
    
\noindent
    \hyperref[framework:split_candidates]{\textbf{(C3)}} \textbf{Split Candidates:} \edit{Data-independent splits have been largely overlooked in central DP settings with effort put into calculating DP quantiles. We advocate it as a good option for the federated setting since the (privacy) cost of finding quantile splits is not repaid in practice. We introduced the Iterative Hessian (IH) approach based on refining candidates over a number of rounds which helps when features are particularly skewed.}
    
\noindent
    \hyperref[framework:feature_interactions]{\textbf{(A1)}} \textbf{Feature Interactions:} \edit{The idea of (maximum) feature interactions generalizes the Explainable Boosting Machines (EBM) method which considers a single feature per tree \cite{nori2021accuracy}.}
    
\noindent
    \hyperref[framework:batched_updates]{\textbf{(A2)}} \textbf{Batched Updates}: \edit{The idea of batching updates has not been studied in the private and federated setting. It can be viewed as boosting individual RFs which is sometimes done in non-private settings. Our focus here is on reducing communication rounds while still maintaining accuracy.}

\subsubsection{Instantiating Related Work}\label{sec:baselines}
In Table~\ref{tab:related_works} we outline how SOTA DP-GBDT models can be expressed in our framework. 
These act as the primary baselines in our experiments. 
We note that many of these methods were originally proposed to use pure $\epsilon$-DP in the central setting and often rely on basic composition results. 
We have re-implemented all methods to use tighter RDP accounting and guarantee $(\epsilon, \delta)$-DP so they are not disadvantaged. To summarise:
\begin{itemize}[leftmargin=*]
    \item \textbf{DP-EBM \cite{nori2021accuracy}} is a DP variant of the EBM model. It uses Gaussian Differential Privacy (GDP) but as this is known to under-report $\epsilon$ values \cite{gopi2021numerical}, we use RDP in our experiments.
    DP-EBM uses TR splits with gradient updates, where each tree only considers a single feature.
    The split candidate method is a central DP histogram that %
    attempts to uniformly distribute observations among bins. We replace this with uniform split candidates in our experiments.
    \item \textbf{DP-RF \cite{fletcher2015tr}} is a central DP method that builds a RF via TR splits. The method was originally proposed for categorical features and later extended to continuous features \cite{fletcher2017differentially}. The Laplace mechanism is used to perturb leaf weights we re-implement this using the Gaussian mechanism under RDP accounting. 
    In our federated framework, DP-RF corresponds to using TR splits, the averaging weight update, and uniform split candidates (with $k=m$, $B=T$).
    \item \textbf{FEVERLESS \cite{wang2021feverless}} corresponds to a Hist split method with Newton weight updates. 
    FEVERLESS uses a quantile sketch which is non-private in our horizontally partitioned setting; we replace this with uniform splits to make FEVERLESS fully private.
\end{itemize}
\edit{In our final comparisons in Section \ref{exp:comp} we compare to an LDP baseline. This baseline has each user add Gaussian noise before releasing their gradient information. Such noise only needs to be scaled by the number of trees ($T$) since each user can release noised gradients at the root node, and the server can %
use this to construct the tree. While LDP does not strictly fall into our framework, it is a useful benchmark to compare against distributed DP counterparts.}
\subsubsection{\edit{Instantiating Other Methods}\label{sec:new}}
\edit{We end Section \ref{sec:experiments} with an end-to-end comparison of the baselines against new combinations expressed under our framework. These methods are:}
\begin{itemize}[leftmargin=*]
    \item \edit{\textbf{DP-EBM Newton,} the DP-EBM method with Newton updates instead of Gradient updates. We also do not train $Tm$ trees but only $T$. The total privacy cost here is $\kappa_c + \kappa_s + \kappa_w = 0 + 0 + T$}
    \item \edit{\textbf{DP-TR Newton,} the TR spit method, uniform split candidate and Newton updates. The privacy cost is the same as DP-EBM.}
    \item \edit{\textbf{DP-TR Newton IH EBM,} a DP-TR Newton with EBM feature interactions (i.e., cyclical $k=1$). The privacy cost is $\kappa_c + \kappa_s + \kappa_w = 0 + 0 + T + sm$ where $s$ is the number of rounds IH is performed.}
    \item \edit{\textbf{DP-TR Batch Newton IH EBM $(p=0.25, p=1)$,} i.e., DP-TR Newton IH EBM with batched updates with $p=0.25$ or $p=1$, the privacy cost is the same as DP-TR Newton IH EBM.}
\end{itemize}
\begin{table}[t]
  \footnotesize
  \caption{Related works under our framework}
  \label{tab:related_works}
  \begin{tabular}{ccccll}
    \toprule
     & DP-EBM \cite{nori2021accuracy} & FEVERLESS \cite{wang2021feverless} &
    DP-RF \cite{fletcher2015differentially} \\
    \midrule
    C1: Split Method & TR & Hist & TR
    \\
    C2: Weight Update & Gradient & Newton & Averaging \\
    C3: Split Candidate & Uniform (DP Hist) & Quantile Sketch & N/A\\
    A1: Feature Interactions & Cyclical $(k=1)$ & $m$-way & $m$-way \\
    A2: Batched Updates & $B=1$ & $B=1$ & $B=T$\\  \midrule
    \edit{$\kappa_c + \kappa_s + \kappa_w$} & \edit{$ 0 + 0 + Tm$} & \edit{$ 0 + Tmd + 0$} & \edit{$ 0 + 0 + T$} \\
    \bottomrule
\end{tabular}
\end{table}

\begin{figure}[t]
\centering
\captionbox{\label{fig:summary}\edit{Snapshot of results for representative  methods studied in Sec \ref{exp:comp}. %
$T \in [25,300], \epsilon =0.5, d=4$.}}{\includegraphics[width=0.9\linewidth]{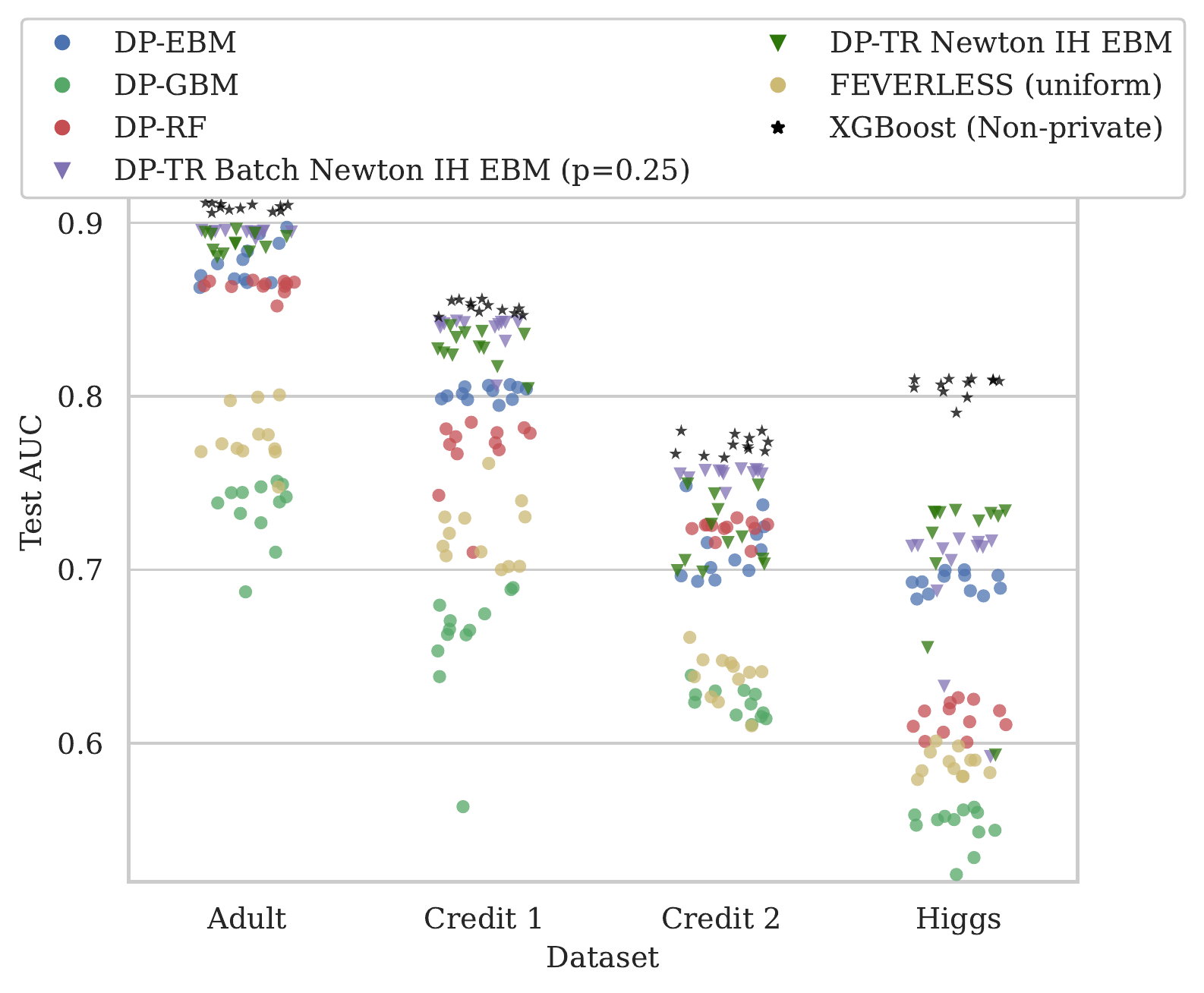}}
\end{figure}

\section{Empirical Evaluation}\label{sec:experiments}
\edit{Sections \ref{sec:core} and \ref{sec:additional} introduced our framework for the private and federated training of GBDT models. In this section we perform a thorough experimental evaluation of the components in our framework. Our main goal is to answer the following questions:}
\begin{figure*}[t!]
\centering
  \subfloat[\label{fig:split-method:vary_t} Varying $T$ with $d=4, \epsilon=1$]{%
       \includegraphics[width=0.3\linewidth]{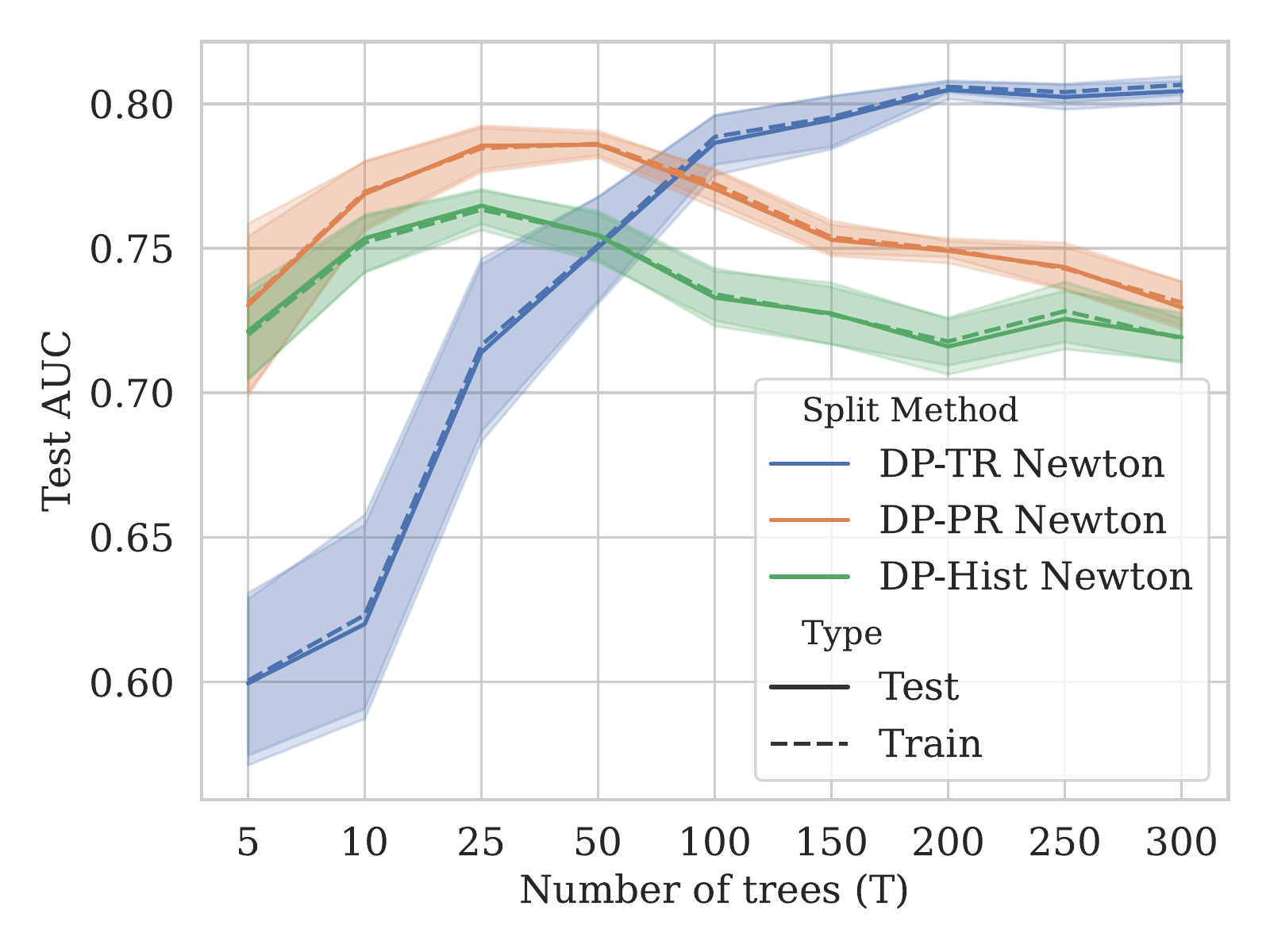}}
  \subfloat[\label{fig:split-method:vary_D} Varying $d$]{%
        \includegraphics[width=0.3\linewidth]{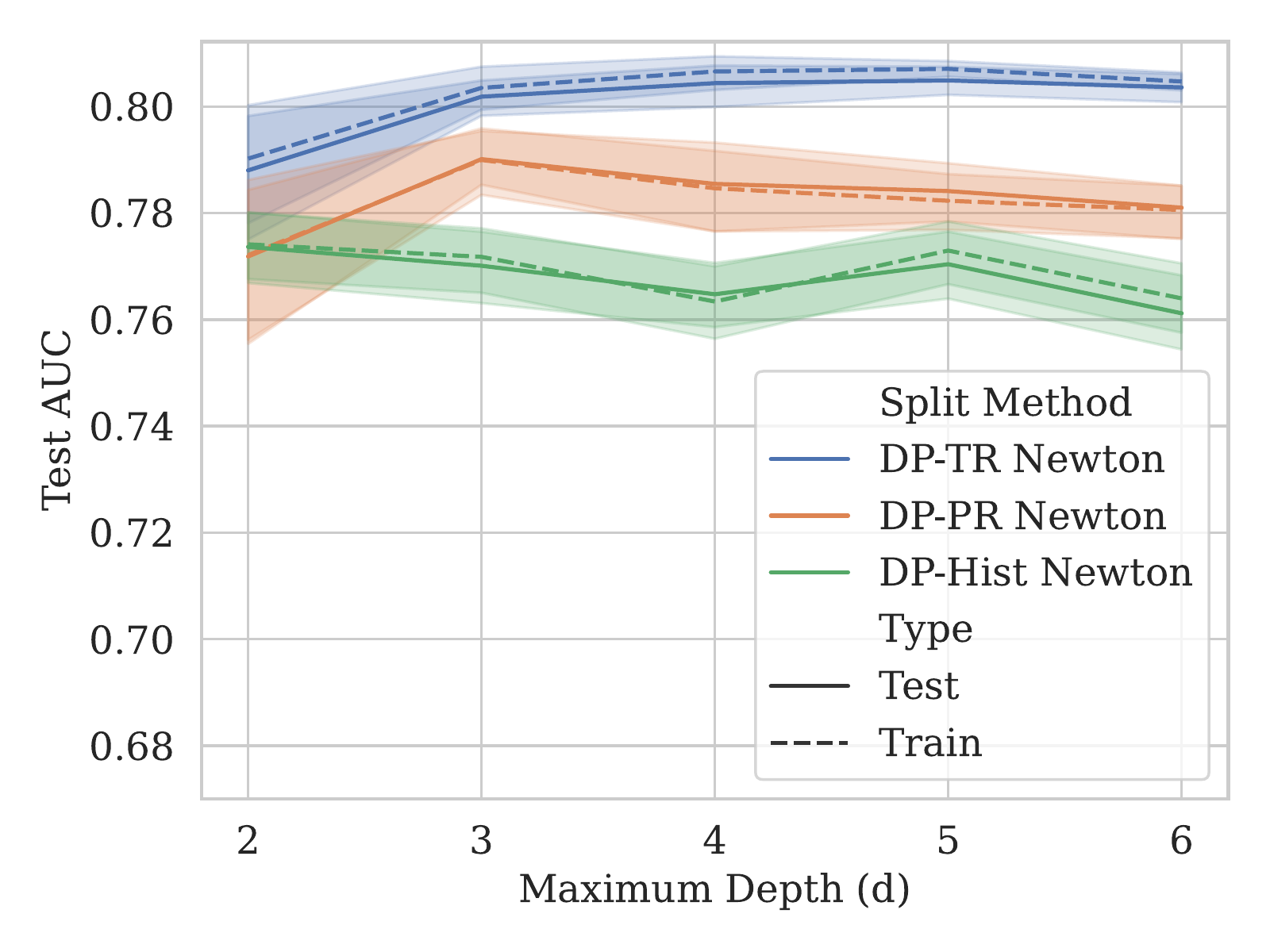}}
  \subfloat[\label{fig:split-method:vary_eps} Varying $\epsilon$]{%
        \includegraphics[width=0.3\linewidth]{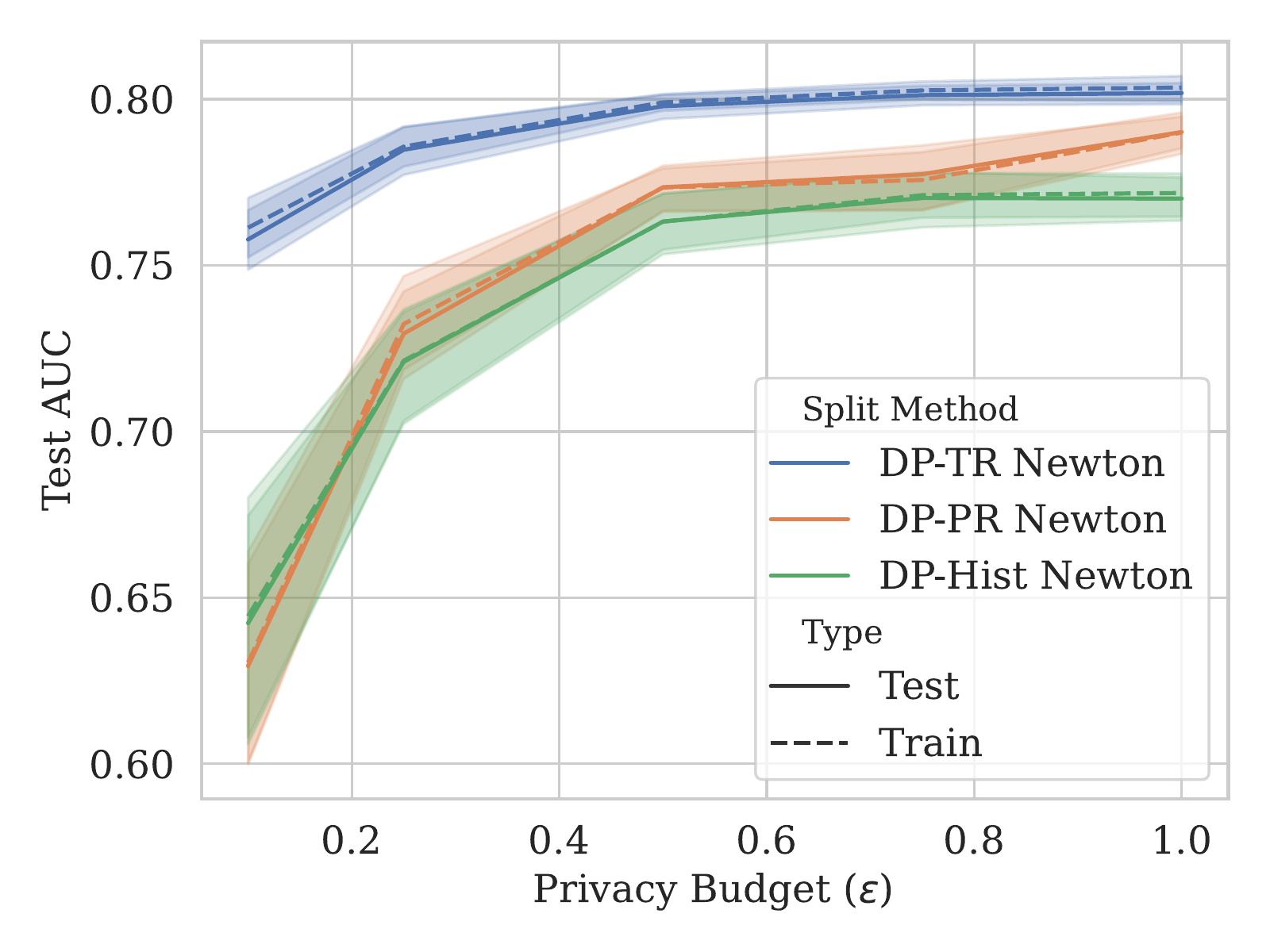}}
  \caption{Split Methods on Credit 1}
  \label{fig:split-method} 
\end{figure*}

\begin{table*}[t]
  \small
  \caption{Weight update methods across the datasets fixing $\epsilon=0.5, d=4$}
  \label{tab:weight_update}
\begin{tabular}{ccccccc}
\toprule
   &        &                       Bank &                   Credit 1 &                   Credit 2 &                      Adult &                      Nomao \\
\midrule
Hist $(T=25)$ & Gradient &           0.6282 +- 0.0688 &           0.5748 +- 0.0852 &           0.6288 +- 0.0569 &           0.6749 +- 0.0524 &           0.8483 +- 0.0138 \\
   & Averaging &           0.7249 +- 0.0274 &            0.6769 +- 0.058 &  \textbf{0.6751 +- 0.0246} &           0.6373 +- 0.0457 &  \textbf{0.8885 +- 0.0038} \\
   & Newton &  \textbf{0.7562 +- 0.0337} &  \textbf{0.7522 +- 0.0162} &           0.6575 +- 0.0486 &  \textbf{0.8013 +- 0.0225} &           0.8758 +- 0.0075 \\
PR $(T=25)$ & Gradient &            0.676 +- 0.0376 &           0.7094 +- 0.0312 &           0.6239 +- 0.0486 &           0.7688 +- 0.0253 &           0.8766 +- 0.0072 \\
   & Averaging &           0.7803 +- 0.0309 &           0.7165 +- 0.0337 &           0.6864 +- 0.0249 &           0.8281 +- 0.0183 &  \textbf{0.8904 +- 0.0055} \\
   & Newton &  \textbf{0.7998 +- 0.0203} &  \textbf{0.7676 +- 0.0196} &  \textbf{0.6882 +- 0.0207} &  \textbf{0.8416 +- 0.0108} &             0.88 +- 0.0072 \\
TR $(T=300)$ & Gradient &  \textbf{0.8508 +- 0.0061} &           0.7847 +- 0.0097 &   \textbf{0.7392 +- 0.008} &  \textbf{0.8737 +- 0.0056} &  \textbf{0.8965 +- 0.0047} \\
   & Averaging &           0.8382 +- 0.0116 &           0.7846 +- 0.0106 &           0.7285 +- 0.0109 &           0.8666 +- 0.0043 &           0.8875 +- 0.0055 \\
   & Newton &           0.8486 +- 0.0075 &  \textbf{0.7983 +- 0.0062} &           0.7344 +- 0.0088 &           0.8718 +- 0.0049 &            0.8883 +- 0.007 \\
\bottomrule
\end{tabular}
\end{table*}
\noindent
    \textbf{1.} \edit{In terms of model performance, what are the best options for each component under our framework?}
    
\noindent    \textbf{2.} \edit{Under privacy, does batching updates improve performance?}
    
\noindent    \textbf{3.} \edit{Can a combination of choices in our framework result in methods that improve over the SOTA baselines discussed in Section \ref{sec:baselines}?}
    
\edit{Figure \ref{fig:summary} shows a snapshot of our findings. We display for a subset of datasets and methods, the average test AUC while fixing the privacy budget $\epsilon = 0.5$. Full results across all datasets are discussed in Section \ref{exp:comp}. We represent baseline methods plotted as circles and new combinations within our framework as triangles. Each point is formed from varying $T \in [25,300]$ in increments of $25$ and is the average test AUC\footnote{\edit{Due to class imbalance, measures such as accuracy are not useful to test performance.}} over 5 runs. We observe that on most datasets we significantly improve over existing methods. In some cases we match the nearest competitor, but often with additional benefits such as reducing the number of rounds of communication. }

\edit{These experimental results, along with others in this section, show that it is possible to train accurate, private and lightweight federated GBDT models with only a small gap behind their non-private counterparts. This conclusion is reached by answering our questions as follows:}

\noindent    \textbf{1.} \edit{In Sections \ref{exp:split_methods}---\ref{exp:split_candidate} we evaluate the choices within each component. We find that the totally-random %
strategy provides a significant reduction in privacy cost and outperforms all other choices. For weight updates we find that utilising Hessian information usually gives better performance with no additional cost, which is similar to the non-private setting. Finally, for split candidates, we find our IH method achieves performance that matches that of (non-private) quantiles with little extra privacy cost.}
    
\noindent    \textbf{2.} \edit{In Section \ref{exp:bb} we study batching updates to help reduce the number of communication rounds. We find this is not the only benefit of batching and in fact, for very high-privacy regimes ($\epsilon < 0.5$), batching updates often gives better model performance than performing boosting for the full $T$ rounds.}
    
\noindent    \textbf{3.} \edit{In Section \ref{exp:comp} we combine the best individual components and compare against our SOTA baselines. We find combining the best options found in each component also results in the best model overall. Specifically, combining batched updates, the IH split candidate method, TR splits and Newton updates often achieves better performance than the most competitive baseline (DP-EBM) and in fewer rounds of communication. %
    }

\subsection{Experimental Setup} 

In our experiments, we use a range of real-world datasets from Kaggle \cite{credit1, credit2} and the UCI repository \cite{Dua:2019}. All datasets are displayed in Table \ref{tab:datasets} in the appendix detailing the number of records ($n$), number of features ($m$), and proportion of the positive class $(p)$. The Higgs dataset has been subsampled to $n=200,000$ for computational reasons. All experiments are repeated 3 times over 5 different 70-30 train-test %
splits resulting in 15 iterations. We measure model performance by the AUC-ROC on the test-set. 
For all boosting experiments we fix the learning rate and regularization parameters $\beta=2, \eta=0.3, \alpha=0$ which generally performed well across all chosen datasets, and do not tune these any further. 
We take $Q=32$ split candidates unless otherwise stated. 
The effect of the number of split candidates is explored in Section~\ref{exp:split_candidate}. In all experiments we use RDP to satisfy $(\epsilon, \delta)$-DP fixing $\delta=1/n$. %
Tests were run with an AMD Ryzen 5 3600 3.6GHz CPU and 16GB of RAM. 
Code for our framework and experiments is open-sourced
\footnote{\url{https://github.com/Samuel-Maddock/federated-boosted-dp-trees}}
\subsection{Split Methods}\label{exp:split_methods}

\begin{figure*}[t!]
\centering
  \subfloat[\label{fig:split_candidates:vary_s} \edit{Varying $s$ with $T=100, d=4$}]{%
       \includegraphics[width=0.3\linewidth]{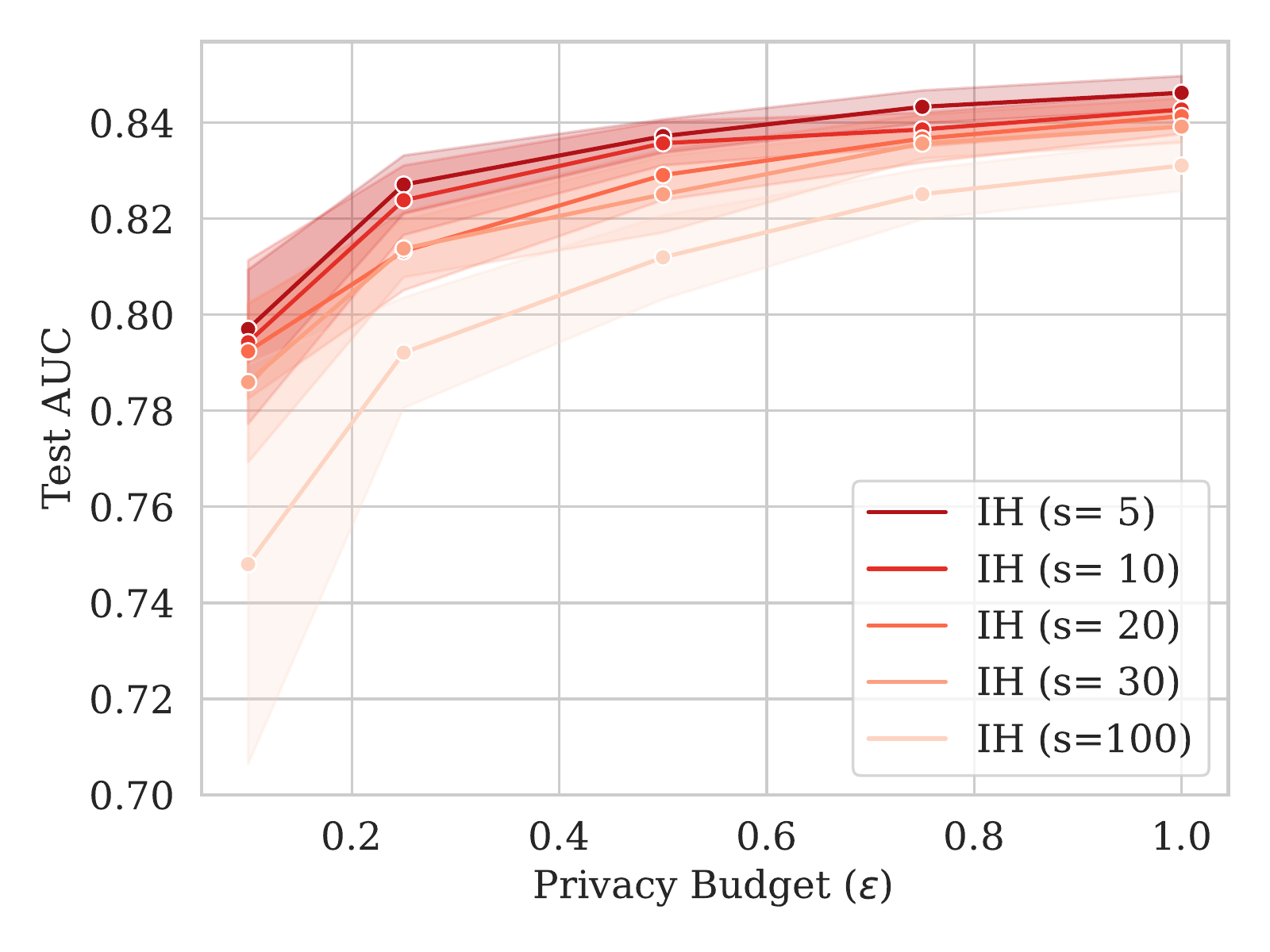}}
  \subfloat[\label{fig:split_candidates:vary_t} \edit{Varying $T$ with $d=4, \epsilon=1$}]{%
        \includegraphics[width=0.3\linewidth]{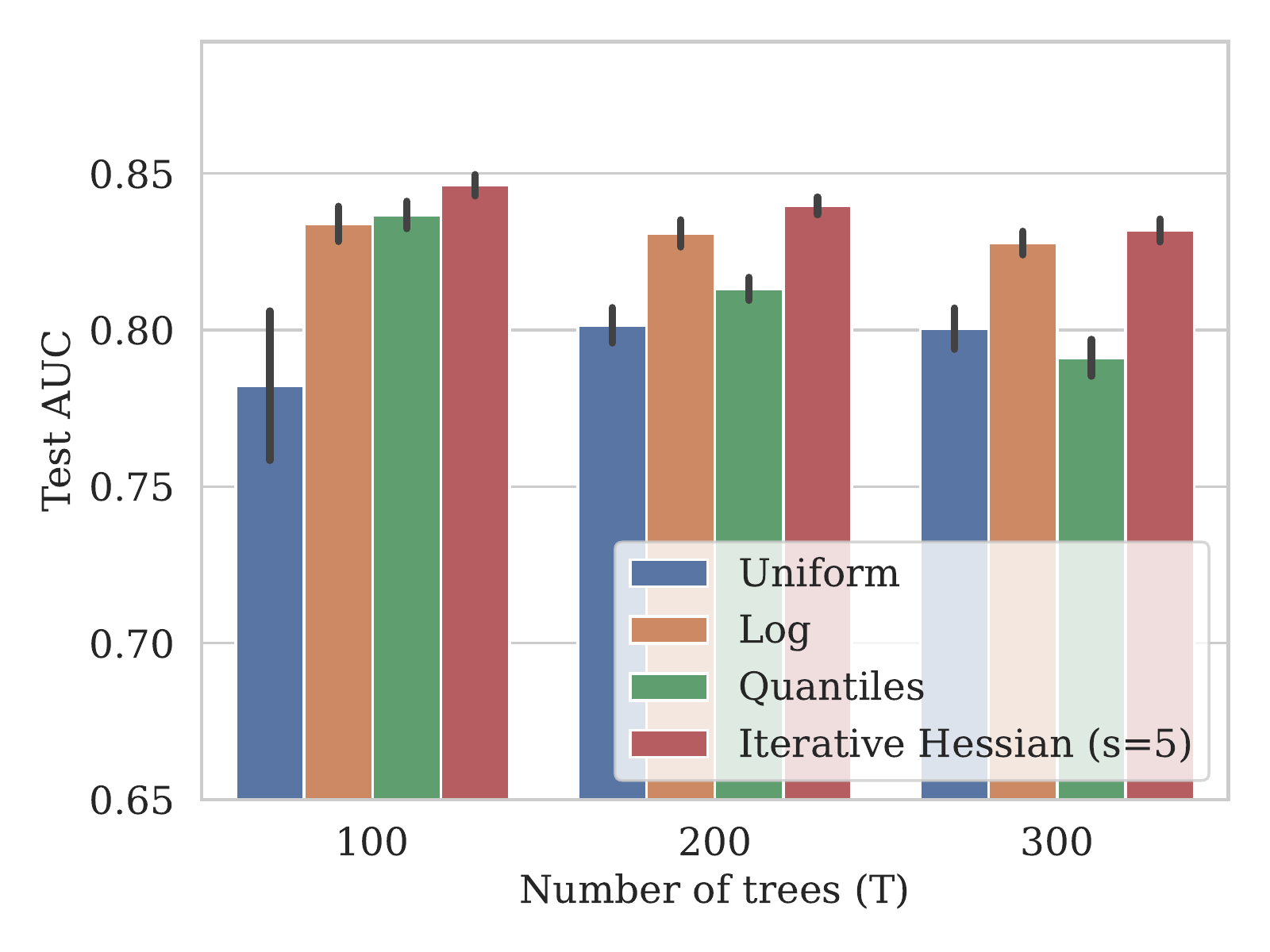}}
  \subfloat[\label{fig:split_candidates:vary_Q} Varying $\edit{Q}$ with $T=100, d=4, \epsilon=1$]{%
        \includegraphics[width=0.3\linewidth]{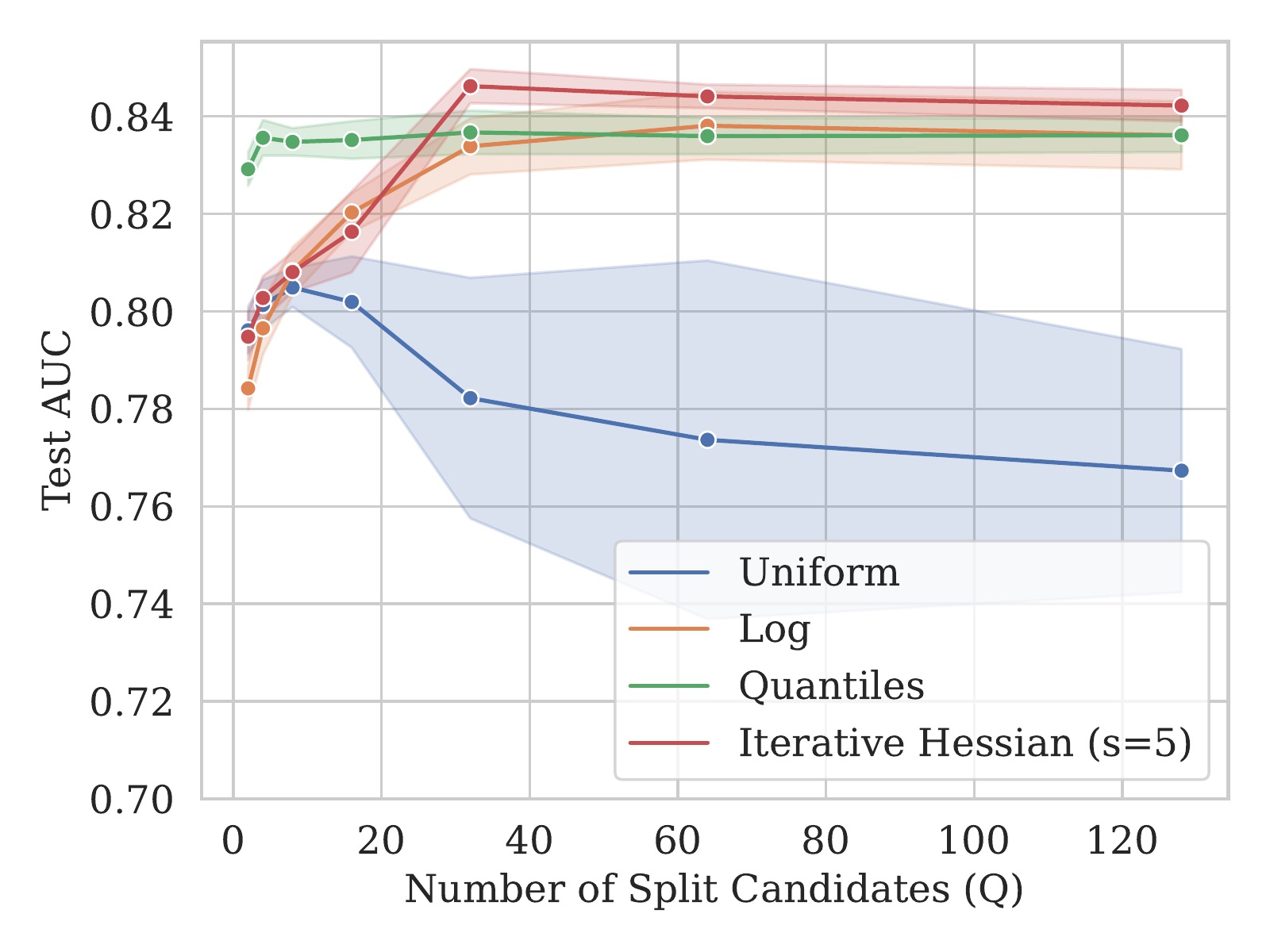}}
  \caption{Split Candidate Methods on Credit 1}
  \label{fig:split_candidates} 
\end{figure*}

We begin by exploring the initial trade-off between the main split-methods: Histogram-Based (Hist), Partially Random (PR), and Totally Random (TR). We study these split methods as we vary parameters that have the largest effect on the AUC of DP-GBDT algorithms: $T$, $d$, and $\epsilon$. 
For now we fix our weight update method to Newton and fix the split candidate method to uniform. We consider the effects of these components separately in Sections~\ref{exp:weight_update} and~\ref{exp:split_candidate}.

Figure \ref{fig:split-method:vary_t} shows the effect of varying the number of trees $T$ while fixing $\epsilon=1, d=4$ on the Credit 1 dataset, and visualises the key differences between the main split methods. Other datasets using the same parameter setup are considered in Appendix \ref{appendix:split_methods}.
Recall that histogram-based and PR are methods that compute split-scores under DP. 
Because they compute split-scores they often \say{converge} to their best test AUC before TR methods in the non-private setting. We can observe that a similar effect occurs in the private setting. 
We see that PR and Hist peak around $T=25-50$ whereas it takes TR $T=300$ trees to achieve its best test AUC.

In the non-private setting this peak is typically caused by overfitting as $T$ gets larger. 
For the private setting this is not quite the case as we can observe little difference in train and test AUC. 
\edit{Instead,} for large $T$ the privacy cost of training a histogram-based GBDT model requires a large amount of noise to be added at each step and this severely impacts performance. 

Recall that both Hist and PR split methods require $Tdm$ queries to train the full model compared to just $T$ for TR. The advantage of TR's minimal privacy cost can be clearly seen from Figure \ref{fig:split-method:vary_t} as it achieves higher AUC than the other two methods. 

In Figure \ref{fig:split-method:vary_D} we fix $\epsilon =1$ and set $T=25$ for Hist/PR and $T=300$ for TR as we vary the maximum depth $d \in \{2,3,4,5,6\}$ on Credit 1. We observe only a small difference in AUC across Hist method and only a minor decrease in performance across TR and PR methods for larger depths. 
For PR and Hist the depth $d$ does increase the privacy cost of each tree but for TR the privacy cost is independent of the depth. 
We observe a small decrease in AUC for TR as $d$ increases and this is likely because training very deep trees can lead to nodes with only a few observations. 
This results in gradient information with magnitude smaller than the noise being added, and hence any meaningful information is lost.

In Figure~\ref{fig:split-method:vary_eps} we vary $\epsilon \in \{0.1, 0.25, 0.5, 0.75, 1\}$ while fixing $d=3$.
We set $T=300$ for TR and $T=25$ for Hist and PR. 
We can immediately make two observations.
First, there is still a clear gap in performance between TR and Hist/PR. 
Second, for large $\epsilon$, PR outperforms the Hist method but for small $\epsilon$ the picture is less clear. 
This is likely due to the additional random variation due to the PR method picking random split candidates.

\noindent
\textbf{Summary.} We recommend using TR splits as it clearly outperforms methods that calculate split scores. This usually results in larger ensembles which can be prohibitive in federated settings. In Section \ref{exp:bb}, we discuss how we can batch updates to get around this.
\subsection{Weight Update Methods}\label{exp:weight_update}
We start this section by asking whether boosted decision trees under DP provide any additional model performance over DP-RFs. 
Table~\ref{tab:weight_update} shows the test AUC across all datasets varying the weight-update method (Gradient, Averaging and Newton) for each split method. 
In these experiments we fix $\epsilon=1, d=4$ and use $T=25$ for PR/Hist and $T=300$ for TR methods. 
The highest AUC for each split method is highlighted in bold.

We can observe that boosting does provide an advantage over the traditional averaging method on these datasets, although it is not completely clear cut. 
Focusing first on the Hist methods we can see that Newton updates perform the best across three of the five datasets -- although results on Credit 2 and Nomao show averaging performs the best. 
However, Newton updates certainly show clearer advantages on Credit 1, Adult, and Bank over both gradient and averaging updates.
This pattern is also present for PR methods with Newton updates performing better than averaging except for Nomao where averaging performs the best.
For TR methods we observe Gradient updates achieve higher AUC on 4 out of 5 of the datasets, although is within random variation of Newton for all datasets except Credit 1, where Newton performs best. 
We also note that the gap in performance between TR and Hist/PR observed in Section~\ref{exp:split_methods} also holds across all the datasets we are considering. 
The impact in performance between Newton and the other weight update methods for TR splits is also less marked than its impact with Hist/PR splits, since the performance of TR with Newton differs by at most $~0.014$ AUC when compared with gradient or averaging. %

\noindent
\textbf{Summary.} We recommend using Newton updates as it exceeds or performs very similarly to Gradient updates and in most cases beats averaging across the split methods. %
We note that averaging methods are certainly still competitive and discuss this further in Section \ref{exp:bb} when we study batched updates.

\subsection{Split Candidate Methods}\label{exp:split_candidate}

\begin{table}[t]
  \small
 \caption{Split candidate methods $T=100, d=4, Q=32, \epsilon=1$}
\begin{tabularx}{\columnwidth}{lXXXX}
{} &                  IH (s=5) &                 Quantiles &                      Log &          Uniform \\
\midrule
Bank         &  \textbf{0.8749} (0.0066) &           0.8695 (0.0087) &          0.8698 (0.0087) &  0.8734 (0.0074) \\
Credit 1     &  \textbf{0.8462} (0.0035) &           0.8367 (0.0045) &          0.8339 (0.0058) &  0.7822 (0.0247) \\
Credit 2     &           0.7377 (0.0084) &            0.738 (0.0083) &  \textbf{0.7495} (0.008) &  0.7461 (0.0092) \\
Adult        &  \textbf{0.8888} (0.0035) &           0.8823 (0.0047) &          0.8848 (0.0054) &  0.8862 (0.0034) \\
Higgs &           0.7211 (0.0181) &  \textbf{0.7352} (0.0082) &           0.688 (0.0141) &  0.6449 (0.0293) \\
Nomao        &  \textbf{0.9026} (0.0041) &           0.8987 (0.0052) &          0.9003 (0.0061) &   0.9021 (0.005) \\
\bottomrule
\end{tabularx}
 \label{tab:split_candidate_methods}
\end{table}

In this section we explore the split candidate methods introduced in Section \ref{framework:split_candidates}. 
We are interested in comparing the Iterative Hessian (IH) method against the private baseline of uniform splitting and the non-private method of quantiles. We mentioned in Section \ref{framework:split_candidates} that Log splits are a viable alternative if we know the skew of features. We will assume that we have prior knowledge about skew and thus Log splits have no extra privacy cost. We will show IH can achieve similar or better results than Log splits with the additional benefit that this prior knowledge is not required. 

\subsubsection{Varying $s$}
One disadvantage of IH splitting when using a TR ensemble is that we must specify the number of split candidate rounds $s$ where budget is spent to produce a Hessian histogram. Figure~\ref{fig:split_candidates:vary_s} shows the effect of $s \in \{5,10,20,30,100\}$ on the Credit 1 dataset with $T=100$ trees while varying $\epsilon \in [0.1,1]$ with DP-TR Newton. 
For higher values of $\epsilon$ there is not so much difference between calculating a Hessian histogram for each round ($s=100$) compared to calculating a Hessian histogram for only $s=5$ rounds. 
Although there is a clear trend that on Credit 1 only $\sim 5$ rounds of IH are needed.
As $\epsilon$ decreases this difference becomes more apparent. 
When $\epsilon = 0.5$ we see a $0.04$ difference in AUC between $s=5$ and $s=100$. At $\epsilon = 0.1$ spreading the already thin privacy budget to compute Hessian information at each tree results in drastically worse performance with $s=100$.
Hence when $\epsilon$ %
is small, spending more of it on
the Hessian histogram results in similar models to using uniform split candidates \edit{and} we lose the benefits of more informed split candidates. 

\subsubsection{Comparison of methods}
In Figure \ref{fig:split_candidates:vary_t} we fix $s=5$ for IH and compare the performance on Credit 1 against the other split candidate methods: Uniform, Log, and Quantiles.  
We vary $T \in \{100, 200, 300\}$ and fix $\epsilon = 1$. 
Consistently across the different parameter settings the Log splits perform well. 
This is because Credit 1 contains many skewed features. 
However, IH with $s=5$ (our private variant) can indeed match and in some cases exceed Log splits. 
This indicates that proposing and refining split candidates around (noisy) Hessian histograms is a useful method when datasets have skewed features. We also note that uniform split candidates perform the worst out of all split candidate methods on Credit 1.
We also observe here that quantiles (the common choice for non-private boosting methods such as XGBoost) do not lead to the best AUC under DP. 
In particular\edit{,} there is \edit{a} large gap for $T=200,300$. 
Yet for $T=100$, quantiles perform similarly to Log and IH candidates.

In Appendix \ref{appendix:split_candidates_vary_eps} we detail an additional experiment varying $\epsilon$ with the split candidate methods. We observe that even if $\epsilon$ is small, IH still outperforms the other methods.

In Table~\ref{tab:split_candidate_methods} we compare the split candidate methods across all the datasets using the same parameter setting. 
Our IH method shows a clear advantage over uniform on Credit 1 and Higgs where features are particularly skewed.
On Credit 2 our IH method achieves the worst performance. However, it does match quantiles in performance. This suggests that quantiles do not produce the best split candidates for Credit 2. It is also likely that because Credit 2 has a large number of categorical features that the repeated splitting in IH serves no additional benefit and could be detrimental to performance.
On other datasets none of the features have any notable skew and all split candidate methods perform \edit{equally} well. %

\subsubsection{Varying $Q$}
The advantage of the IH method is its more informed split candidates for very skewed features. One may think that we can circumvent the issues of uniform splitting by increasing the number of split candidates, thus considering more fine-grained candidates. In Figure~\ref{fig:split_candidates:vary_Q}, we fix $T=100, d=4, \epsilon=1$ and vary $Q \in \{2,4,8,16,32,64,128\}$ on Credit 1. 
\edit{The same experiment is detailed on other datasets in Appendix \ref{appendix:split_candidates_vary_Q}.} We can immediately observe further issues with uniform split candidates when combined with TR splits. While proposing more candidates results in fine-grained split choices, the variance from choosing such splits completely at random results in very variable performance when using $> 32$ split candidates. The experiment supports our choice of $Q=32$ in other experiments, and also shows that the IH method is relatively robust to the initial number of split candidates.

\noindent
\textbf{Summary:} We recommend using the IH method to iteratively refine split candidates over a small number of rounds, finding that $s=5$ usually works the best. Other private split methods like Uniform and Log are competitive depending on the dataset.

\begin{figure}[t]
\centering
\captionbox{\label{fig:bb:low_eps}Batched updates, $T=200, d=4$}{\includegraphics[width=0.8\linewidth]{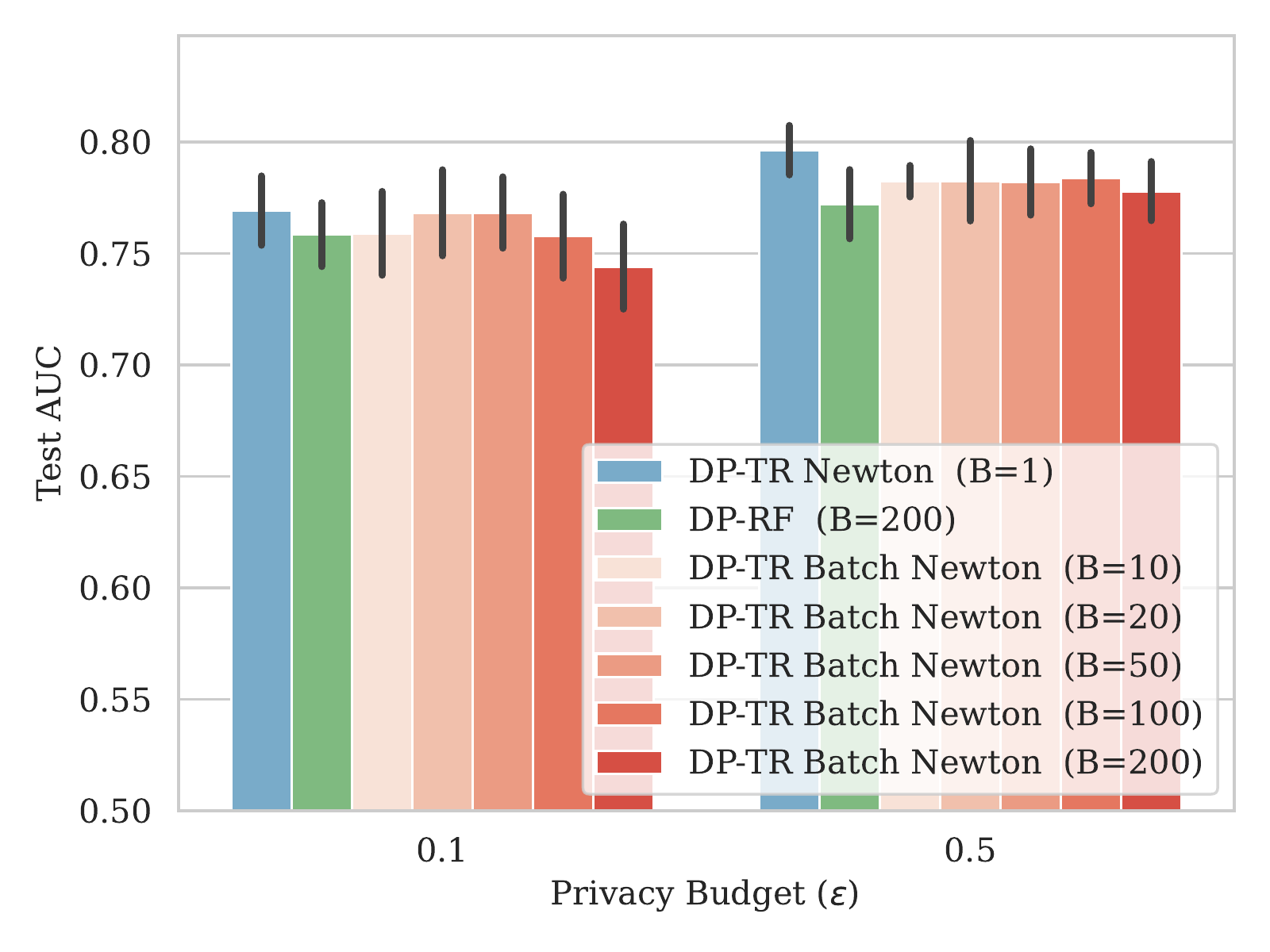}}
\end{figure}

\begin{table}[t]
    \centering
    \small
    \caption{Batched updates fixing $T=200, \epsilon=0.1, d=4$.}
    \label{tab:bb}
    \begin{tabularx}{\columnwidth}{lXXXXX}
\toprule
{} &                     Bank &                 Credit 1 &                  Credit 2 &                   Adult &                     Nomao \\
\midrule
Batch (B=10)  &          0.7876 (0.0233) &          0.7585 (0.0147) &            0.719 (0.0156) &         0.8438 (0.0086) &           0.8859 (0.0056) \\
Batch (B=20)  &  \textbf{0.819} (0.0108) &          0.7591 (0.0194) &  \textbf{0.7199} (0.0164) &  \textbf{0.86} (0.0057) &  \textbf{0.8929} (0.0055) \\
Batch (B=200) &          0.7752 (0.0143) &          0.7578 (0.0194) &             0.71 (0.0146) &         0.8443 (0.0065) &           0.8858 (0.0061) \\
DP-RF (B=200) &          0.7663 (0.0127) &          0.7441 (0.0196) &           0.7106 (0.0106) &         0.8382 (0.0105) &           0.8852 (0.0058) \\
Newton (B=1)  &          0.7866 (0.0224) &  \textbf{0.7693} (0.016) &            0.695 (0.0134) &         0.8371 (0.0148) &           0.8669 (0.0088) \\
\bottomrule
    \end{tabularx}
\end{table}

\begin{figure}[t]
  \centering
  \includegraphics[width=0.5\textwidth]{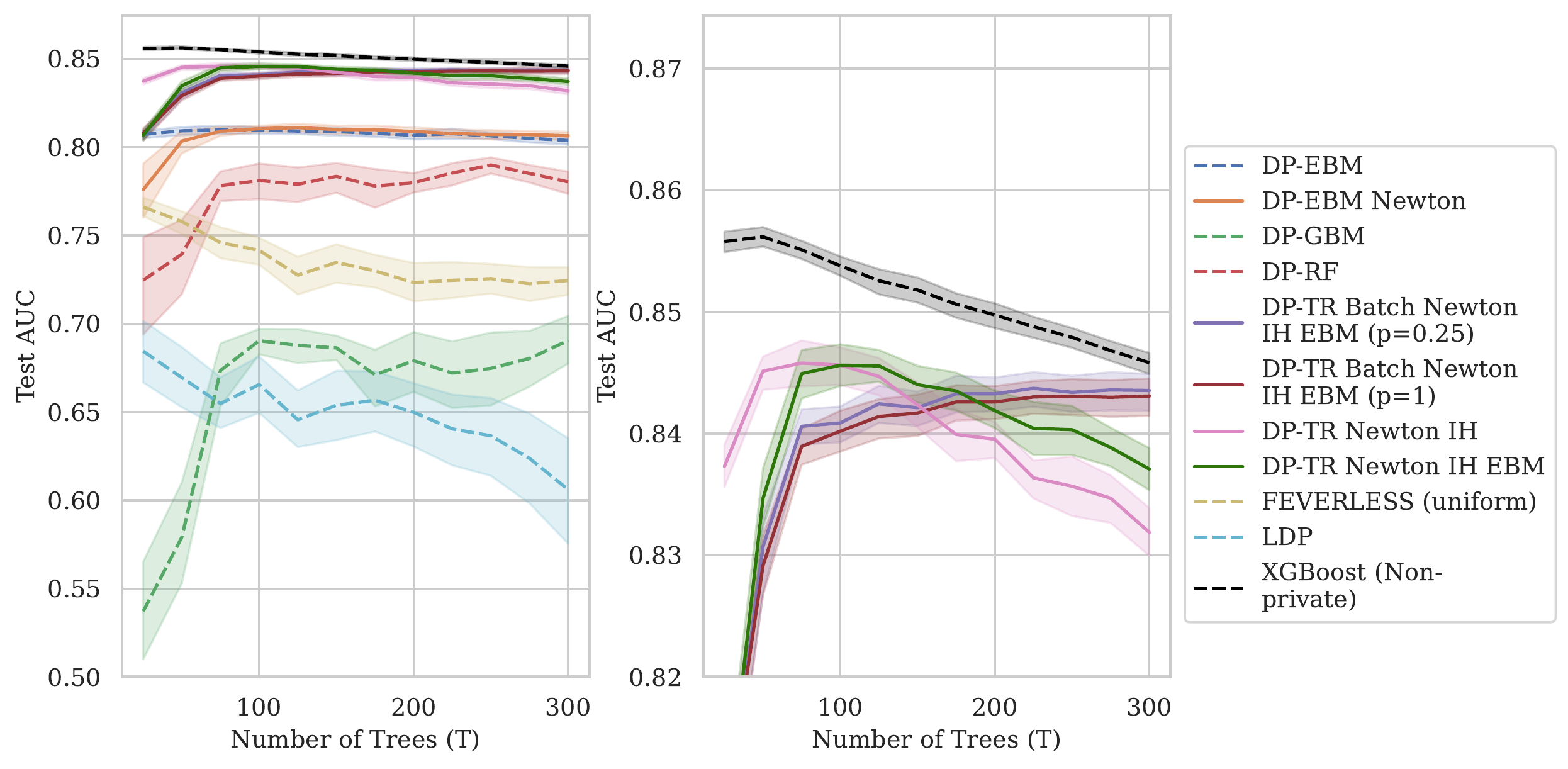}  \caption{\edit{Comparison of DP-GBDT methods and LDP baseline on Credit 1, $d=4, \epsilon=1$; Left (all methods), Right (zoomed)}}
  \label{fig:comparisons}
\end{figure}

\begin{table}[t]
\small
\caption{\edit{Average rank of methods across datasets---rank 1 for highest AUC. $(*)$ indicates new methods in our framework}} %
\begin{tabularx}{\columnwidth}{lXXX}
\toprule
& & $\epsilon$ & \\ 
Methods &            0.1 &            0.5 &            1.0 \\
\midrule
DP-EBM                             &           5.83 &            4.5 &            3.5 \\
DP-EBM Newton $(\ast)$                      &            4.0 &           3.33 &           3.17 \\
DP-GBM                             &            9.0 &            9.0 &            9.0 \\
DP-RF                              &            4.5 &           6.67 &            7.0 \\
DP-TR Batch Newton IH EBM (p=0.25) $(\ast)$ &  \textbf{1.17} &  \textbf{2.33} &           3.33 \\
DP-TR Batch Newton IH EBM (p=1) $(\ast)$  &            2.0 &            3.5 &           4.67 \\
DP-TR Newton IH $(*)$                    &           5.33 &            4.5 &           3.83 \\
DP-TR Newton IH EBM $(\ast)$             &           5.17 &           3.17 &  \textbf{2.67} \\
FEVERLESS (uniform)                &            8.0 &            8.0 &           7.83 \\
\bottomrule
\end{tabularx}
\label{tab:rank}
\end{table}

\subsection{Batched Updates}\label{exp:bb}
In Section~\ref{framework:batched_updates} we discussed that boosting is an inherently sequential process and so can take a large number of communication rounds in distributed settings. 
This is exacerbated by the TR method that often requires a large number of trees (rounds) to achieve good performance. 
We proposed the idea of batching updates by averaging weights across multiple trees before performing a boosting round. 
In Figure~\ref{fig:bb:low_eps} we vary $\epsilon = 0.1, 0.5$ and fix $T=200, d=4$ on the Credit 1 dataset. The same experiment on other datasets is presented in Appendix \ref{appendix:bb}. 
We compare the Newton method which takes $T=200$ rounds and the averaging method which only takes 1 round. 
We then consider batched updates, varying the size of the batch as $B=p\cdot T$ for $p \in \{0.05, 0.1, 0.25, 0.5, 1\}$. 

Focusing first when $\epsilon = 1$ we observe that the Newton model achieves the best performance. 
This is followed by batched updates that perform some amount of boosting (i.e, $B < 200$). As an example taking $B=100$ results in only 2 rounds of boosting. 
A surprising observation is that limiting to 2 rounds of communication achieves a very similar performance to the full Newton model that requires 200 rounds of boosting. When $\epsilon = 0.1$ Newton boosting still performs the best but we observe batched updates with $B=20,50$ and thus only a small number of boosting rounds perform very similarly.

To study this more closely, we present a similar experiment in Table \ref{tab:bb}, fixing $\epsilon=0.1$ 
We vary the batch size $B$ and compare to averaging and Newton boosting across all the datasets. We consider TR trees, uniform split candidates, and $T=200, d=4$. 
We still observe that batched updates is a surprisingly competitive alternative to the full boosting procedure across all datasets. 
We note as in Figure~\ref{fig:bb:low_eps} that all methods on Credit 1 are roughly within random variation of one another. The difference in methods is more striking on other datasets with batched updates of size $B=20$ performing better than Newton. This suggests that under a setting where more noise is added to the training process, boosting is a more unreliable method as it attempts to correct mistakes from previous rounds and can lead to overcompensating for noise. By batching updates we help to average out noise and boost a smoothed update. 
Generally, batched methods with $B=20$ or $B=50$ achieve the best performance with $10$ and $5$ rounds of boosting respectively. 
In most cases taking $B=200$, resulting in a single round of communication (and no actual boosting) only loses at most $0.04$ AUC compared to other batched update methods.

\noindent
\textbf{Summary.} We recommend batching Newton updates to reduce communication rounds and have shown it loses little in performance. Under high privacy, small batches ($p=0.25-0.5$) seem to give the best performance and even beat private Newton boosting.

\noindent
\subsection{End-to-end Comparisons}\label{exp:comp}
We conclude with comparisons between baseline methods \edit{and those formed from selecting} the best options \edit{found in previous sections}.

\noindent
\textbf{Summary across datasets} %
\edit{In Table \ref{tab:rank} we display the average rank of a method across each of the 6 datasets when ranked in terms of their mean test AUC, where a rank of 1 indicates the highest AUC. We fix $T=100$ and vary $\epsilon \in \{0.1, 0.5, 1\}$. We observe that most baseline methods underperform and rank consistently in the lower half. The closest competitor DP-EBM performs well when $\epsilon=1$ but is beaten by DP-TR Newton IH EBM which consistently ranks higher across datasets. When $\epsilon$ is small, our batch boosting variant consistently ranks the best across all  datasets. }

\textbf{Discussion on Credit 1} \edit{To investigate further,} we fix $\epsilon =1, d=4$ and vary $T$ on Credit 1 in Figure \ref{fig:comparisons}. We present comparisons on other datasets in Appendix \ref{appendix:comparisons}. 
These results are best broken down into four main observations which reflect conclusions from previous sections. 
The first observation is the performance of histogram-based methods. DP-GBM performs the worst followed by FEVERLESS. This shows (as in Section \ref{exp:weight_update}) that Newton updates when combined with histogram-based methods do increase model performance over normal gradient updates, but in either case, training a large tree with many features entails adding a large amount of noise into the training process and generally poor models.

The second remark concerns the performance of the TR methods. 
We see a clear performance gap between the DP-RF and DP-TR Newton methods which indicates that boosting does enhance performance when compared to DP-RF. 
This was confirmed in Section~\ref{exp:weight_update} where we observed Newton updates under DP generally provided better performance than gradient and averaging updates.

Thirdly, while DP-EBM is very competitive, we can achieve similar AUC by using Newton updates and not training for the full $Tm$ rounds as in \cite{nori2021accuracy}. 
In Figure \ref{fig:comparisons} DP-EBM trains $Tm$ trees, corresponding to $10T$ on Credit 1. 
Instead our DP-EBM Newton variant uses Newton updates and trains $T$ trees. 
This shows that we can get the same performance as DP-EBM with far fewer trees when using Newton updates, while
reducing communication rounds.

Finally, we note the performance of batched methods when combined with EBM and IH split candidates. 
We see batched methods with $p=0.25, 1.0$ essentially match the performance of DP-TR Newton IH and achieve similar performance to the top methods on this dataset. 
When compared to the full 200 rounds needed for DP-TR Newton IH there is a negligible loss in performance ($< 0.01$ AUC) but a dramatic reduction in communication rounds.

\noindent
\textbf{Summary.} \edit{By} combining the best options in each component (TR, Newton updates, IH, EBM, and batches with $p=0.25$) we achieve competitive performance that \edit{often} outperforms our baselines.

\section{Conclusion}\label{sec:conclusion}
We have proposed a framework for the differentially private training of GBDT models in the federated setting. %
By %
evaluating different options at each stage of our framework we have found a dominant approach based on random splits, Newton updates, cyclical training and our iterative Hessian (IH) method. Our approach often outperforms SOTA methods on a range of datasets and \edit{results in models close in performance to non-private counterparts}. %
When combined with batching updates, one can train models in only a small number of communication rounds for little loss in performance.

\begin{acks}
This work is supported by the UKRI Engineering and Physical Sciences Research Council (EPSRC) under grants EP/W523793/1, EP/R007195/1, EP/V056883/1, EP/N510129/1 EP/W037211/1 and EP/S035362/1. This material is also based on work supported by DARPA under agreement number 885000, Air Force Grant FA9550-18-1-0166, ARO with grant W911NF-17-1-0405 and the National Science Foundation (NSF) with grants 1646392, 2039445, CNS-2220433, CNS-2213700, CCF-2217071, CCF-FMitF-1836978, SaTC-Frontiers-1804648 and CCF-1652140.
\end{acks}

\newpage
\bibliographystyle{ACM-Reference-Format}
\bibliography{references}

\appendix

\begin{table}[h]
  \caption{Notation\label{tab:notation}}
  \begin{tabularx}{\columnwidth}{lX}
    \toprule
    \textbf{Notation} & Meaning \\
    \midrule
    $D$ & Dataset\\
    $\mathcal{T}, T, d$ &     Forest $\mathcal{T}$ of $T$ trees, each of max depth $d$\\
    $ i \in [n]$ & Observation $i$ in the full dataset of $n$ records \\
    $j \in [m]$ & Feature index, $m$ features\\
    $k$ & Maximum number of feature interactions \\
    $l \in [L]$ & Number of leaf nodes \\
    $q \in [Q]$ & $Q$ is the number of split candidates and $q$ is the index \\
    $S_j = \{ s_1^j, \dots s_Q^j\}$ & $s_q^j$ is $q$-th split candidate for feature $j$ \\
    $g_i^{(t)}, h_i^{(t)}$ & Gradient and hessian of observation $i$ \emph{at the start} of step $t$ \\
    $\eta, \beta, \gamma$ & Regularisation parameters \\
    $\ell(y_i, \hat{y}_i)$ & Loss function \\ 
    \bottomrule
\end{tabularx}
\end{table}

\begin{algorithm}[h]
\caption{Iterative Hessian Splitting}\label{alg:hess}
\begin{algorithmic}[1]
    \Input $Q$ - target number of candidates, $H^{(t)}_j$ - Hessian histogram for each feature calculated at the previous tree
    \For{each feature $j \in [m]$}
        \State $S^{(t+1)}_j = \emptyset, \forall j=1,\dots,m$
        \State Compute the target per-bin Hessian $\theta = \frac{\sum_{q \in [Q]}H_{jq}^{(t)}}{Q}$
        \For{each bin $q \in H^{(t)}_j$}
            \If{$H^{(t)}_{j,q} < \theta$}
                \State Merge bin $q$ with bin $q+1$ via 
                $
                    H^{(t)}_{j, q+1} = H^{(t)}_{j,q} + H^{(t)}_{j, q+1}
                $
            \Else 
                \State Calculate the bin midpoint 
                $
                    s^* = (s_{q-1}^{j}+ s_{q}^{j})/2
                $
                \State Add $s_{q-1}^j, s_{q}^j, s^{*}$ to $S^{(t+1)}_j$
            \EndIf
        \EndFor
        \State Calculate the number of empty bins $b = Q/|S^{(t+1)}_j|$
        \For{each $i \in S_j^{(t+1)}$}
            \State Add $b$ splits to $S^{(t+1)}_j$ uniformly over  $[s_{q-1}^{j}, s_{q}^{j}]$
        \EndFor
    \EndFor
    \State \Return new split candidates $\{S_1^{(t+1)}, \dots, S_m^{(t+1)}\}$
\end{algorithmic}
\end{algorithm}

\begin{table}[h]
  \caption{Datasets}
  \label{tab:datasets}
  \begin{tabular}{cccc}
    \toprule
    \textbf{Dataset} & \textbf{n} & \textbf{m} & \textbf{p}\\
    \midrule
    Credit 1 \cite{credit1} & 120,269 & 10 & 0.07 \\
    Credit 2 \cite{credit2} & 30,000 & 23 & 0.22 \\
    Adult \cite{adult} & 32,651 & 14 & 0.24 \\
    Nomao \cite{nomao} & 34,465 & 10 & 0.28 \\
    Bank Marketing \cite{bank} & 45,211 & 16 & 0.11 \\
    Higgs (subset) \cite{baldi2014searching} & 200,000 &  28 & 0.47  \\
    \bottomrule
\end{tabular}
\end{table}

\section{Further Experiments}

In this appendix we detail additional experiments that were omitted from the main text. 
In Section \ref{sec:experiments} we presented experiments for each component in our framework. To do so, we often showed results on the Credit 1 dataset before varying methods across all of the datasets. This appendix contains plots in the same style of Section \ref{sec:experiments} but on datasets other than Credit 1. \edit{We present the full details of all datasets used in our experiments in Table \ref{tab:datasets}}.

\subsection{Split Methods: Other Datasets}\label{appendix:split_methods}
\begin{figure*}[t]
\centering
  \subfloat[Varying $T$]{%
       \includegraphics[width=0.25\linewidth]{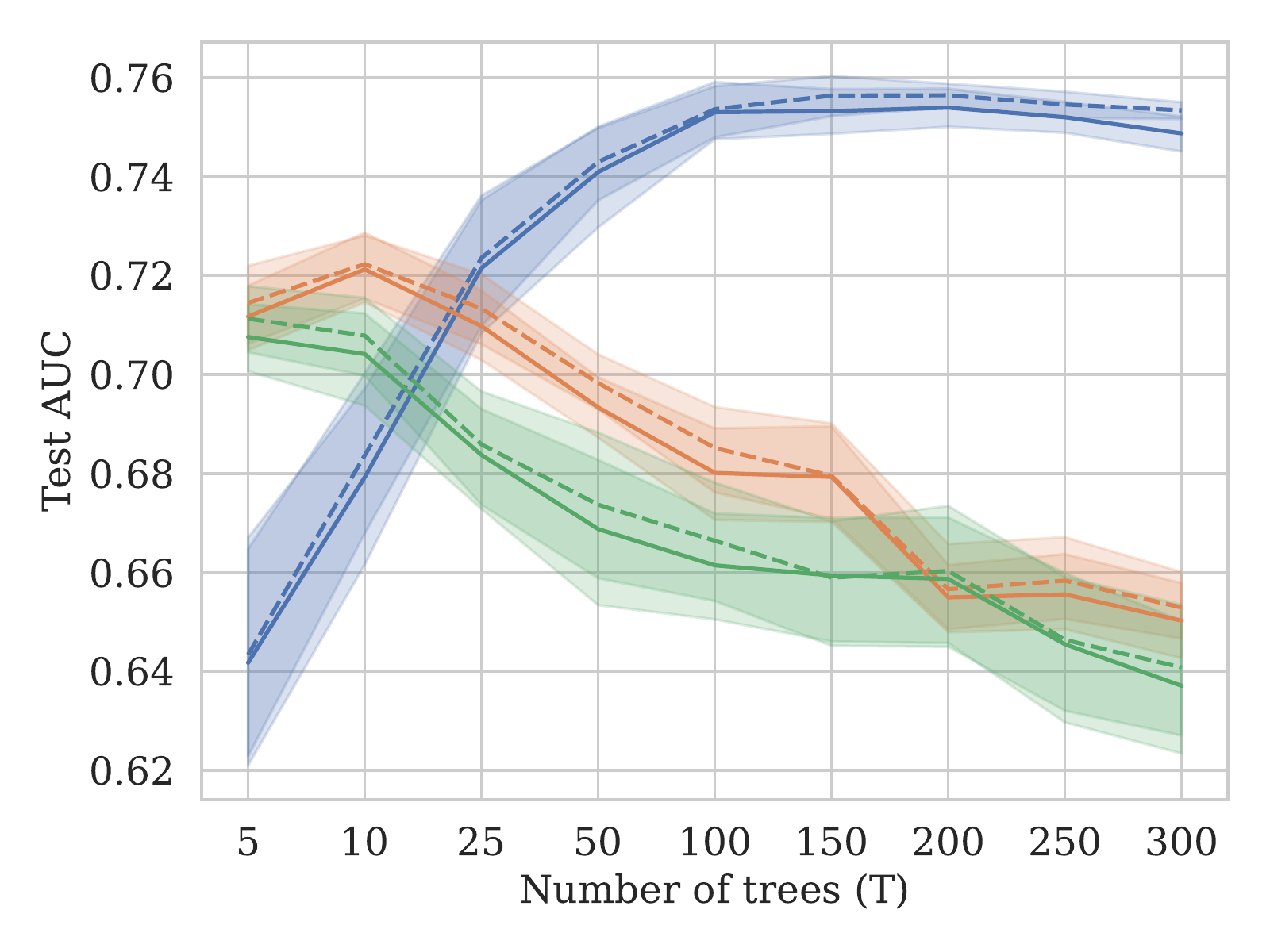}}
  \subfloat[Varying $D$]{%
        \includegraphics[width=0.25\linewidth]{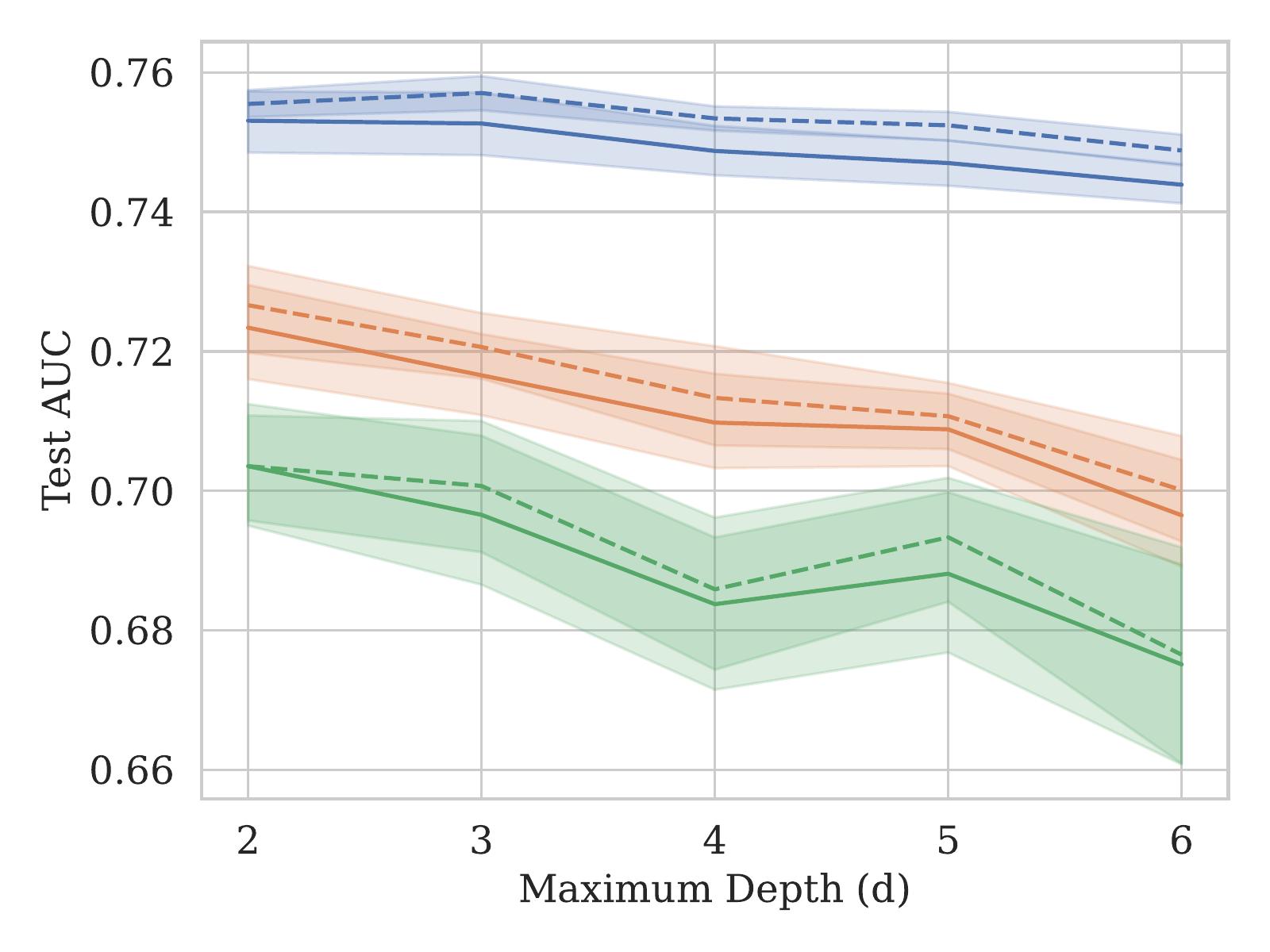}}
  \subfloat[ Varying $\epsilon$]{%
        \includegraphics[width=0.25\linewidth]{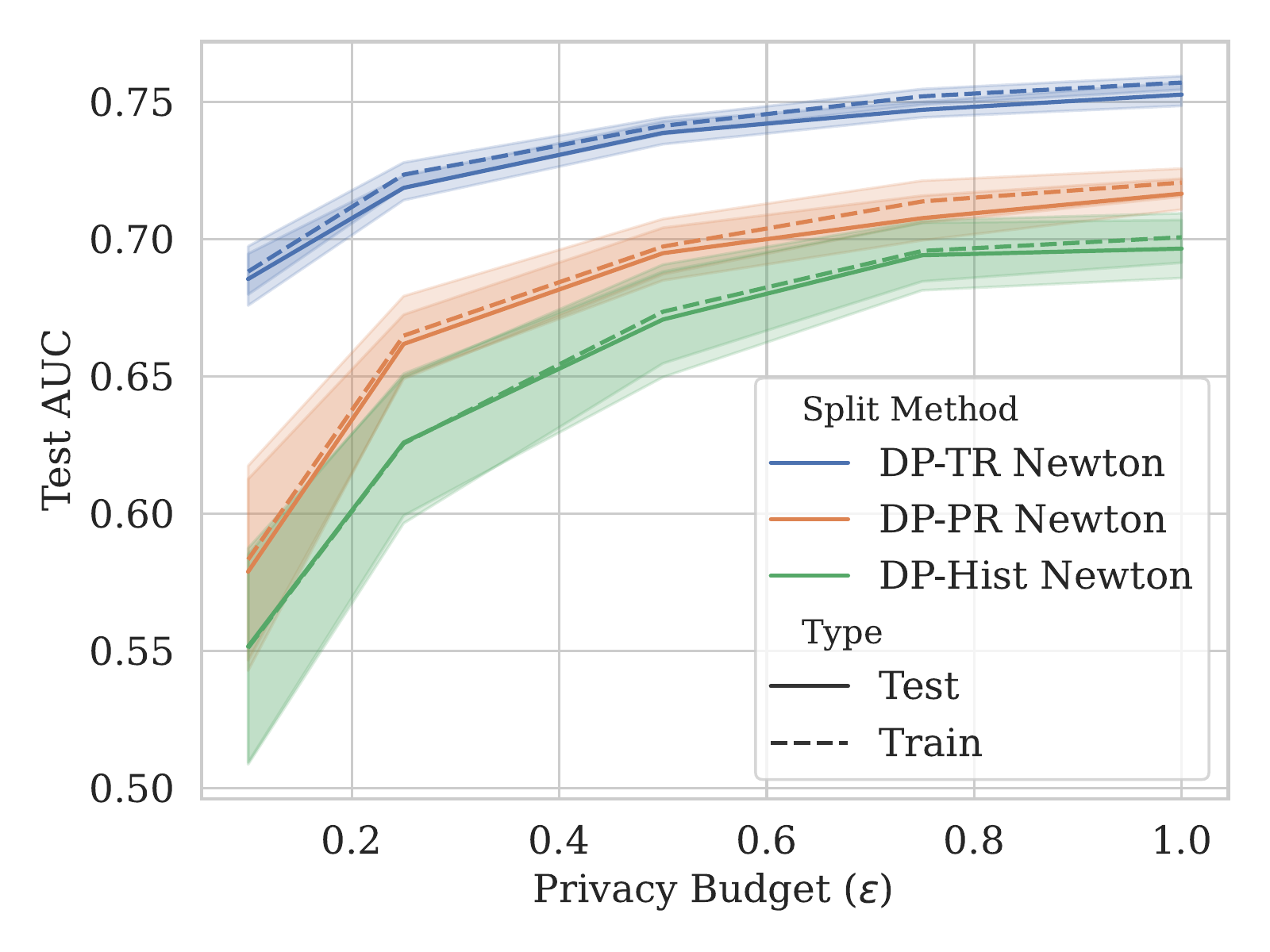}}
  \caption{Split Methods on Credit 2\label{fig:c1_cred2}}
\end{figure*}

\begin{figure*}[t]
\centering
  \subfloat[Varying $T$]{%
       \includegraphics[width=0.25\linewidth]{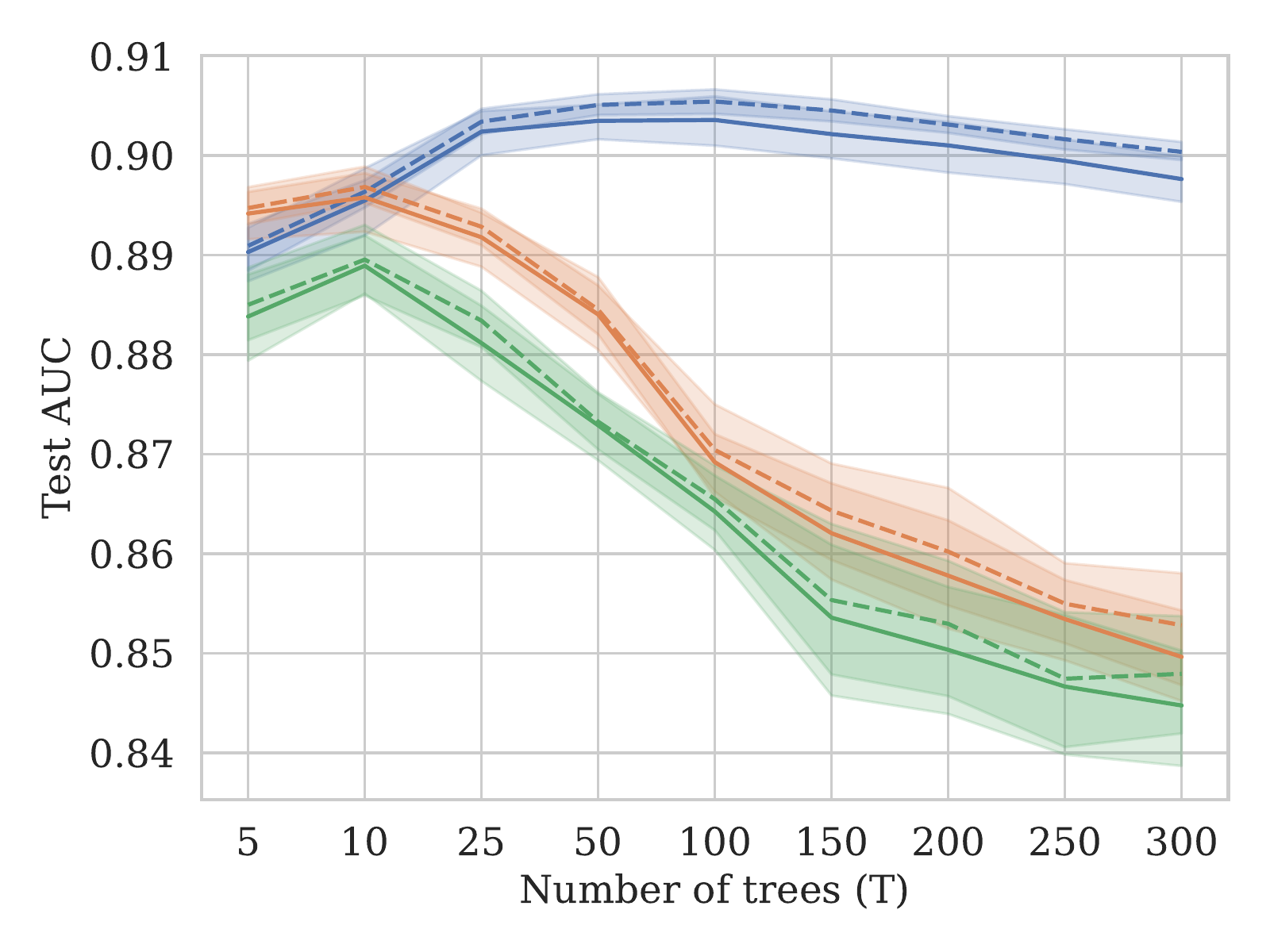}}
  \subfloat[Varying $D$]{%
        \includegraphics[width=0.25\linewidth]{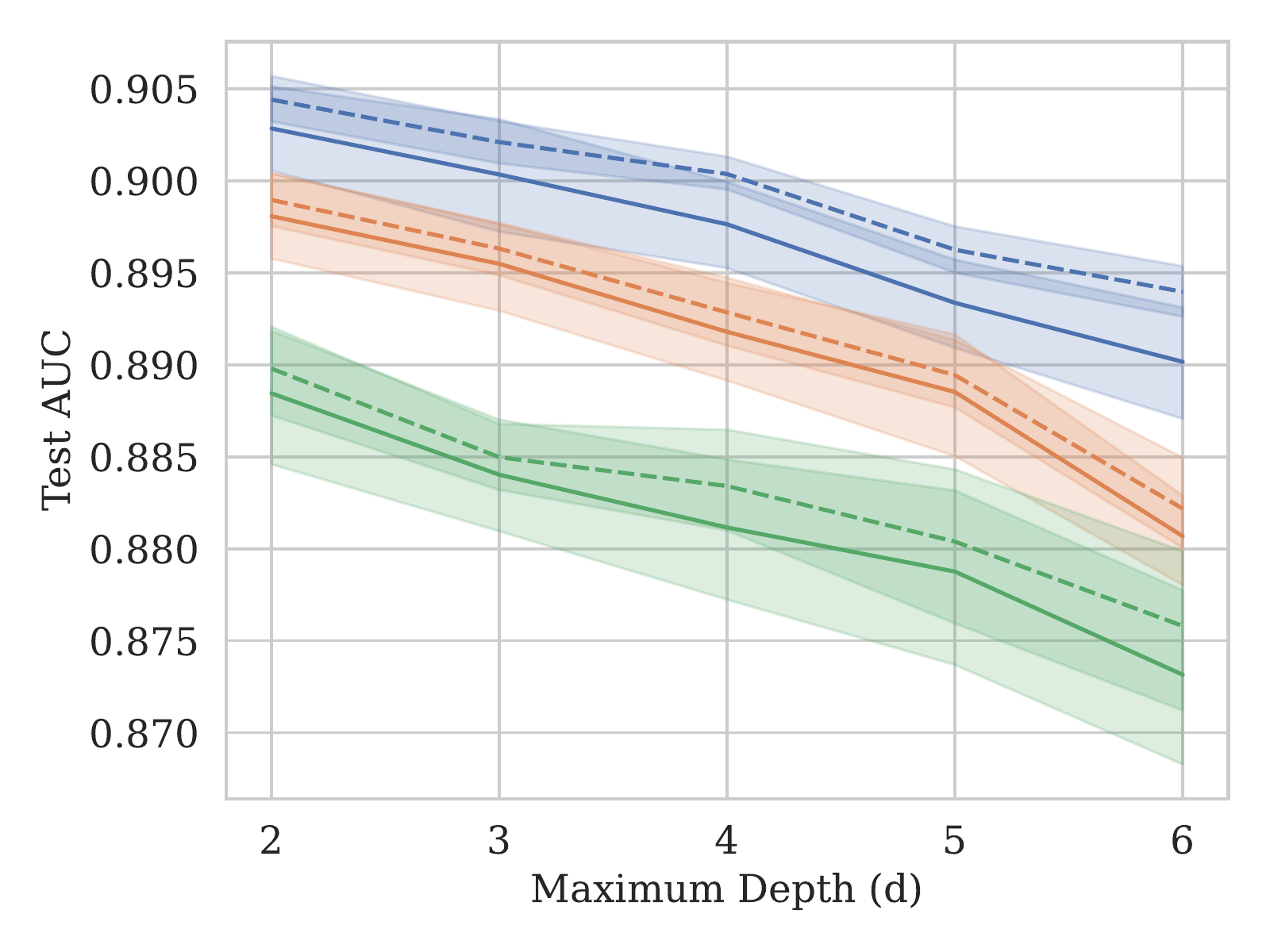}}
  \subfloat[ Varying $\epsilon$]{%
        \includegraphics[width=0.25\linewidth]{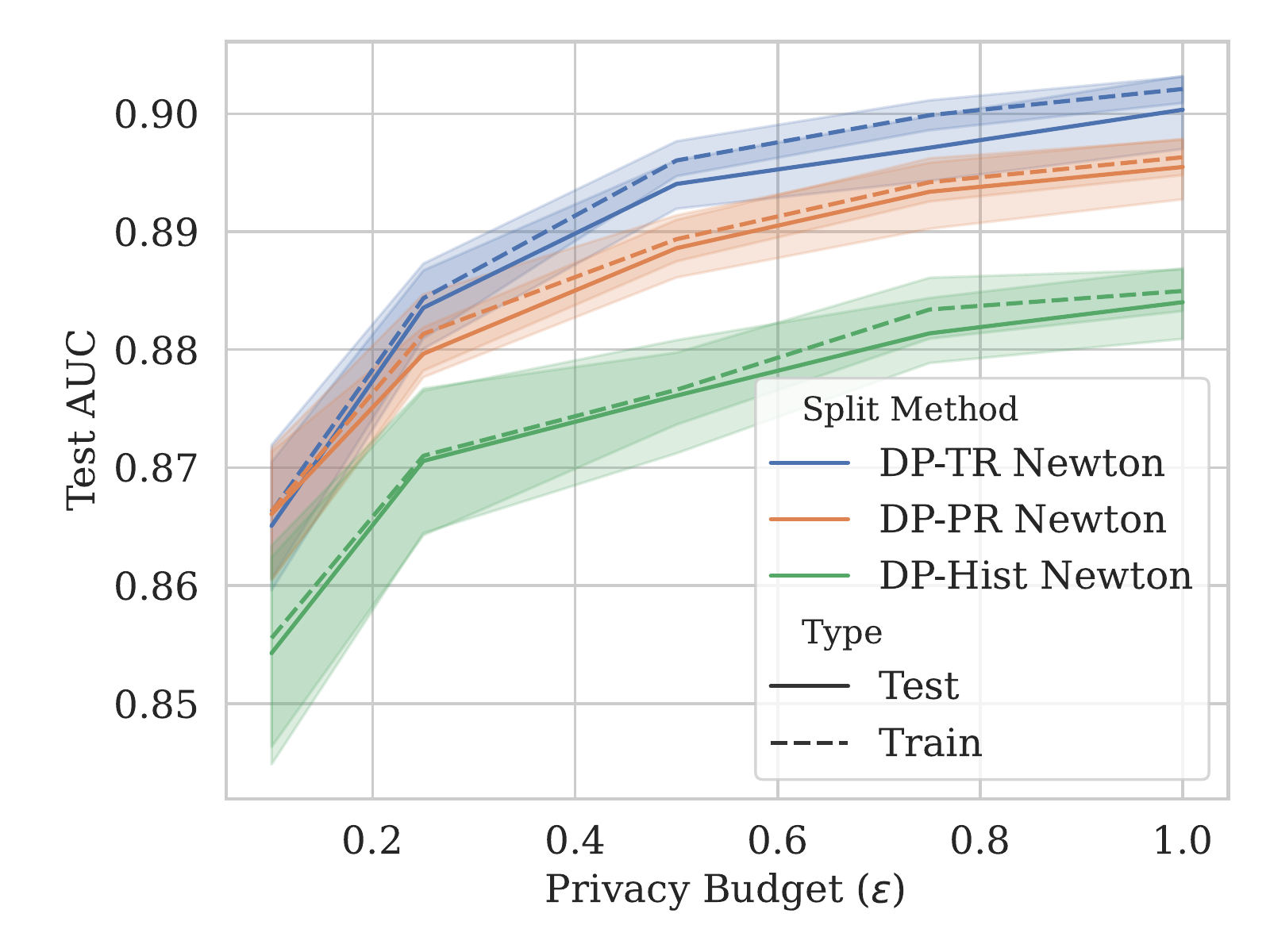}}
  \caption{Split Methods on Nomao\label{fig:c1_nomao}}
\end{figure*}

In Section \ref{exp:split_methods} we presented Figure \ref{fig:split-method} which looked at split methods while varying $T$, $d$ and $\epsilon$ on Credit 1. Here we present the same setup but for the Credit 2, Adult, Bank and Nomao datasets. These are presented in Figures \ref{fig:c1_cred2}---\ref{fig:c1_nomao}. To summarise, the main conclusions from Section \ref{exp:split_methods} also hold across the other datasets. We still observe the same differences when varying $T$ with totally random (TR) splits obtaining the best performance followed by partially random (PR) and then histogram-based (Hist). The gap in AUC performance between Hist and TR is also consistent across the datasets. Furthermore, Hist and PR tend to achieve their best results when $T$ is small (i.e $T \in [10, 50]$) which further justifies our choice of $T=25$ for Hist/PR in Table \ref{tab:weight_update}.

When varying the maximum depth $(d)$ we again find a consistent pattern across the datasets with a performance decrease as we increase the maximum depth. As in Section \ref{exp:split_methods} this is likely due to the fact that deeper trees result in leaf nodes with a smaller number of observations and adding noise to this can destroy the weight update information. For some datasets (Adult, Bank and Nomao) there is some evidence of overfitting as $d$ increases which also explains the decrease in performance. Finally, when varying $\epsilon$ the trend is again consistent with Section \ref{exp:split_methods} and across other datasets. One observation of interest is that for very small $\epsilon$ on the Nomao dataset all three split methods perform very similarly.

\subsection{Weight Updates: Varying $\epsilon$}\label{appendix:weight_updates}
In Section \ref{exp:weight_update} we presented Table \ref{tab:weight_update} which varied the choice of weight updates across the datasets fixing $d=4, \epsilon=0.5$. 
In this appendix we also present similar tables but varying the epsilon values $\epsilon \in \{0.1, 0.25, 0.75, 1\}$. 
These are shown in Tables \ref{tab:weight_update_eps=0.1}---\ref{tab:weight_update_eps=1}.

In Section \ref{exp:weight_update} (Table \ref{tab:weight_update}) we observed that, for $\epsilon=0.5$, Newton updates generally performed the best across the split methods although were occasionally beaten by gradient and averaging updates. 
We make two further observations here. The first is that when we have a larger privacy budget $(\epsilon = 0.75, \epsilon =1)$ and thus less noise the Newton updates more clearly outperform averaging and gradient updates. 
The second is to observe that under a higher-privacy setting $(\epsilon =0.1, \epsilon=0.25)$ averaging updates start to outperform both gradient and Newton updates. This implies that under a setting with more noise, averaging the weights across trees is more effective than boosting. This also provides more evidence for our results with batched boosting in Section \ref{exp:bb} and in Appendix \ref{appendix:bb} where we observe a similar effect in high-privacy regimes.

\subsection{Feature Interactions}\label{appendix:interactions}

\begin{figure*}[t]
\centering
\captionbox{\label{fig:ebm:vary_k} Cyclical $k$-way on Credit 1}{     \includegraphics[width=0.3\linewidth]{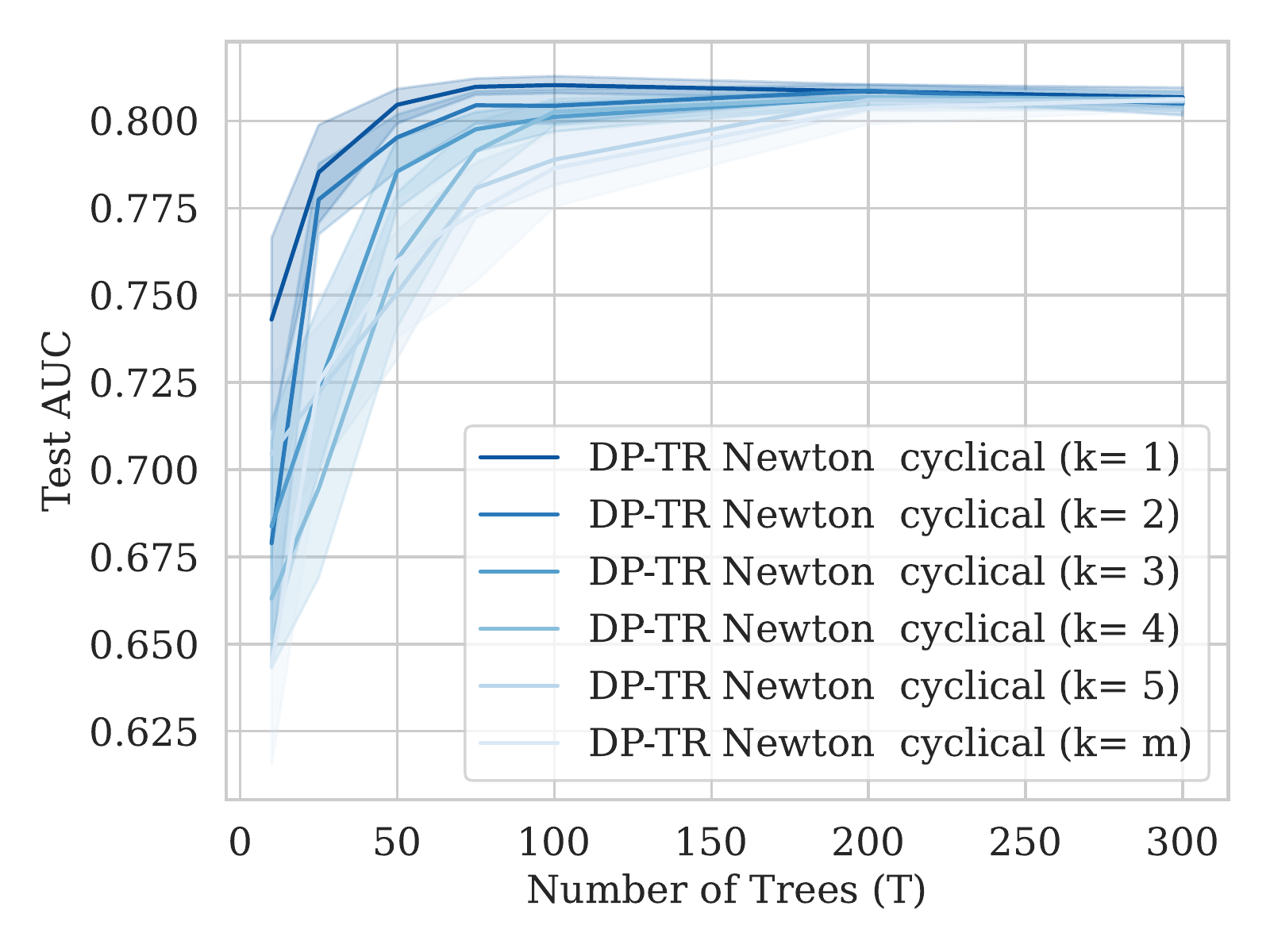}}
\captionbox{\label{fig:ebm:dp} $k=1$ vs $k=m$ on Credit 1}{\includegraphics[width=0.3\linewidth]{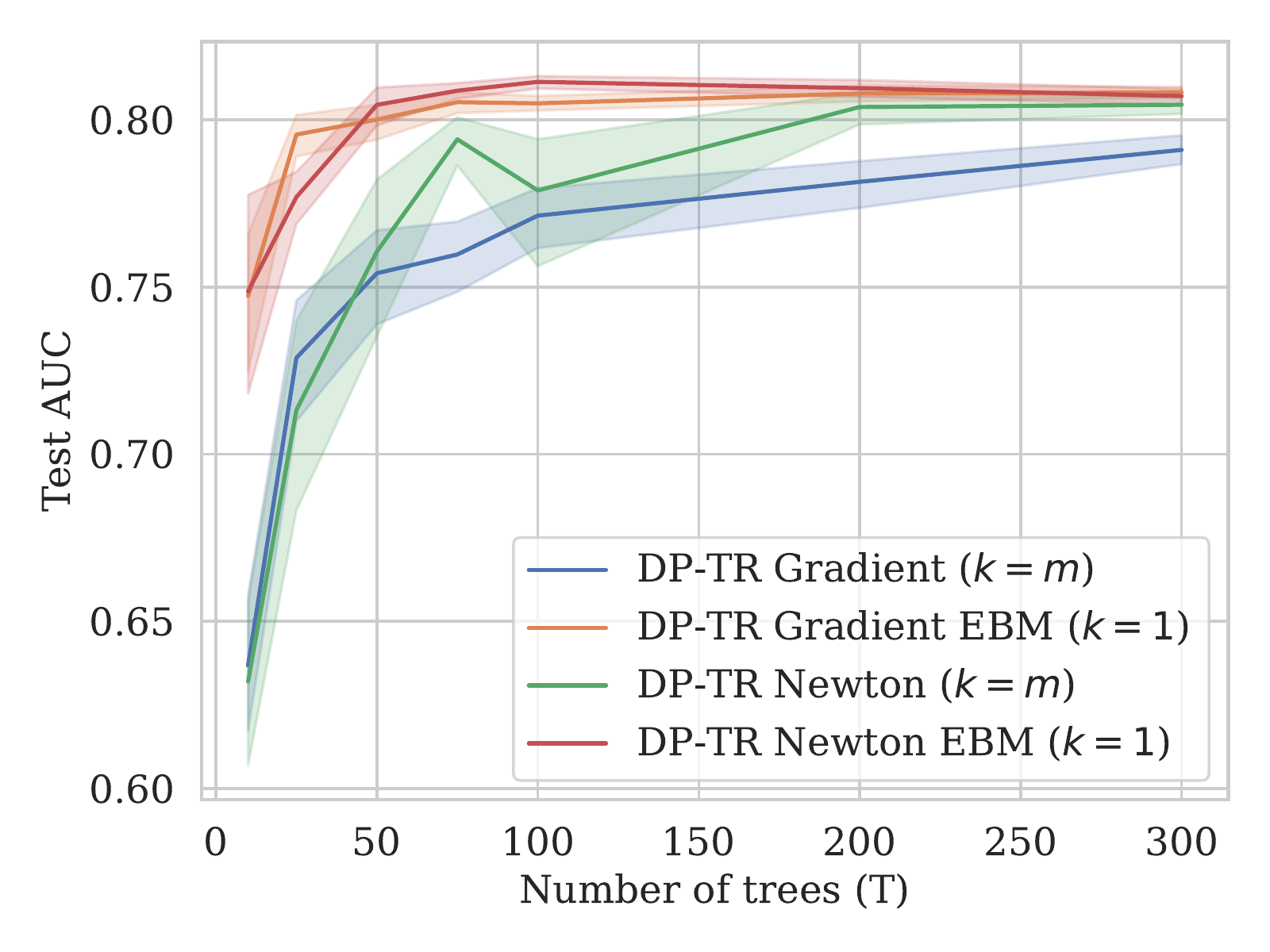}}
\captionbox{Varying $\epsilon$ and split candidate methods on Credit 1 with TR Newton \label{fig:split_candidates:vary_eps}}{     \includegraphics[width=0.3\linewidth]{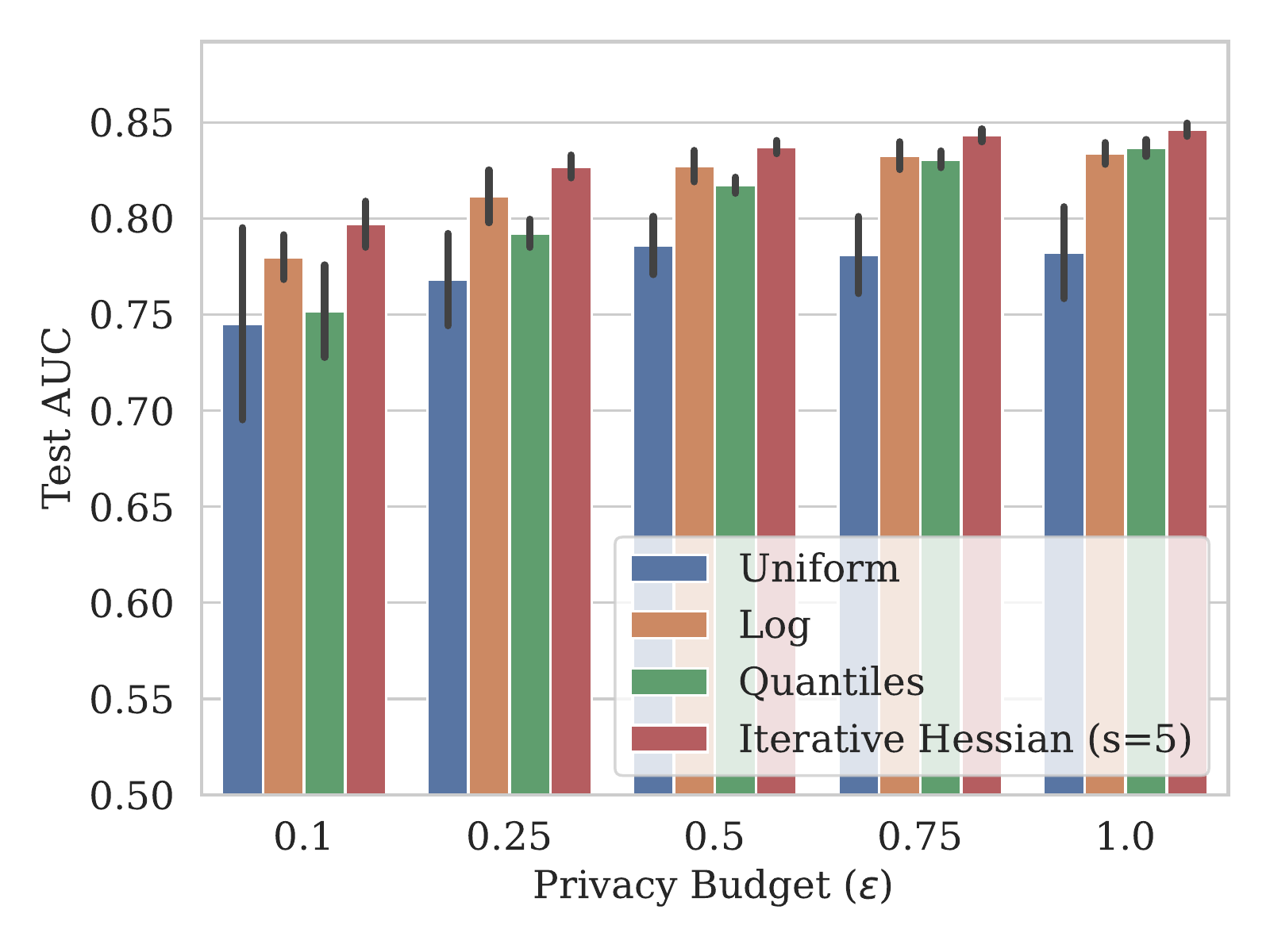}}
\end{figure*}

\edit{We mentioned in Section \ref{framework:feature_interactions} that cyclical $k=1$ (i.e, EBM) performed the best and thus we use this setting in our experimental study. In this section we study the value of $k$ and its effect on performance. In Figure \ref{fig:ebm:vary_k} we vary $T$ with the TR Newton method, fixing $\epsilon = 1$ and study the cyclical training method with $k \in \{1,2,3,4,5,m\}$. 
We note that varying $k$ has no effect on the privacy budget being spent for TR methods. 
We observe a clear pattern when $T$ is small, that methods which split only on a small subset of features per tree obtain the highest AUC. 
When $k=1$, the cyclical method corresponds exactly to the EBM training method. 
When $T$ gets large there is little difference between the feature interaction methods as all  essentially converge to the highest test AUC they can obtain under DP with totally random (TR) splits. We also observed a very similar pattern when studying the random feature interaction method that takes $k$ random features per tree, but omit this plot.}

\edit{We next look at the effect that cyclical training with $k=1$ (i.e., EBM) has on model performance when compared to the standard ($k=m$) method which is free to split on any feature when building a tree. Figure~\ref{fig:ebm:dp} shows the result of an experiment fixing $\epsilon=1$.
There is a clear gap between cyclical and non-cyclical methods in AUC performance and a gap between Newton and Gradient updates, but this gap is lessened when using cyclical training. 
We also observe that cyclical models seem to achieve the highest test AUC in a smaller number of trees than their non-cyclical counterparts. 
It remains the case that Newton updates provide slightly better performance than GBM updates (as noted for Credit 1 in Section \ref{exp:weight_update} and other datasets in Appendix \ref{appendix:weight_updates}) but this gap narrows considerably when using cyclical training. }

\subsection{Split Candidates: Varying $\epsilon$ on Credit 1}\label{appendix:split_candidates_vary_eps}

In Section \ref{exp:split_candidate} we studied the different split candidate methods when using DP-TR Newton. In Figure~\ref{fig:split_candidates:vary_eps} we present the same methods while varying $\epsilon \in \{0.1, 0.25, 0.5, 0.75, 1\}$ and fixing $T=100$. As mentioned in the main text we still observe that the IH method performs well across the different choices of $\epsilon$ and provides a clear advantage on Credit 1 over uniform and even quantiles for small $\epsilon$.

\subsection{Split Candidates: Other Datasets}\label{appendix:split_candidates_vary_Q}
\begin{figure*}[b]
\centering
  \subfloat[Higgs]{%
       \includegraphics[width=0.3\linewidth]{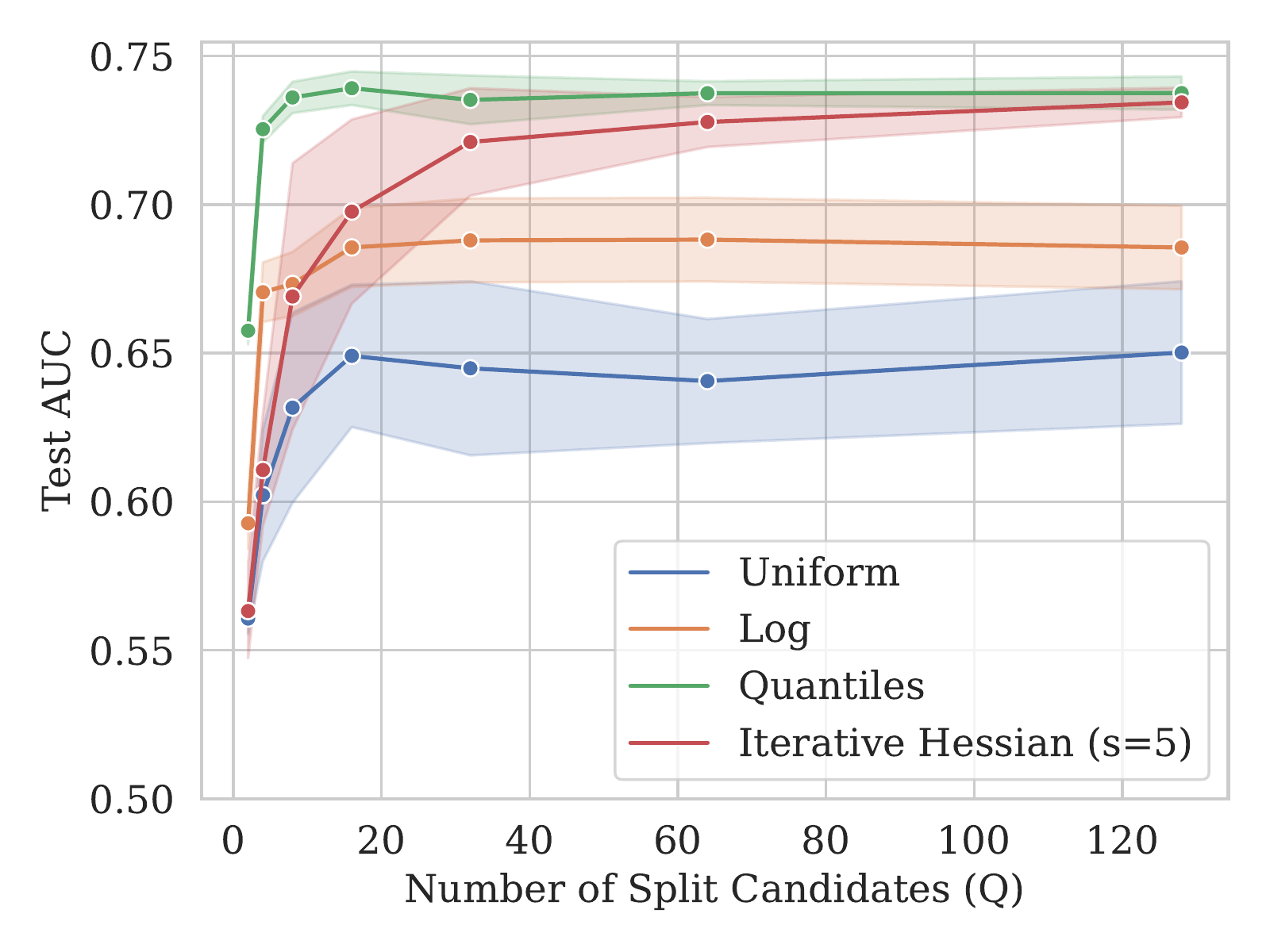}}
  \subfloat[Credit 2]{%
        \includegraphics[width=0.3\linewidth]{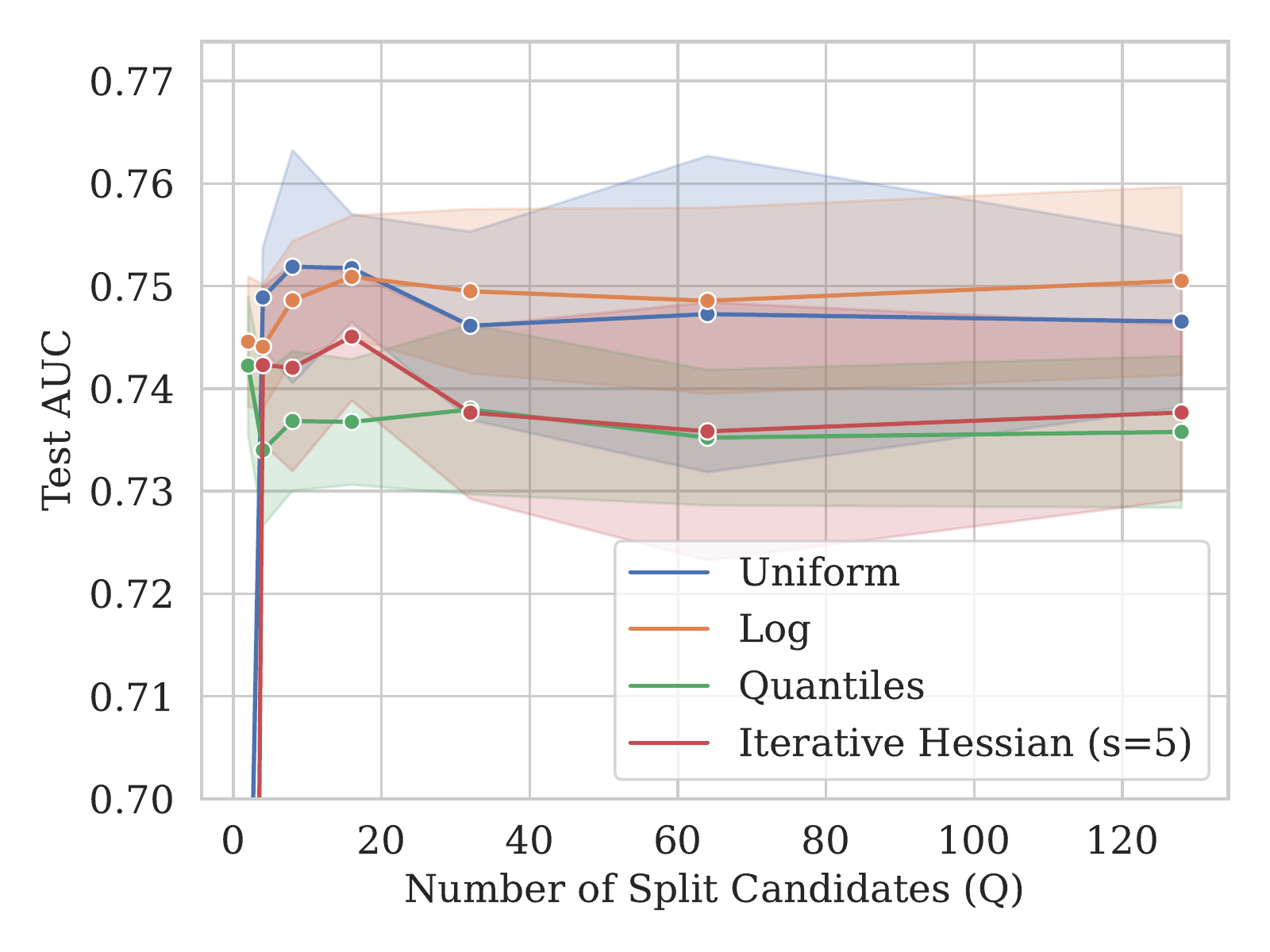}}
  \subfloat[Adult]{%
        \includegraphics[width=0.3\linewidth]{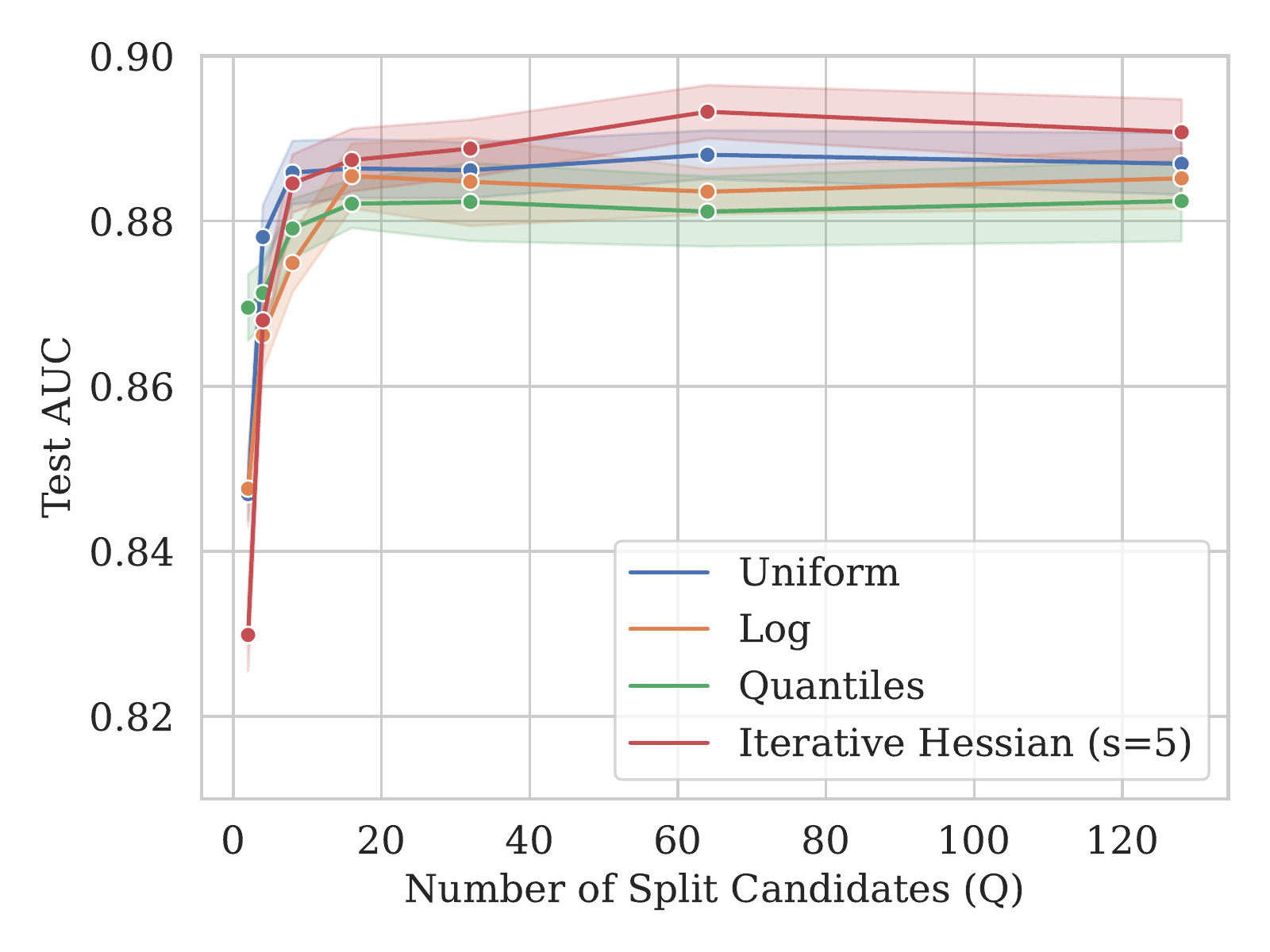}} \\
  \subfloat[Nomao]{%
       \includegraphics[width=0.3\linewidth]{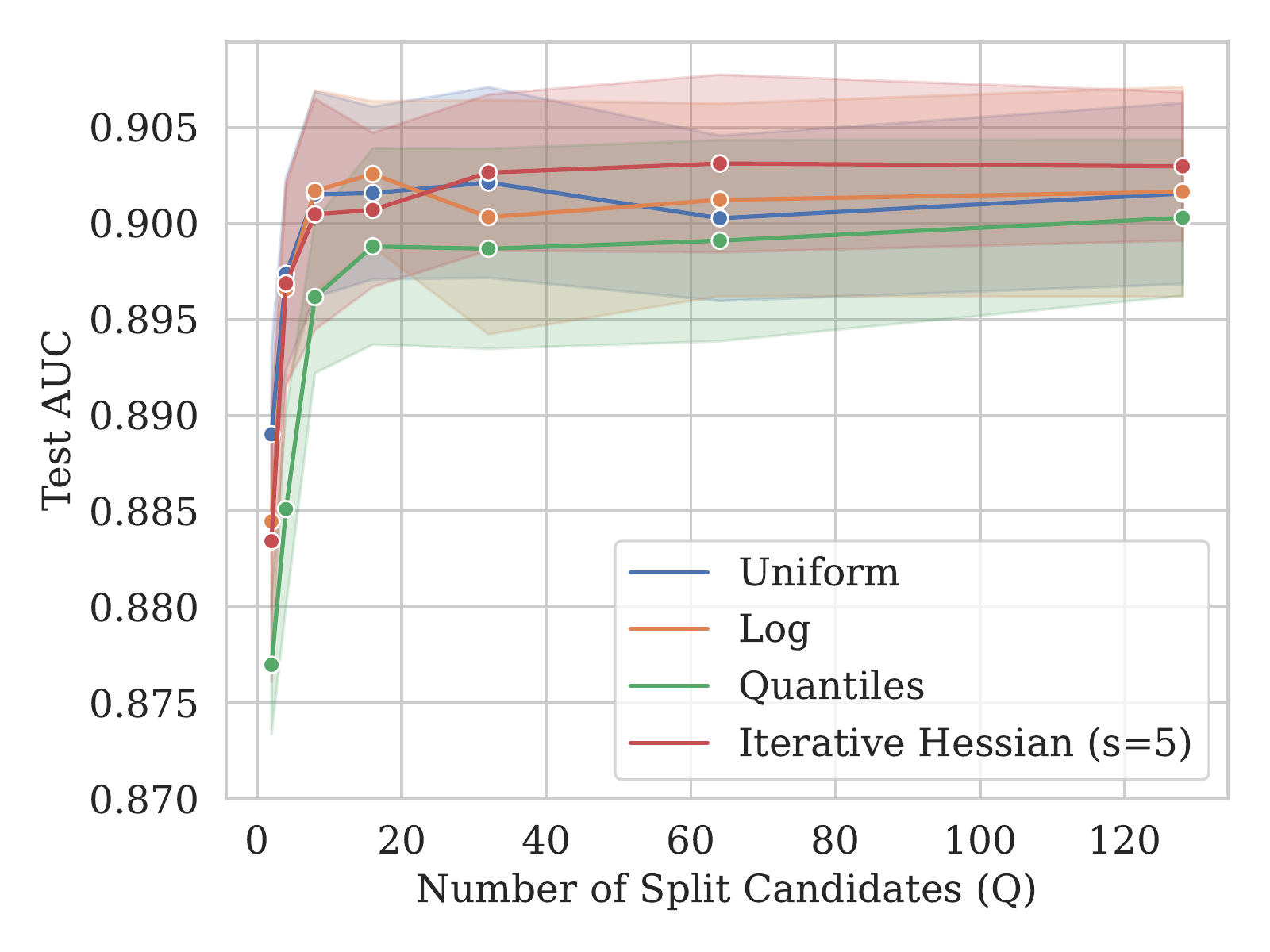}}
  \subfloat[Bank]{%
        \includegraphics[width=0.3\linewidth]{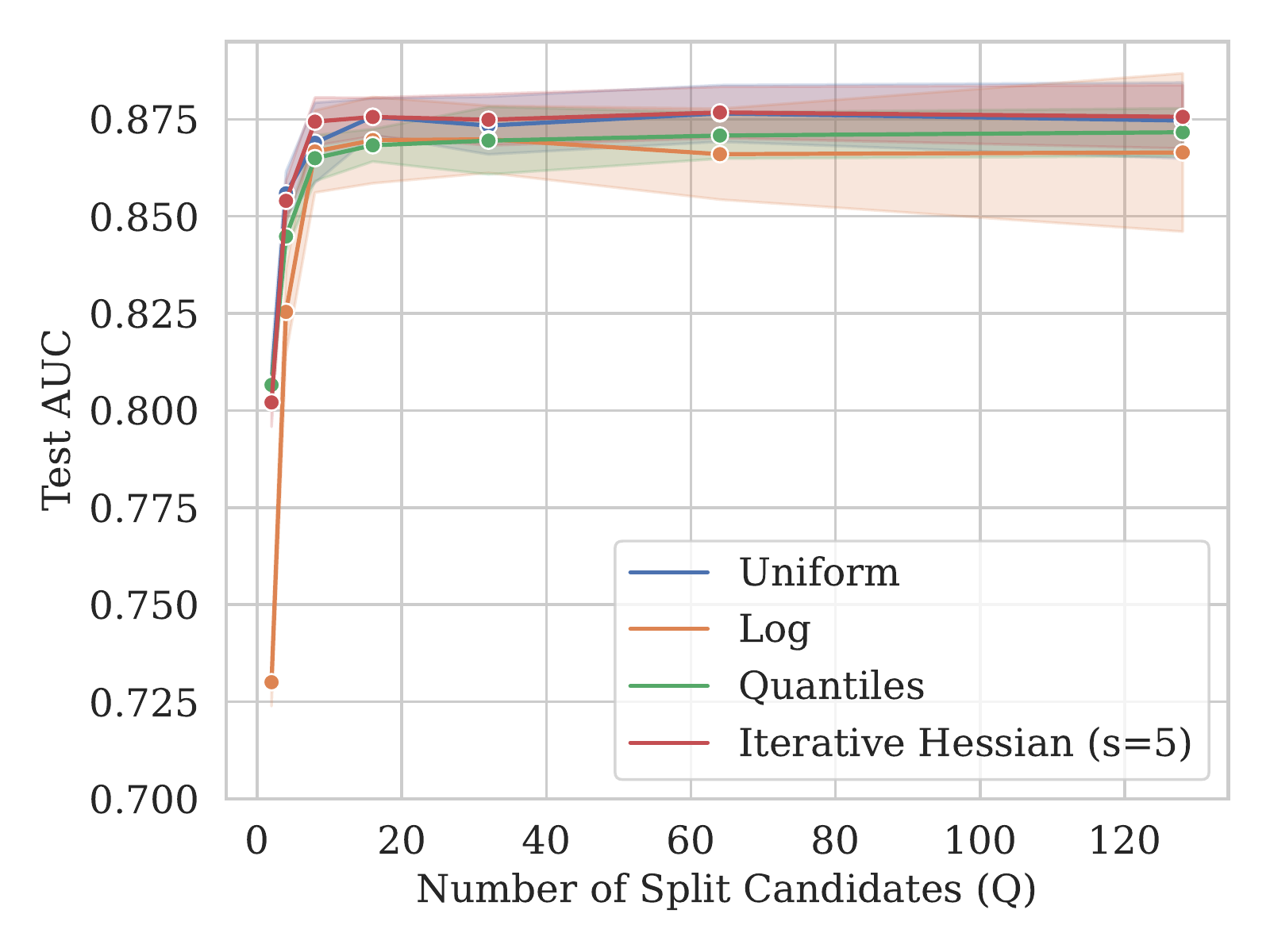}}
    \caption{Split Candidate Methods: Varying $Q \in \{2,4,,16,32,64,128\}$ \label{fig:appendix:vary_q}}
\end{figure*}

In Section \ref{exp:split_candidate} we presented Figure \ref{fig:split_candidates:vary_Q} which varied the number of split candidates $Q$ on the Credit 1 dataset. We present similar figures here for Higgs, Credit 2, Adult, Nomao and Bank in Figure \ref{fig:appendix:vary_q}. These results provide further justification for our choice of $Q=32$ \edit{in our main experiments, since it performs reasonably well across all datasets. Note however, that $Q=32$ is not the optimal value for each dataset, but does not hinder performance significantly compared to optimal values of $Q$. See also Appendix \ref{appendix:num_clients} which further shows robustness to the choice of $Q$}. 

Figure \ref{fig:appendix:vary_q} also provide further evidence of our conclusions in Section \ref{exp:split_candidate} (more specifically Table \ref{tab:split_candidate_methods}). For the Higgs dataset we can see clearly the advantage of the IH split candidate method and that as $Q$ increases we approach the (best) accuracy of using quantiles. Recall that the Higgs dataset has a large number of highly skewed features and this is where IH excels. For Nomao, Bank and Adult we observe that the split candidate methods all perform very similarly. As noted in Section \ref{exp:split_candidate}, IH performed poorly on the Credit 2 dataset and we can observe here that as we increase $Q$ it approaches the AUC of quantiles. Hence IH is still capturing the distribution of features well via split candidates, but that quantiles is not optimal for Credit~2.

\subsection{Batched Updates: Other Datasets}\label{appendix:bb}
\begin{figure*}[t!]
\centering
  \subfloat[Adult]{%
       \includegraphics[width=0.33\linewidth]{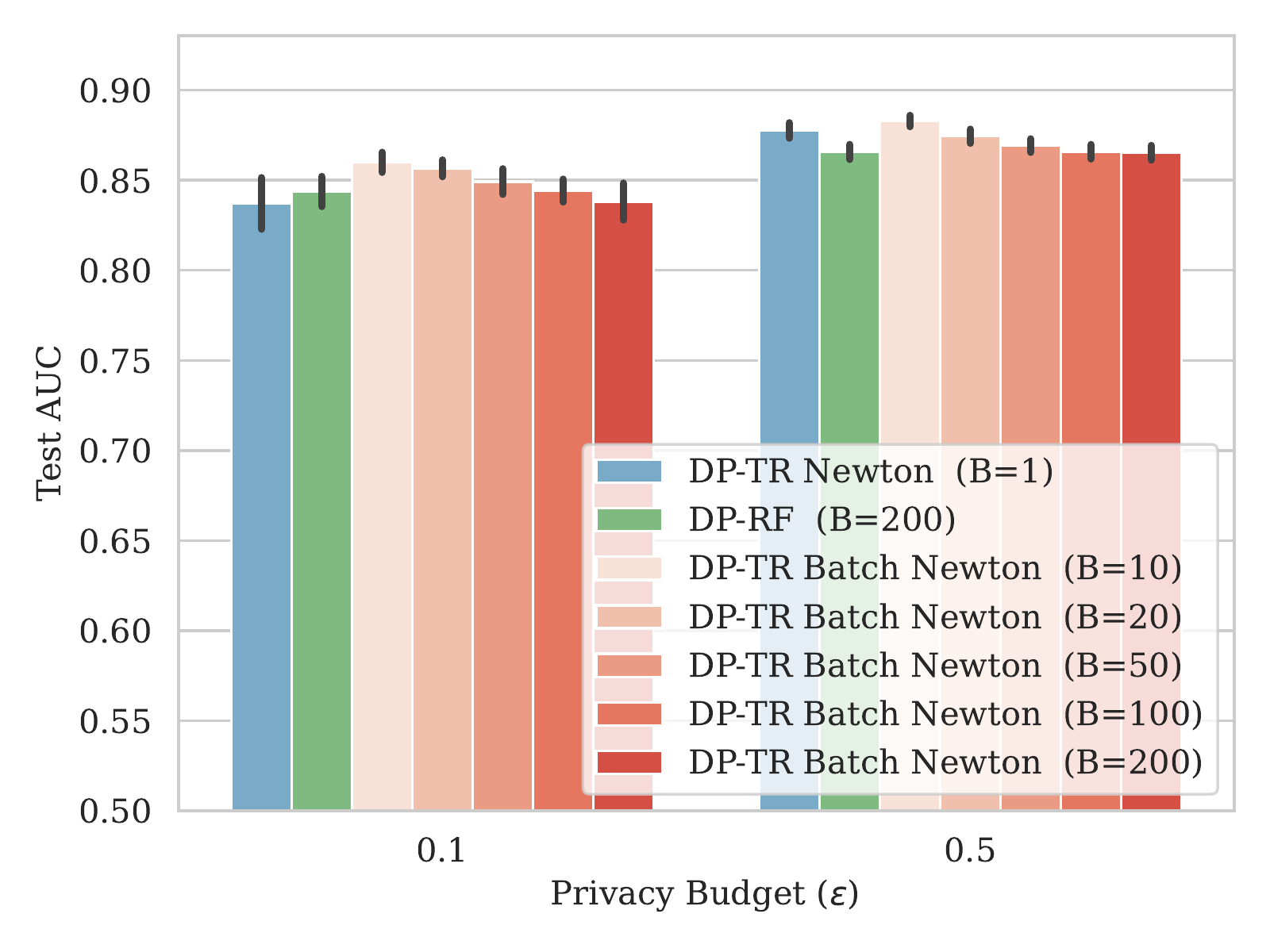}}
  \subfloat[Bank]{%
        \includegraphics[width=0.33\linewidth]{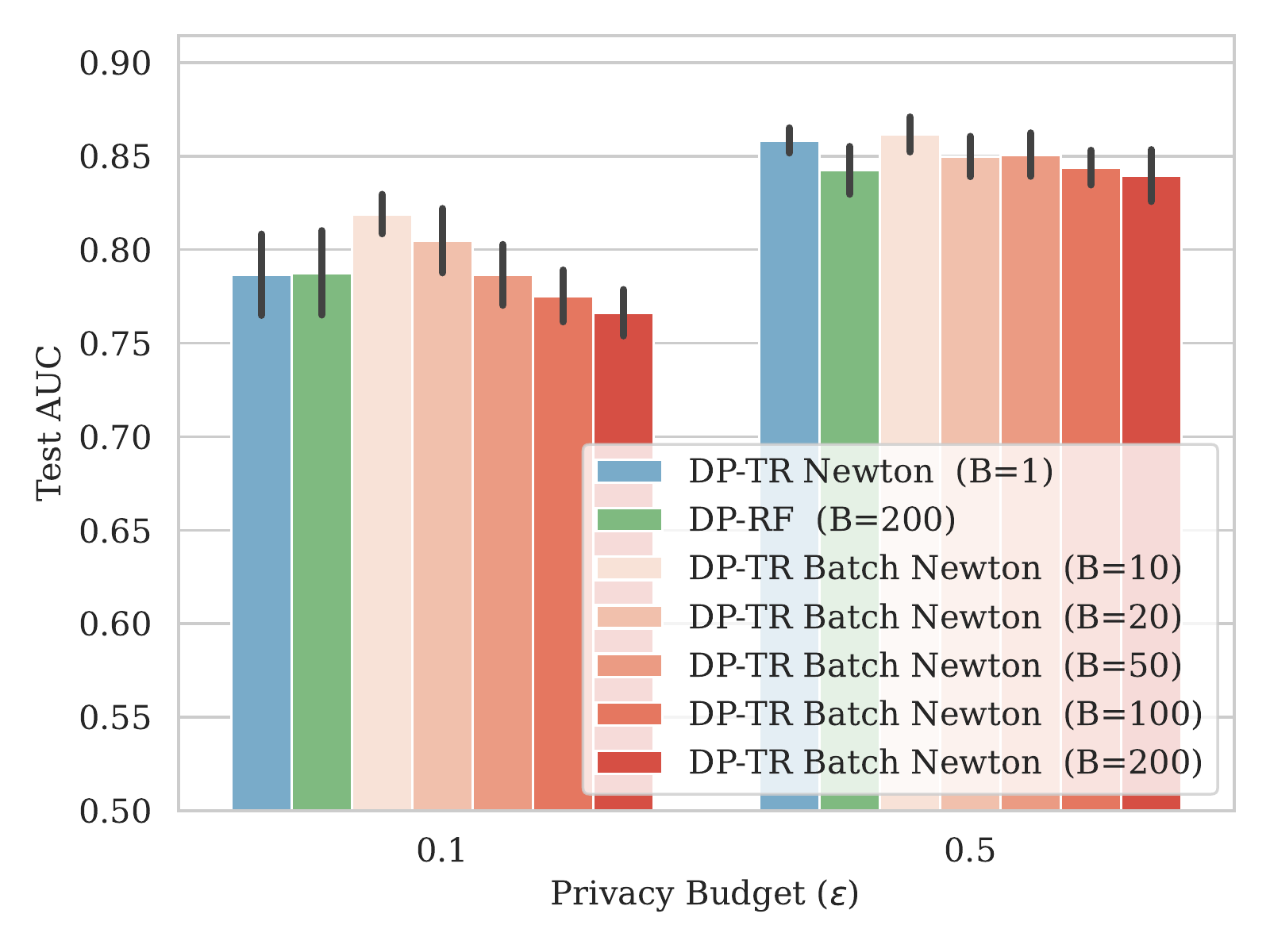}} \\
  \subfloat[Credit 2]{%
        \includegraphics[width=0.33\linewidth]{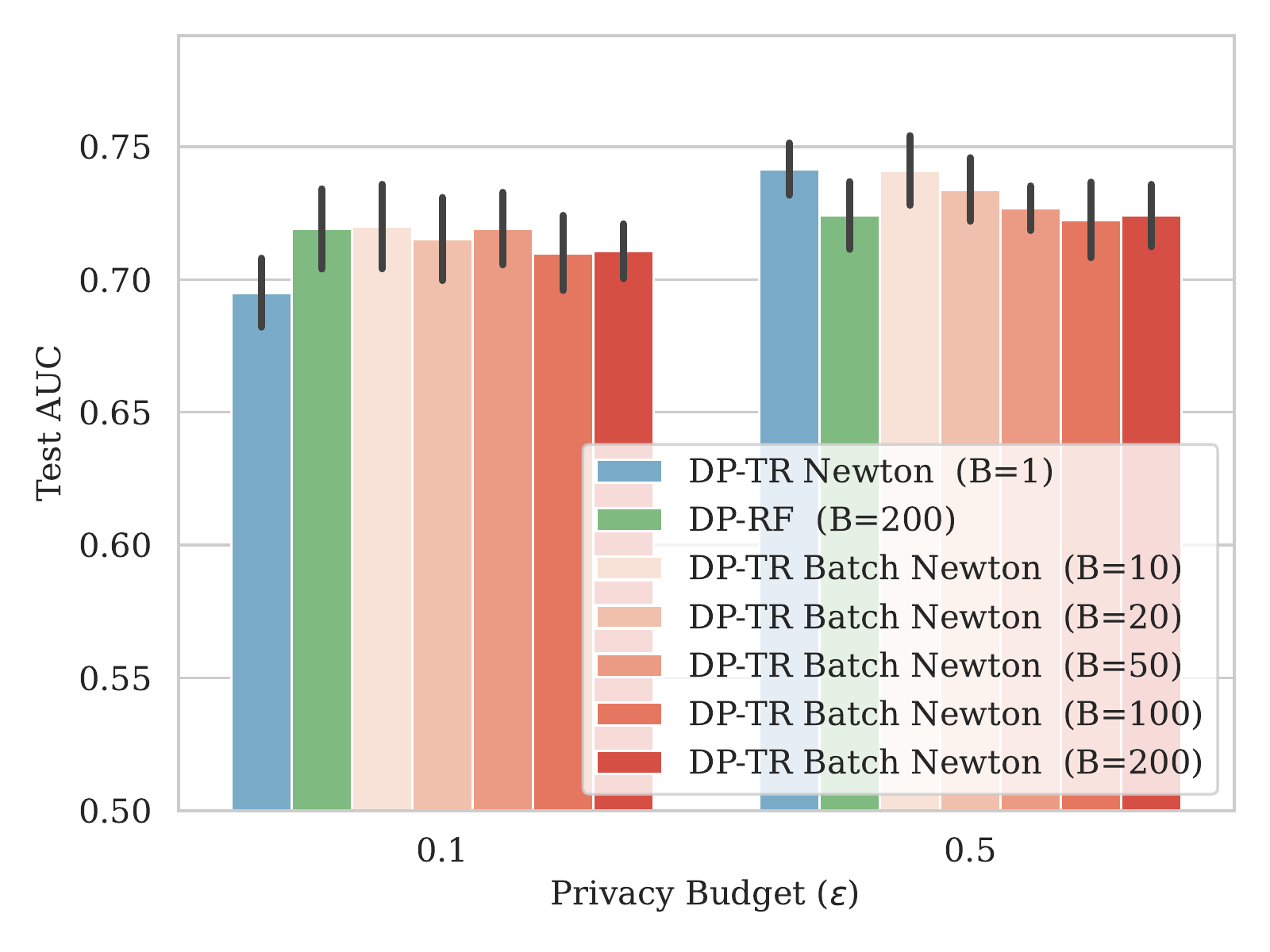}} 
  \subfloat[Nomao]{%
       \includegraphics[width=0.33\linewidth]{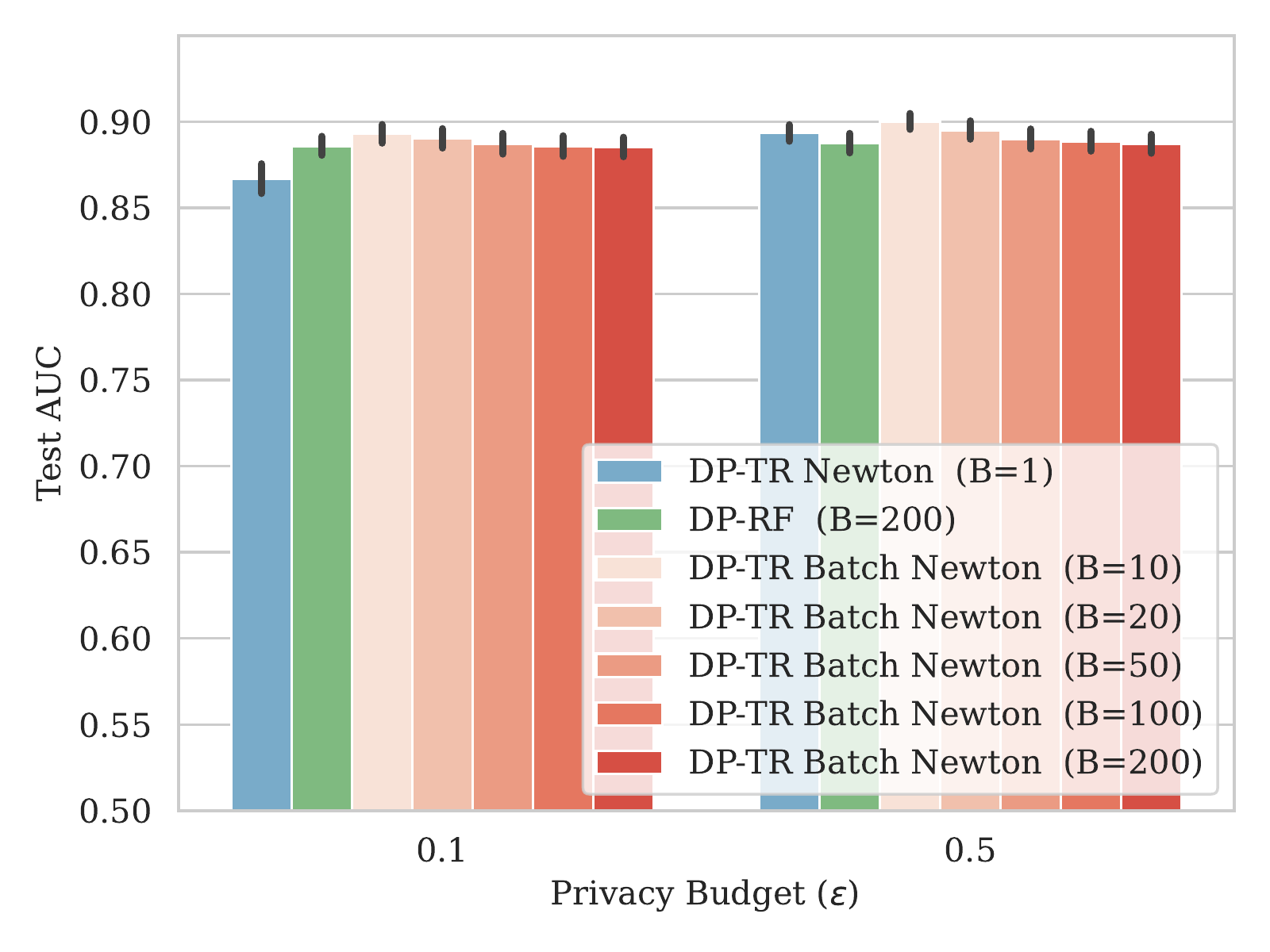}}
    \caption{Batched updates across the datasets $(T=200, d=4)$ \label{fig:appendix:bb}} 
\end{figure*}

In Section \ref{exp:bb} we presented Figure \ref{fig:bb:low_eps} on the Credit 1 dataset when studying the effect of batching updates. Here we present similar figures but for other datasets: Adult, Bank, Credit 2 and Nomao. This is displayed in Figure \ref{fig:appendix:bb}.

In general, we can draw the same conclusions as in Section~\ref{exp:bb}. 
In particular, batching updates are certainly competitive when compared to DP-TR Newton and that it is a useful way to reduce the number of communication rounds needed to train the model. Two new observations are present when we perform this experiment on these other datasets, specifically for $\epsilon=0.1$. We find DP-RF to be more competitive with DP-TR Newton and also find that batching updates can sometimes perform better than DP-TR Newton. This shows in general that under higher privacy some form of averaging across trees in the ensemble can help with the larger amount of noise that is being added. This again corroborates our insights from Section \ref{exp:bb} and Appendix \ref{appendix:weight_updates}. 

\subsection{Comparisons: Other Datasets}\label{appendix:comparisons}
\begin{figure*}[t]
  \centering
  \includegraphics[width=0.6\linewidth]{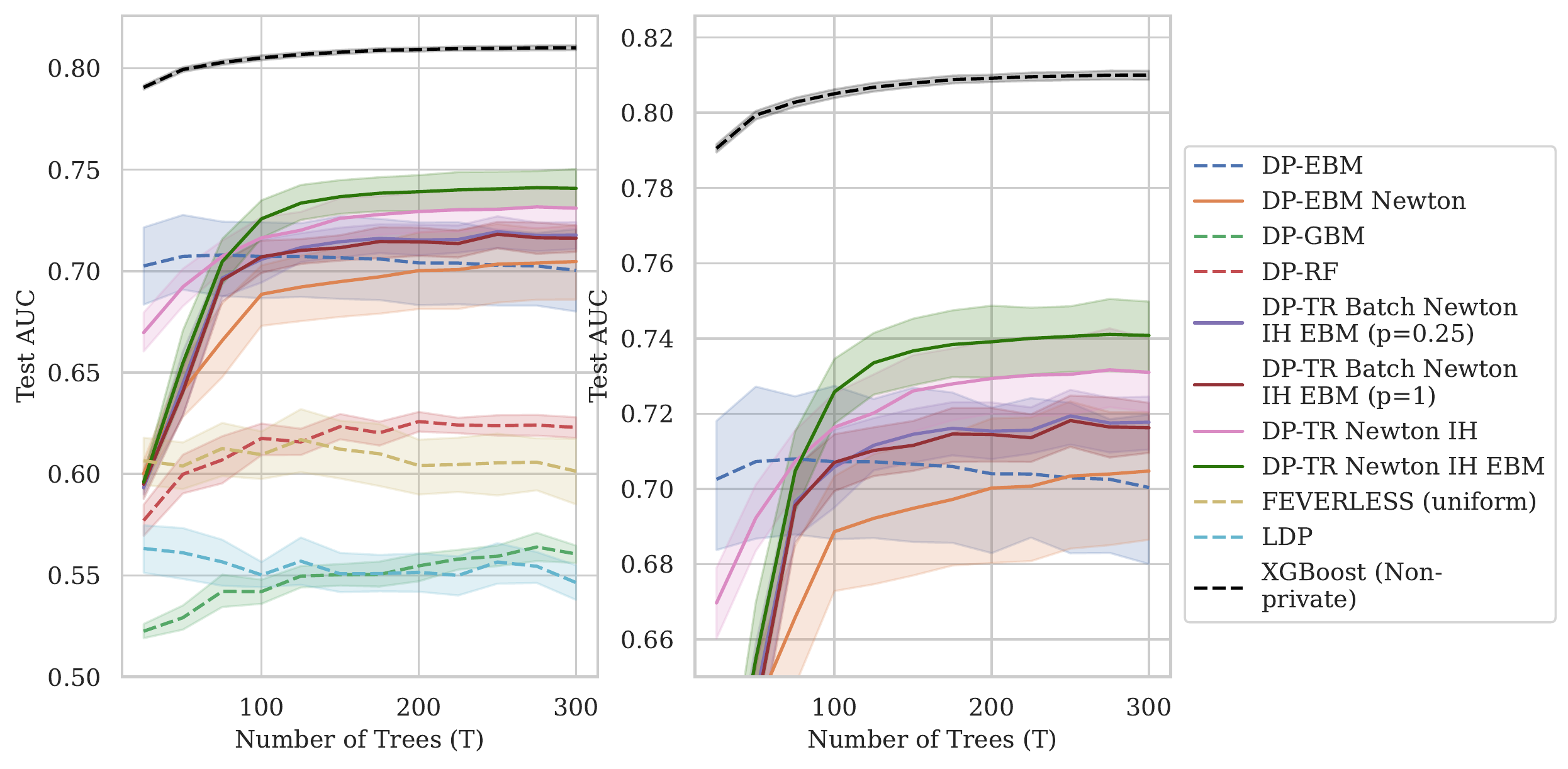}
  \caption{Comparison of methods: Higgs $(d=4, \epsilon=1)$ \label{fig:appendix:comparison_higgs}}
\end{figure*}

\begin{figure*}[t]
  \centering
  \includegraphics[width=0.47\linewidth]{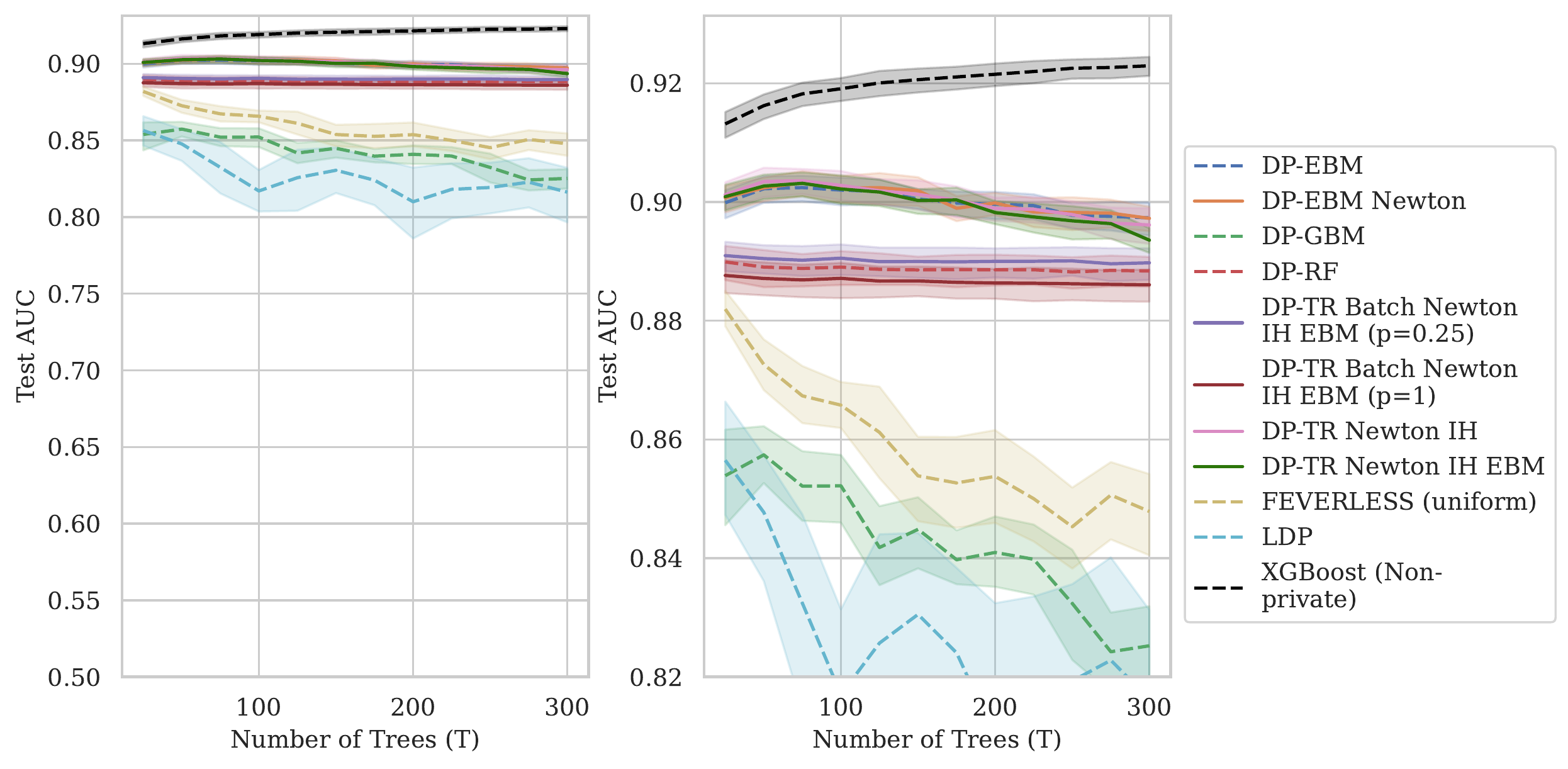}
  \caption{Comparison of methods: Nomao $(d=4, \epsilon =1)$\label{fig:appendix:comparison_nomao}}
\end{figure*}

In Section \ref{exp:comp} we presented an end-to-end comparison of methods in our framework against existing baselines. Figure \ref{fig:comparisons} details these results on the Credit 1 dataset. We present similar experiments here but for other datasets: Higgs, Credit 2, Adult, Bank and Nomao. These are shown in Figures \ref{fig:appendix:comparison_higgs}---\ref{fig:appendix:comparison_nomao}.

The four main observations detailed in Section \ref{exp:comp} on the Credit~1 dataset also largely still hold across the different datasets. These can be summarised by noting that the histogram-based methods DP-GBM and FEVERLESS perform the worst with FEVERLESS usually outperforming DP-GBM. This is followed by DP-RF and then DP-TR Newton with IH. This is common across most of the datasets (including Credit 1 as noted in Section \ref{exp:comp}). 

What differs slightly across datasets is the best performing methods. For some datasets (like Higgs) DP-TR Newton with IH and batched updates clearly perform the best and outperform our baseline methods. 
On datasets like Credit 2 and Adult where IH split candidates give less of a benefit, the DP-EBM baseline and its variant DP-EBM Newton remain very competitive. \edit{The results discussed here are summarised in the rankings presented in Table \ref{tab:rank}.}

\edit{\section{Computation and Communication}}
\label{sec:computationcommunicationcost}
\edit{\subsection{Communication rounds} The total number of communication rounds depends on the choice of split-method as follows: 
\begin{itemize}
    \item \textbf{Histogram-based (Hist).} At every level in the tree, clients must aggregate the required gradient histograms in order to make split decisions for the next level. Hence, the number of communication rounds are of order $O(Td)$, and the size of each message is $O(Qk)$ where $k \leq m$ is the maximum number of features being considered in a tree. 
    \item \textbf{Totally random (TR).} For totally random methods, the random trees can be pre-computed by the server and clients are required to aggregate gradient information for leaf nodes of a tree. Thus, the total number of communication rounds is of order $O(T)$, and the size of each message is $O(2^d)$. When using the IH method number of communication rounds is increased to $O(T + s)$ where the message size of each round of IH is $O(Qk)$.
    \item \textbf{Batched updates.} For batched updates with TR, the number of communication rounds is of the order $O(T/B)$, but the messages are $B$ times larger, $O(B2^d)$. If using IH, the number of rounds becomes $O(T/B + s)$.
\end{itemize}}
\edit{In the main text, when we refer to batched boosting reducing rounds of communication, this is without considering any secure-aggregation overhead. In practical implementation such as \cite{bell2020secure}, the overhead of secure-aggregation typically requires $3$ rounds of communication under an honest-but-curious threat model. Hence, methods like batched boosting are advantageous since reducing a single round of communication results in a 3x reduction under secure-aggregation.}

\edit{\subsection{Communication Cost}}
\edit{The per-round communication cost of methods will vary depending on how the tree is built. We will assume that the server builds a tree a level at a time, which means that gradient information is batched. For example, histogram-based methods at level $i$ will send $O(2^iQk)$ gradient information corresponding to all gradient histograms at each node in level $i$. Depending on the federated setting (i.e., availability of devices, communication bandwidth), the server could instead build the tree a node at a time incurring an increase in the number of rounds but a smaller per-round communication cost.}

\edit{In Figure \ref{fig:c&c:comm} we present an experiment on synthetic data. 
This plot is illustrative in the sense that real-world communication costs will be somewhat higher (due to omitting the overheads of secure-aggregation and communication packets). 
However, the experiment is run on real GBDT models and reveals that the communication cost is often less than the worst-case presented above. 
For example, many methods that calculate gradient histograms can utilise Hessian information to stop growing the tree early which reduces the overall communication cost.}

\edit{The experiment in Figure \ref{fig:c&c:comm} varied over $T \in [25, 300]$, $m \in \{10,20,30,40\}$, $d \in {3,4,5}$ and $Q \in \{4, 8, 16, 32, 64, 128\}$ as these are the main parameters that effect communication cost. While Figure \ref{fig:c&c:comm} only displays the total communication cost that needs to be sent over the full training of $T$ trees, the overall per-round communication cost is often much smaller. Even for FEVERLESS while training $T=1000$ trees results in $\approx 100$MB communication overall, the per-round communication would be of order $\approx 0.025$MB which is reasonable for federated clients (i.e., mobile devices) which may choose to not participate in all rounds.
Meanwhile, the methods we advocate incur a total communication cost of under 1MB to build the full model for a client participating in every round.}

\edit{We note that the total size of data received from the aggregating sever is the same across all methods except for those that use IH, since there is an additional (but small) cost in receiving split candidates over $s$ rounds.}

\edit{\subsection{Computation Cost}}

\edit{ Since participating devices in the federated setting are often limited in computation, one may wonder how intensive private GBDT protocols are. 
In Figures~\ref{fig:c&c:client_comp} and~\ref{fig:c&c:server_comp} we provide approximate benchmarks for the client and server costs of methods studied in Section \ref{exp:comp}. 
These experiments are run without secure-aggregation overheads. 
We assume a setup where clients have $10,000$ samples in their local datasets and we vary parameters $T \in [75, 150], d \in \{3,4,5\}, m \in \{10, 20, 30, 40, 50\}, Q \in [16, 32, 64]$ as these all affect the total computation time for the full protocol.}

\edit{In general, all private GBDT methods are lightweight for both clients and servers. For all methods, clients must compute gradient and Hessian information of their samples. For some methods like FEVERLESS or those that use IH they must also form aggregated gradient histograms. The computation scales linearly in the size of a client's local dataset. 
For methods that use histograms, clients must also partition their data into $Q$ bins before aggregating.}

\edit{In practice, the computation overhead of using secure-aggregation would not be large but would dominate most of the local computation costs of the GBDT algorithm. For example, in \cite{bell2020secure}, Bell et al. benchmark client computation costs for $n=10^5, l=10^5$ where $l$ is the dimension of the quantity the server is aggregating. Their secure-aggregation algorithm takes $\approx 0.35$ seconds for clients. In the worst-case their parameter settings match ours and the per-round computation overhead for clients when using secure-aggregation will be no more than a second under our honest-but-curious threat model.}

\section{Varying the number of clients} \label{appendix:num_clients}

\edit{One may ask whether the private GBDT protocols in our framework are sensitive to the number of participating clients. Every method in our framework relies on aggregating gradient information of the form $q(I) = (\sum_{i \in I} g_i, \sum_{i \in I} h_i)$ from a number of clients, where $I$ is the set of samples at a (leaf or internal) node. Thus, the total number of clients is somewhat unimportant to the overall utility of the algorithm. Instead, the most important factor is the total number of data samples. In our experiments in Section \ref{sec:experiments} we assumed that one client holds one data item, and hence varying the number of clients is equivalent to varying the number of samples. But in scenarios where each client may hold more than one data item, the number of participating clients is only important if they are providing more data to the algorithm. In practice, we would hope that there is enough participating data to ensure that the aggregated gradient information will still be meaningful under noise.}

\edit{The number of samples at a given node affects both the construction of the tree (internal splits and leaf weights) and our IH split candidate method as it depends on Hessian histograms. Thus we are interested in analysing two effects: First, how sensitive our methods are to the number of participating clients (when compared to non-private XGBoost) and secondly, the sensitivity of the number of split candidates $Q$. In particular, we are interested in how sensitive the IH method is to the value of $Q$ since it relies on refining split candidates around Hessian information, which is itself dependent on the total number of participating samples.}

\begin{figure*}[t!]
\centering
  \subfloat[\label{fig:c&c:comm} \edit{Communication - $d \in [3,4,5], m \in [10,20,30,40]$}]{%
       \includegraphics[width=0.3\linewidth]{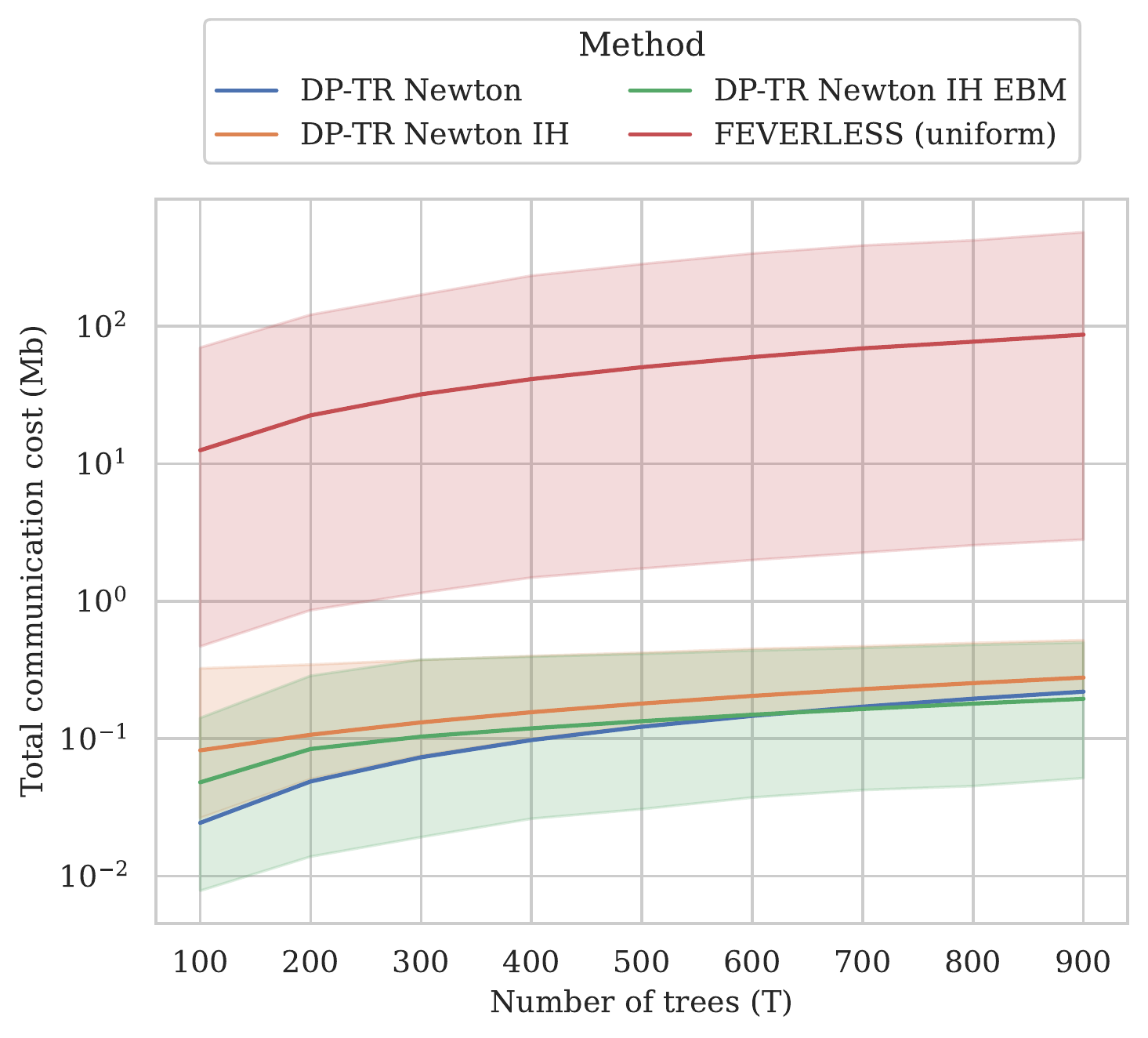}}
  \subfloat[\label{fig:c&c:client} \edit{Varying clients}]{%
        \includegraphics[width=0.3\linewidth]{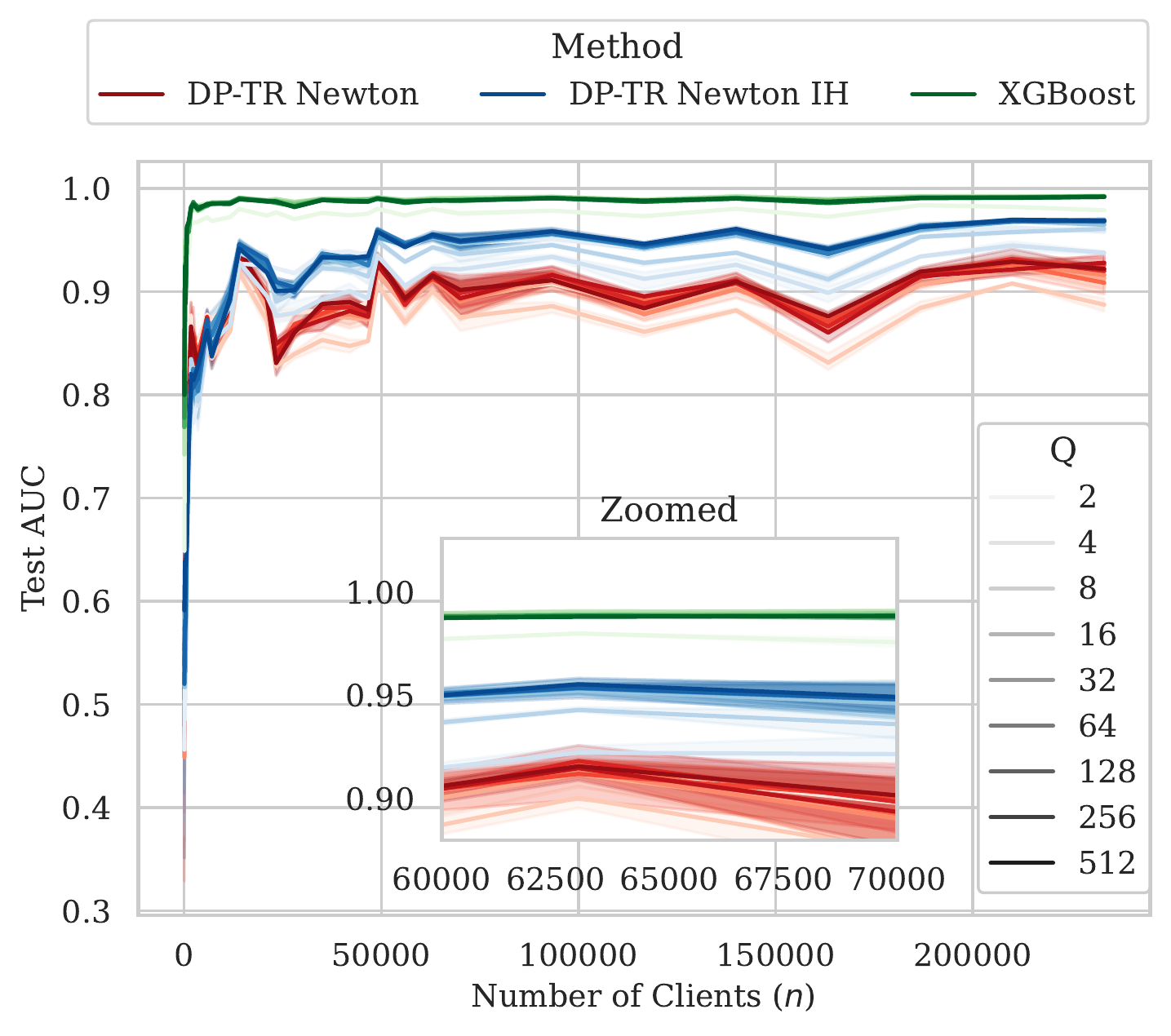}} \\
  \subfloat[\label{fig:c&c:client_comp} \edit{Client computation}]{%
        \includegraphics[width=0.3\linewidth]{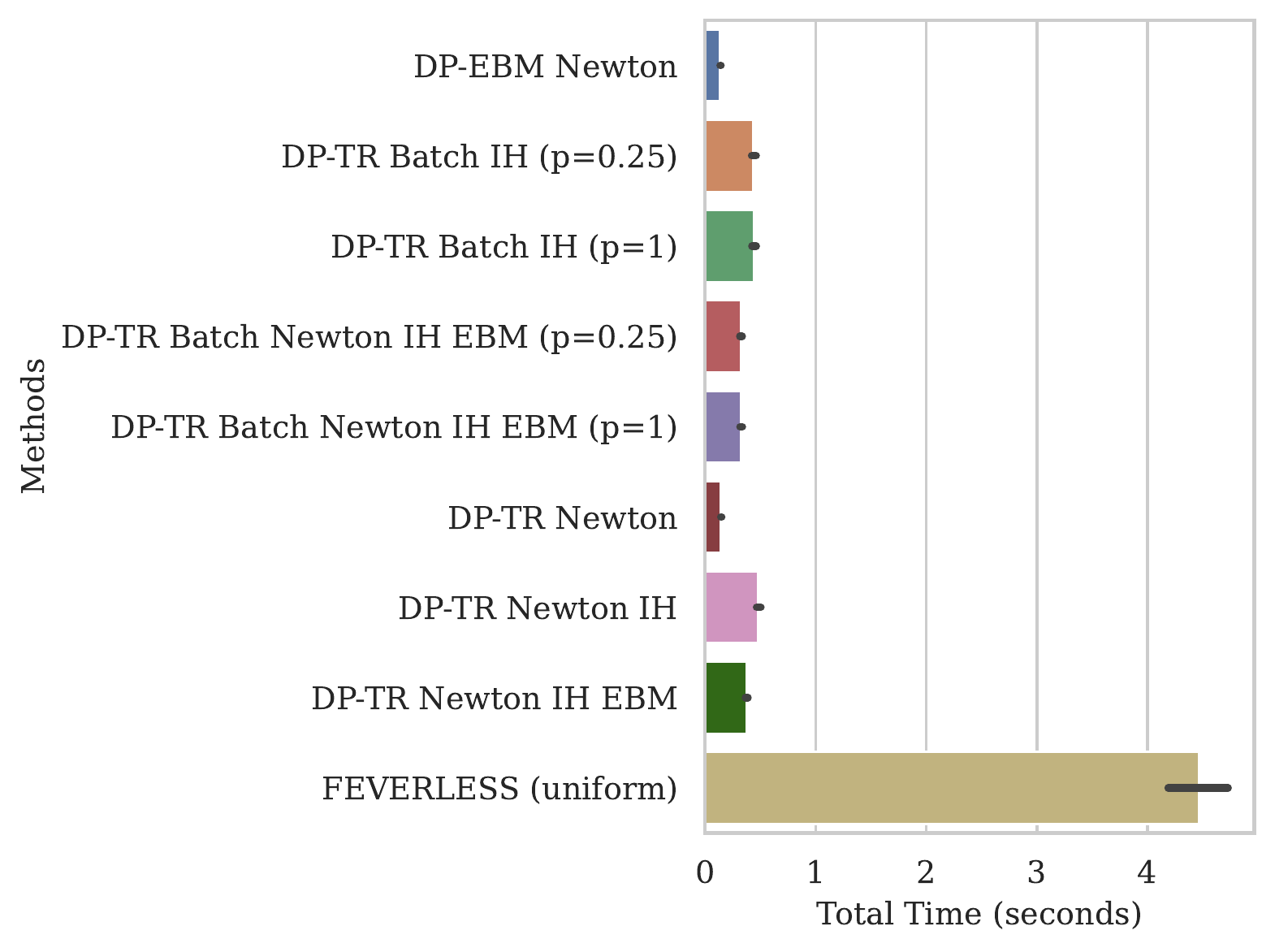}} 
  \subfloat[\label{fig:c&c:server_comp} \edit{Server computation}]{%
       \includegraphics[width=0.3\linewidth]{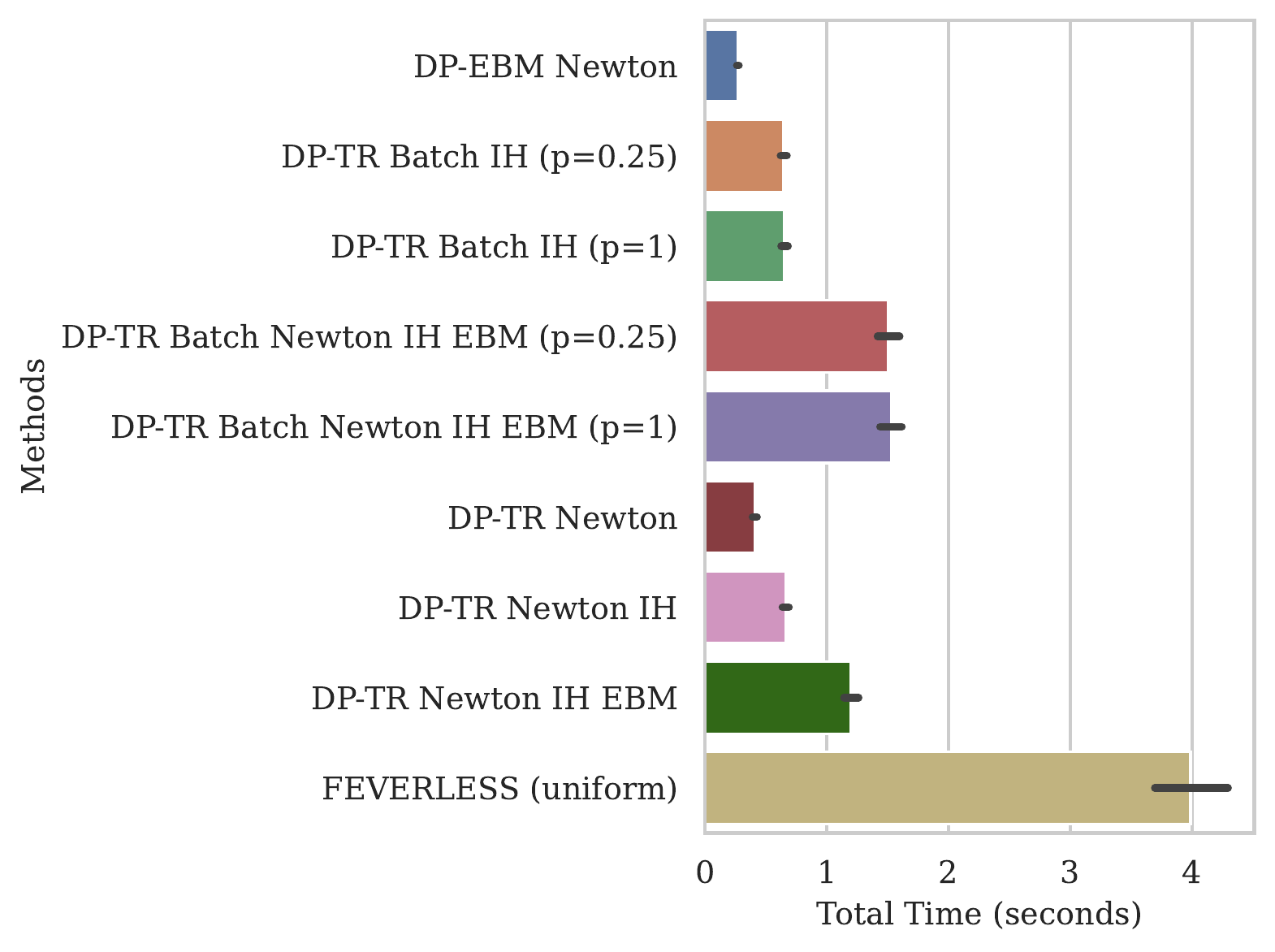}}
    \caption{Computation Benchmarks} 
    \label{fig:cc}
\end{figure*}

\edit{To understand how this affects utility we run an experiment, shown in Figure \ref{fig:c&c:client} on synthetic data generated with $m=30$ and the number of clients (equivalently, the number of samples) $n \in [70, 200000]$\footnote{See \url{https://scikit-learn.org/stable/modules/generated/sklearn.datasets.make_classification.html}}. We choose to use a synthetic dataset which is ``easy'' to classify in the sense that in the non-private setting we only need $n \approx 1000$ samples to get almost perfect ($> 0.95$) AUC. This allows us to see more clearly the effect that DP noise has when varying the number of clients. 
We compare our two most competitive methods DP-TR Newton with uniform splits and DP-TR Newton with IH splits while varying $n$ and the number of split candidates $Q$. We fix $d=4, \epsilon = 1$.}

\edit{We observe that when $n < 2000$ the results of our DP methods are poor and the performance gap between non-private XGBoost and our private DP-GBDT methods is large. We find that when $n \geq 10,000$ the DP methods stabilise and achieve a reasonable AUC of around $0.95$. 
This is suitable for the federated setting where $n \geq 10,000$ is expected. }

\edit{For the number of split candidates $Q$, we first note that this is not a particularly sensitive parameter in the non-private setting where the only significant gap in performance occurs when $Q=2$. For the DP methods, the value of $Q$ is understandably more sensitive. We remark that DP-TR Newton (uniform) and DP-TR Newton IH perform very similarly when $n \leq 10,000$ and when $n \geq 10,000$ there is a large gap in performance as long as $Q$ is not too small. This is expected, as when $n$ is small, the Hessian histograms will likely have more noise than signal and thus the split candidates formed from IH are likely to be very similar to uniform. When $n$ is larger, the IH method has more accurate Hessian histograms and so more refined split candidates and thus better performance.} 

\edit{We find in this experiment that values of $Q=2,4,8$ usually result in the lowest AUC (by at most $0.05$) and that values between $Q=32-512$ perform very similarly. The general conclusion here is that the choice of $Q$ is robust as long as you do not choose it too small. This also agrees with our conclusions from Section \ref{exp:split_candidate} and Appendix \ref{appendix:split_candidates_vary_Q}.}

\begin{figure*}[t]
\centering
  \subfloat[Varying $T$]{%
       \includegraphics[width=0.25\linewidth]{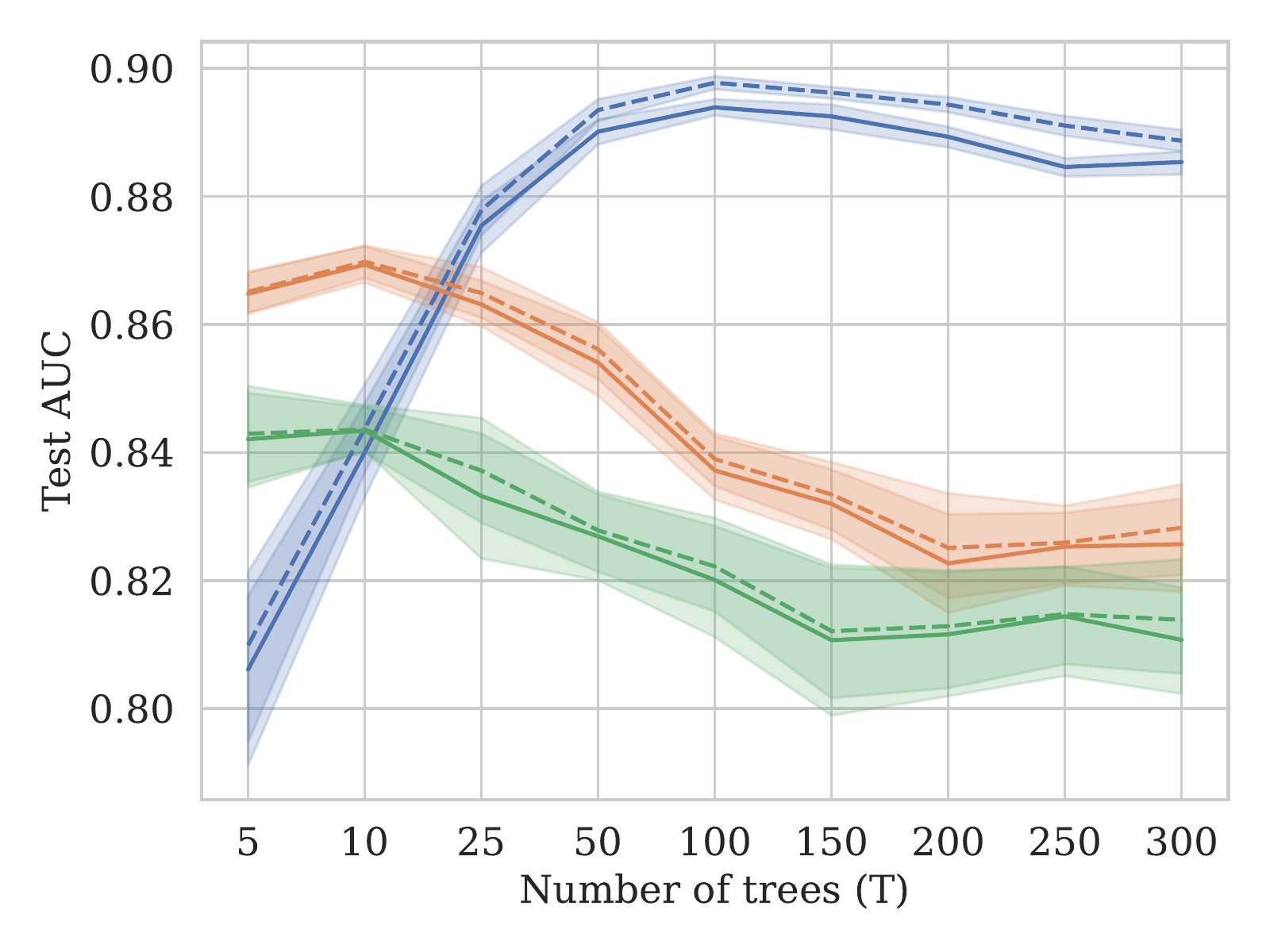}}
  \subfloat[Varying $D$]{%
        \includegraphics[width=0.25\linewidth]{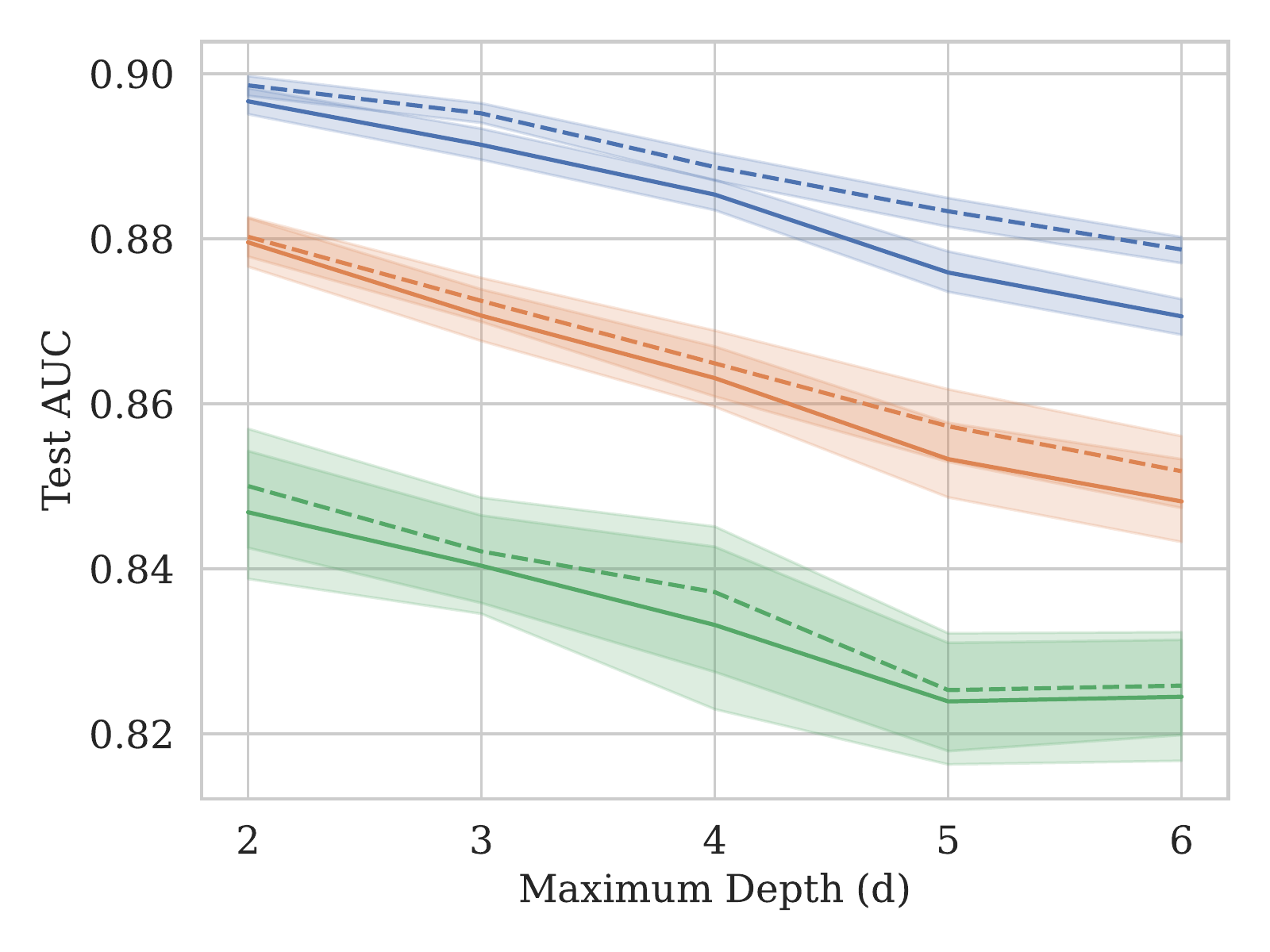}}
  \subfloat[ Varying $\epsilon$]{%
        \includegraphics[width=0.25\linewidth]{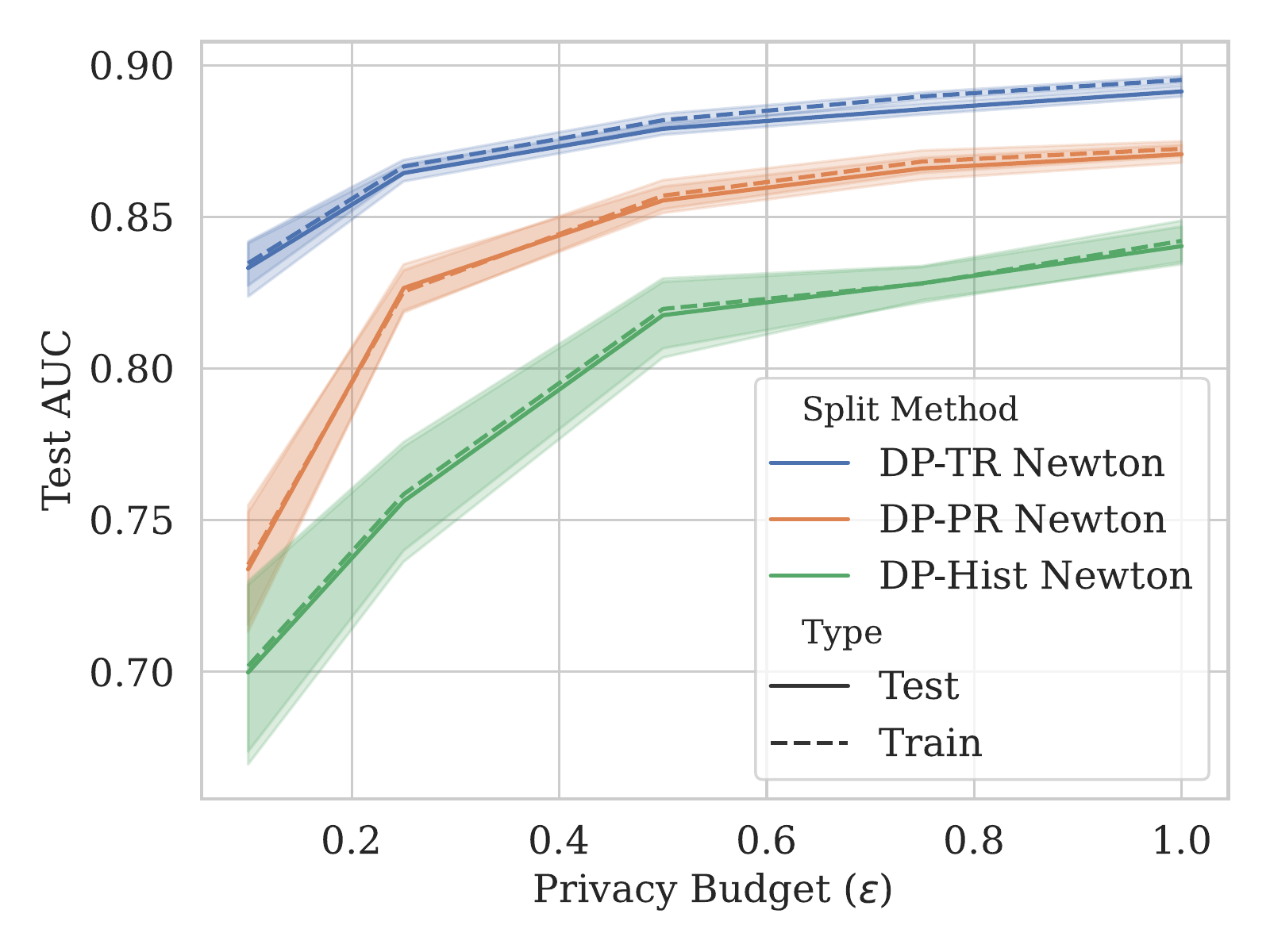}}
  \caption{Split Methods on Adult}
\end{figure*}

\begin{figure*}[t]
\centering
  \subfloat[Varying $T$]{%
       \includegraphics[width=0.25\linewidth]{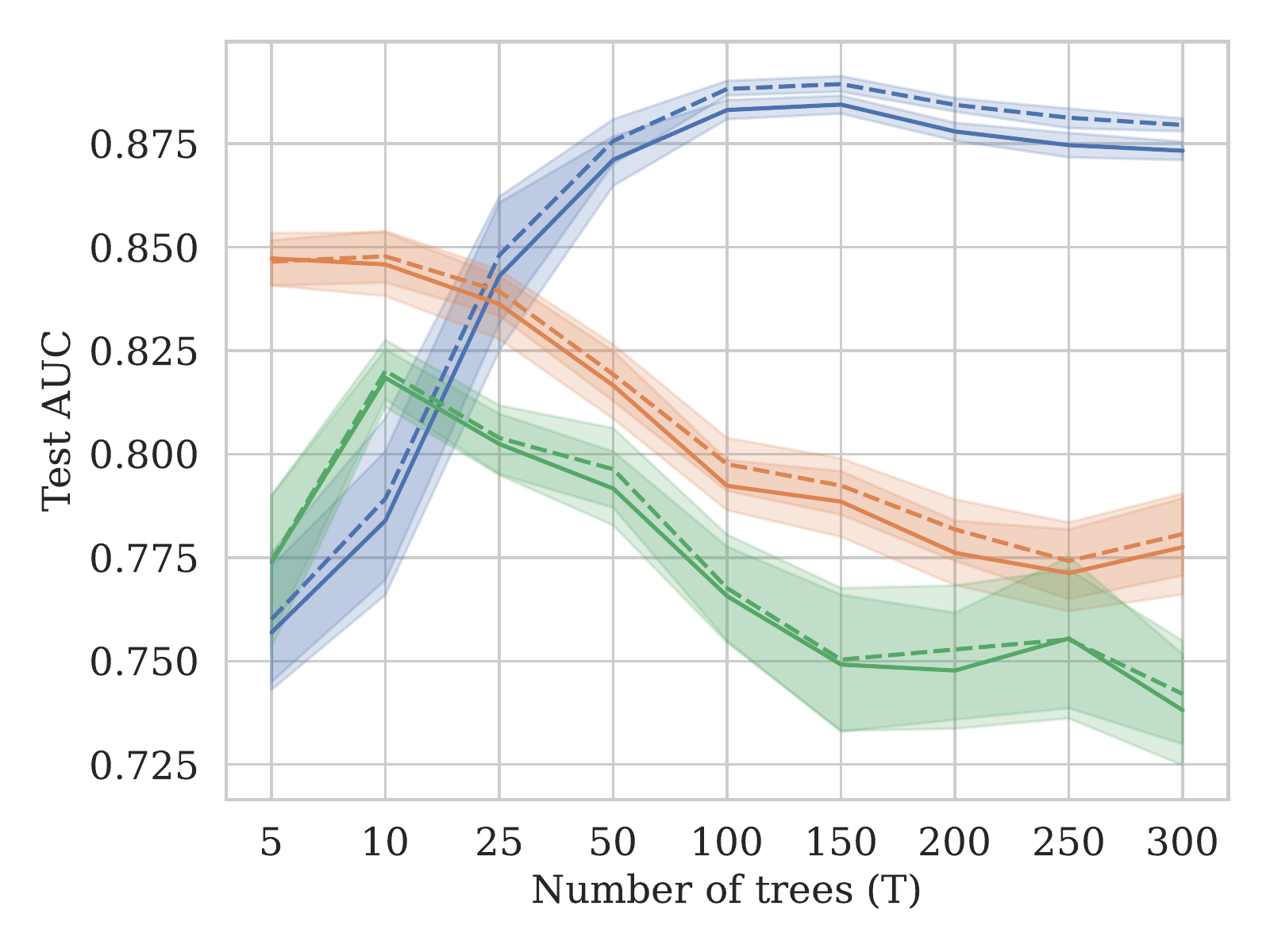}}
  \subfloat[Varying $D$]{%
        \includegraphics[width=0.25\linewidth]{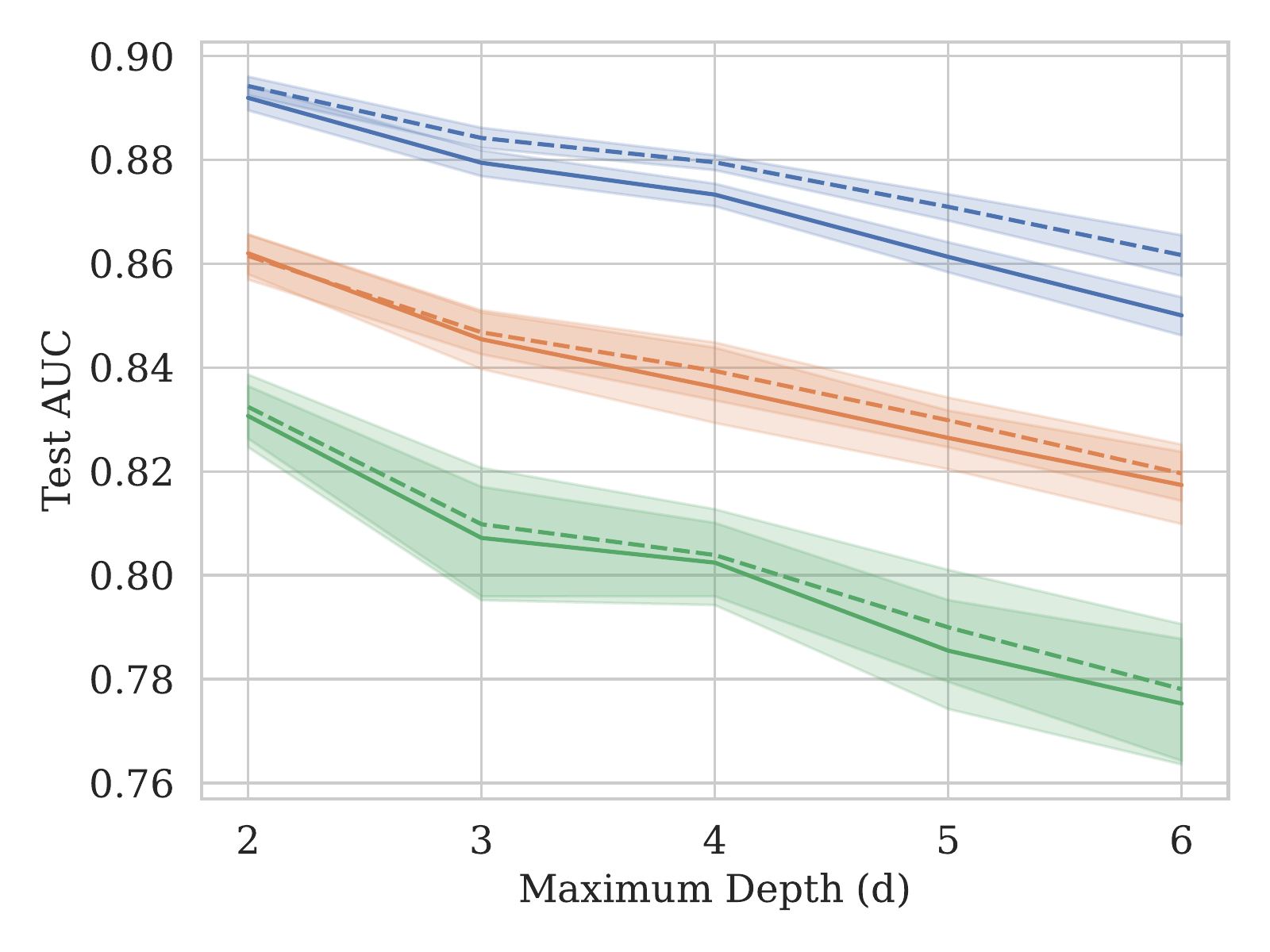}}
  \subfloat[ Varying $\epsilon$]{%
        \includegraphics[width=0.25\linewidth]{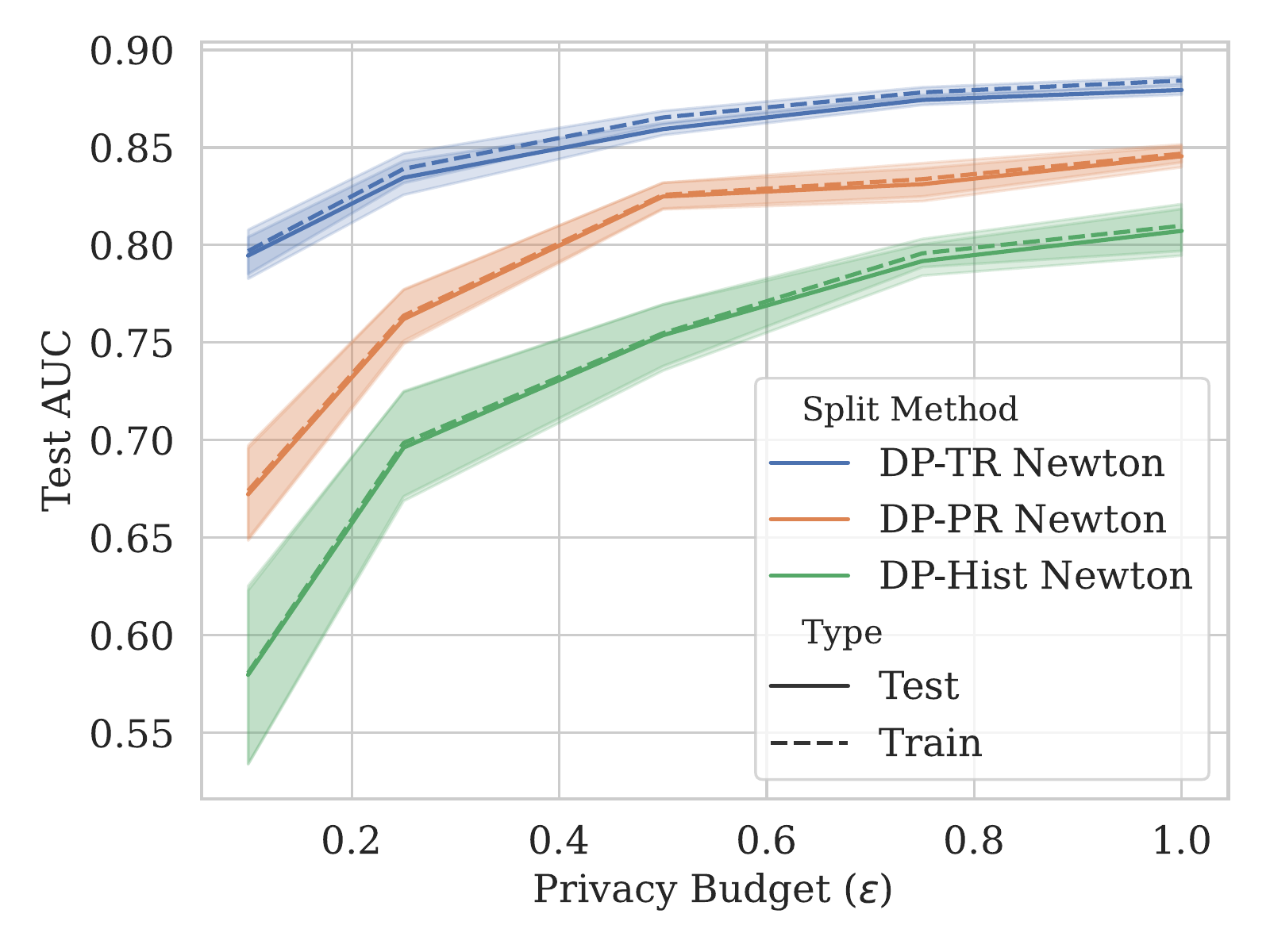}}
  \caption{Split Methods on Bank}
\end{figure*}

\begin{table*}[t]
  \small
  \caption{Weight update methods across the datasets fixing $\epsilon=0.1$\label{tab:weight_update_eps=0.1}}
\begin{tabular}{ccccccc}
\toprule
   &        &                       Bank &                   Credit 1 &                   Credit 2 &                      adult &                      nomao \\
\midrule
Hist & Gradient &   \textbf{0.607 +- 0.0698} &           0.5415 +- 0.1124 &            0.5444 +- 0.089 &           0.6385 +- 0.0892 &           0.8382 +- 0.0312 \\
   & Averaging &           0.6002 +- 0.0554 &  \textbf{0.7018 +- 0.0498} &           0.5383 +- 0.0596 &           0.6518 +- 0.0984 &  \textbf{0.8731 +- 0.0141} \\
   & Newton &           0.6032 +- 0.0805 &           0.5879 +- 0.0949 &  \textbf{0.5729 +- 0.0808} &  \textbf{0.6842 +- 0.0572} &           0.8525 +- 0.0264 \\
PR & Gradient &            0.5993 +- 0.052 &           0.5678 +- 0.0909 &            0.5922 +- 0.074 &           0.6512 +- 0.0517 &           0.8504 +- 0.0121 \\
   & Averaging &  \textbf{0.6597 +- 0.0506} &  \textbf{0.6351 +- 0.0618} &  \textbf{0.6457 +- 0.0438} &            0.7036 +- 0.066 &  \textbf{0.8828 +- 0.0065} \\
   & Newton &            0.6413 +- 0.056 &           0.6264 +- 0.0486 &            0.5869 +- 0.049 &  \textbf{0.7207 +- 0.0413} &             0.86 +- 0.0146 \\
TR & Gradient &           0.7413 +- 0.0257 &           0.7404 +- 0.0296 &            0.6917 +- 0.017 &           0.8173 +- 0.0167 &           0.8743 +- 0.0056 \\
   & Averaging &    \textbf{0.78 +- 0.0181} &           0.7669 +- 0.0139 &  \textbf{0.7133 +- 0.0112} &  \textbf{0.8471 +- 0.0086} &  \textbf{0.8862 +- 0.0052} \\
   & Newton &           0.7742 +- 0.0223 &   \textbf{0.7694 +- 0.015} &           0.6921 +- 0.0156 &            0.8385 +- 0.008 &            0.8665 +- 0.011 \\
\bottomrule
\end{tabular}
\end{table*}
\begin{table*}[ht]
  \small
  \caption{Weight update methods across the datasets fixing $\epsilon=0.25$}
\begin{tabular}{ccccccc}
\toprule
   &        &                       Bank &                   Credit 1 &                   Credit 2 &                      adult &                      nomao \\
\midrule
Hist & Gradient &           0.6505 +- 0.0427 &           0.5959 +- 0.0695 &           0.5879 +- 0.0453 &           0.6927 +- 0.0522 &           0.8308 +- 0.0319 \\
   & Averaging &  \textbf{0.7078 +- 0.0418} &           0.6917 +- 0.0539 &  \textbf{0.6383 +- 0.0343} &           0.6631 +- 0.0626 &  \textbf{0.8858 +- 0.0043} \\
   & Newton &           0.6878 +- 0.0381 &  \textbf{0.6935 +- 0.0501} &           0.6331 +- 0.0476 &  \textbf{0.7547 +- 0.0384} &           0.8696 +- 0.0155 \\
PR & Gradient &           0.6207 +- 0.0396 &           0.6273 +- 0.0598 &           0.6127 +- 0.0562 &           0.7094 +- 0.0266 &           0.8645 +- 0.0087 \\
   & Averaging &            0.7128 +- 0.045 &           0.6735 +- 0.0551 &  \textbf{0.6811 +- 0.0235} &           0.7699 +- 0.0494 &  \textbf{0.8883 +- 0.0055} \\
   & Newton &  \textbf{0.7436 +- 0.0158} &  \textbf{0.7356 +- 0.0322} &           0.6594 +- 0.0273 &  \textbf{0.8103 +- 0.0183} &            0.8706 +- 0.009 \\
TR & Gradient &           0.8175 +- 0.0128 &           0.7698 +- 0.0186 &  \textbf{0.7226 +- 0.0077} &           0.8551 +- 0.0085 &  \textbf{0.8917 +- 0.0077} \\
   & Averaging &           0.8268 +- 0.0167 &           0.7742 +- 0.0183 &           0.7206 +- 0.0132 &           0.8608 +- 0.0036 &           0.8871 +- 0.0055 \\
   & Newton &  \textbf{0.8322 +- 0.0095} &  \textbf{0.7836 +- 0.0109} &           0.7212 +- 0.0096 &  \textbf{0.8619 +- 0.0067} &           0.8838 +- 0.0056 \\
\bottomrule
\end{tabular}
\end{table*}
\begin{table*}[ht]
  \small
  \caption{Weight update methods across the datasets fixing $\epsilon=0.75$}
\begin{tabular}{ccccccc}
\toprule
   &        &                       Bank &                   Credit 1 &                   Credit 2 &                      adult &                      nomao \\
\midrule
Hist & Gradient &           0.6386 +- 0.0502 &           0.5547 +- 0.0535 &           0.6242 +- 0.0387 &           0.6913 +- 0.0252 &           0.8499 +- 0.0117 \\
   & Averaging &            0.732 +- 0.0182 &           0.6398 +- 0.0576 &  \textbf{0.6827 +- 0.0219} &           0.6421 +- 0.0265 &  \textbf{0.8869 +- 0.0075} \\
   & Newton &  \textbf{0.7789 +- 0.0349} &  \textbf{0.7616 +- 0.0137} &           0.6813 +- 0.0232 &  \textbf{0.8264 +- 0.0164} &           0.8798 +- 0.0084 \\
PR & Gradient &           0.7112 +- 0.0311 &            0.6921 +- 0.057 &           0.6381 +- 0.0477 &           0.7874 +- 0.0233 &            0.8806 +- 0.006 \\
   & Averaging &           0.8168 +- 0.0155 &           0.7583 +- 0.0335 &           0.6968 +- 0.0262 &           0.8419 +- 0.0117 &   \textbf{0.8908 +- 0.006} \\
   & Newton &  \textbf{0.8192 +- 0.0121} &   \textbf{0.7764 +- 0.016} &   \textbf{0.7062 +- 0.018} &  \textbf{0.8558 +- 0.0097} &           0.8865 +- 0.0056 \\
TR & Gradient &           0.8649 +- 0.0103 &           0.7871 +- 0.0103 &           0.7446 +- 0.0108 &             0.88 +- 0.0049 &  \textbf{0.9008 +- 0.0047} \\
   & Averaging &           0.8506 +- 0.0103 &            0.776 +- 0.0156 &           0.7311 +- 0.0111 &           0.8701 +- 0.0031 &            0.888 +- 0.0052 \\
   & Newton &  \textbf{0.8711 +- 0.0071} &  \textbf{0.7993 +- 0.0113} &  \textbf{0.7459 +- 0.0062} &  \textbf{0.8845 +- 0.0046} &            0.8999 +- 0.005 \\
\bottomrule
\end{tabular}
\end{table*}
\begin{table*}[ht]
  \small
  \caption{Weight update methods across the datasets fixing $\epsilon=1$\label{tab:weight_update_eps=1}}
\begin{tabular}{ccccccc}
\toprule
   &        &                       Bank &                   Credit 1 &                   Credit 2 &                      adult &                      nomao \\
\midrule
Hist & Gradient &           0.6407 +- 0.0538 &            0.542 +- 0.0525 &           0.6288 +- 0.0506 &           0.6803 +- 0.0288 &            0.8492 +- 0.016 \\
   & Averaging &           0.7219 +- 0.0275 &           0.6132 +- 0.0543 &           0.6831 +- 0.0191 &           0.6597 +- 0.0294 &  \textbf{0.8881 +- 0.0053} \\
   & Newton &  \textbf{0.8024 +- 0.0158} &  \textbf{0.7647 +- 0.0124} &  \textbf{0.6837 +- 0.0217} &  \textbf{0.8332 +- 0.0204} &           0.8812 +- 0.0079 \\
PR & Gradient &           0.7459 +- 0.0295 &           0.7301 +- 0.0379 &            0.6426 +- 0.037 &            0.8196 +- 0.008 &            0.885 +- 0.0056 \\
   & Averaging &           0.8247 +- 0.0151 &           0.7508 +- 0.0521 &             0.69 +- 0.0218 &           0.8546 +- 0.0065 &  \textbf{0.8923 +- 0.0059} \\
   & Newton &  \textbf{0.8362 +- 0.0155} &   \textbf{0.7855 +- 0.017} &  \textbf{0.7098 +- 0.0146} &  \textbf{0.8631 +- 0.0076} &           0.8918 +- 0.0055 \\
TR & Gradient &            0.8676 +- 0.008 &           0.7835 +- 0.0141 &           0.7445 +- 0.0088 &           0.8855 +- 0.0031 &  \textbf{0.9019 +- 0.0043} \\
   & Averaging &           0.8508 +- 0.0116 &           0.7757 +- 0.0178 &           0.7296 +- 0.0127 &            0.869 +- 0.0057 &           0.8884 +- 0.0054 \\
   & Newton &   \textbf{0.878 +- 0.0047} &  \textbf{0.8048 +- 0.0065} &  \textbf{0.7539 +- 0.0076} &  \textbf{0.8893 +- 0.0033} &            0.901 +- 0.0055 \\
\bottomrule
\end{tabular}
\end{table*}

\begin{figure*}[t]
  \centering
  \includegraphics[width=0.6\linewidth]{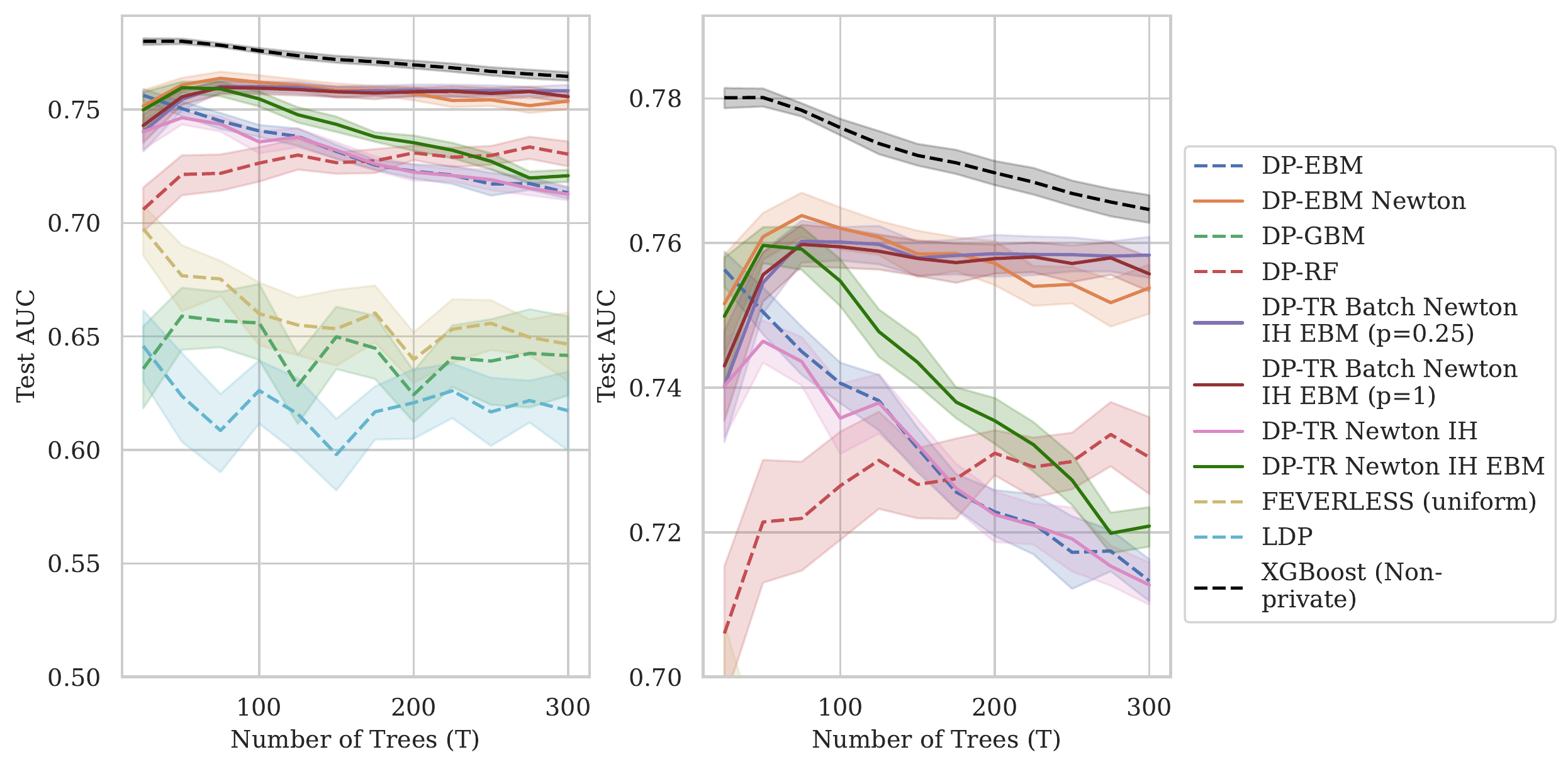}
  \caption{Comparison of methods: Credit 2 $(d=4, \epsilon=1)$}
\end{figure*}

\begin{figure*}[t]
  \centering
  \includegraphics[width=0.6\linewidth]{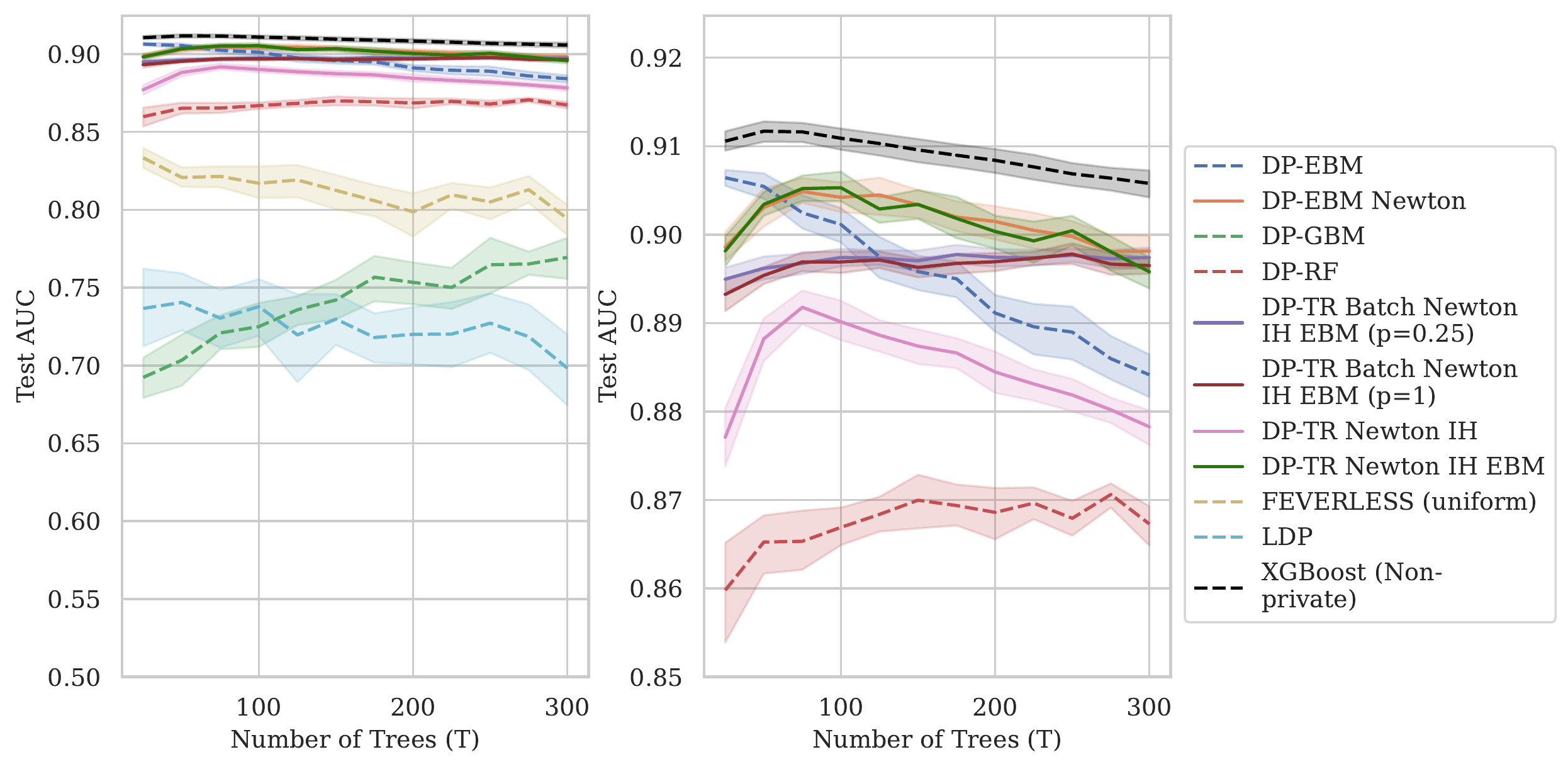}
  \caption{Comparison of methods: Adult $(d=4, \epsilon = 1)$}
\end{figure*}

\begin{figure*}[t]
  \centering
  \includegraphics[width=0.47\linewidth]{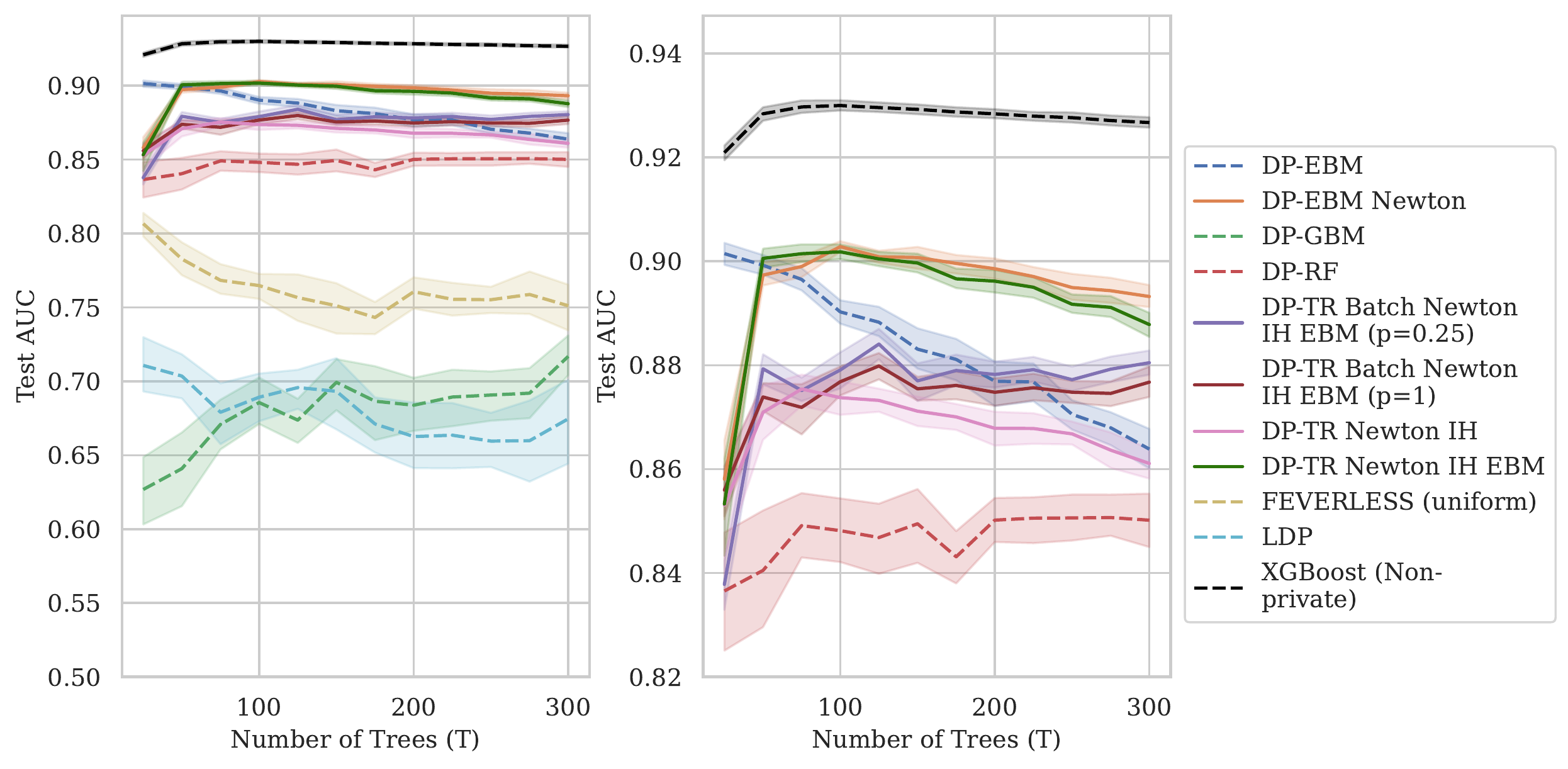}
  \caption{Comparison of methods: Bank $(d=4, \epsilon =1)$}
\end{figure*}

\end{document}